\documentclass[11pt]{article}
\usepackage{amsfonts}
\usepackage[dvipsnames]{xcolor}
\usepackage{dsfont}
\usepackage{amssymb}
\usepackage{amsmath}
\usepackage{amsxtra}
\usepackage{graphicx}
\usepackage{psfrag}
\usepackage{mathrsfs}
\usepackage{multirow}
\usepackage[sort, numbers]{natbib}
\usepackage{stmaryrd}
\usepackage{subfigure}
\usepackage{tikz}
\usepackage{empheq}
\usepackage[normalem]{ulem}
\usepackage{pdfpages}
\usepackage{color}
\usepackage{longtable}
\usepackage{footmisc}
\usepackage{hyperref}
\usepackage{lineno}

\begin{document}
\def\BGamma{\mbox{\boldmath$\Gamma$}}
\def\BDelta{\mbox{\boldmath$\Delta$}}
\def\BTheta{\mbox{\boldmath$\Theta$}}
\def\Btheta{\mbox{\boldmath$\theta$}}
\def\BLambda{\mbox{\boldmath$\Lambda$}}
\def\BXi{\mbox{\boldmath$\Xi$}}
\def\BPi{\mbox{\boldmath$\Pi$}}
\def\BSigma{\mbox{\boldmath$\Sigma$}}
\def\BUpsilon{\mbox{\boldmath$\Upsilon$}}
\def\BPhi{\mbox{\boldmath$\Phi$}}
\def\BPsi{\mbox{\boldmath$\Psi$}}
\def\Btheta{\mbox{\boldmath$\Omega$}}
\def\Balpha{\mbox{\boldmath$\alpha$}}
\def\Bbeta{\mbox{\boldmath$\beta$}}
\def\Bgamma{\mbox{\boldmath$\gamma$}}
\def\Bdelta{\mbox{\boldmath$\delta$}}
\def\Bepsilon{\mbox{\boldmath$\epsilon$}}
\def\Bzeta{\mbox{\boldmath$\zeta$}}
\def\Beta{\mbox{\boldmath$\eta$}}
\def\Btheta{\mbox{\boldmath$\theta$}}
\def\Biota{\mbox{\boldmath$\iota$}}
\def\Bkappa{\mbox{\boldmath$\kappa$}}
\def\Blambda{\mbox{\boldmath$\lambda$}}
\def\Bmu{\mbox{\boldmath$\mu$}}
\def\Bnu{\mbox{\boldmath$\nu$}}
\def\Bxi{\mbox{\boldmath$\xi$}}
\def\Bpi{\mbox{\boldmath$\pi$}}
\def\Brho{\mbox{\boldmath$\rho$}}
\def\Bsigma{\mbox{\boldmath$\sigma$}}
\def\Btau{\mbox{\boldmath$\tau$}}
\def\Bupsilon{\mbox{\boldmath$\upsilon$}}
\def\Bphi{\mbox{\boldmath$\phi$}}
\def\Bchi{\mbox{\boldmath$\chi$}}
\def\Bpsi{\mbox{\boldmath$\psi$}}
\def\Bomega{\mbox{\boldmath$\omega$}}
\def\Bvarepsilon{\mbox{\boldmath$\varepsilon$}}
\def\Bvartheta{\mbox{\boldmath$\vartheta$}}
\def\Bvarpi{\mbox{\boldmath$\varpi$}}
\def\Bvarrho{\mbox{\boldmath$\varrho$}}
\def\Bvarsigma{\mbox{\boldmath$\varsigma$}}
\def\Bvarphi{\mbox{\boldmath$\varphi$}}
\def\bone{\mbox{\boldmath$1$}}
\def\bzero{\mbox{\boldmath$0$}}
\def\bnabla{\mbox{\boldmath$\nabla$}}
\def\bvarepsilon{\mbox{\boldmath$\varepsilon$}}
\def\bA{\mbox{\boldmath$ A$}}
\def\bB{\mbox{\boldmath$ B$}}
\def\bC{\mbox{\boldmath$ C$}}
\def\bD{\mbox{\boldmath$ D$}}
\def\bE{\mbox{\boldmath$ E$}}
\def\bF{\mbox{\boldmath$ F$}}
\def\bG{\mbox{\boldmath$ G$}}
\def\bH{\mbox{\boldmath$ H$}}
\def\bI{\mbox{\boldmath$ I$}}
\def\bJ{\mbox{\boldmath$ J$}}
\def\bK{\mbox{\boldmath$ K$}}
\def\bL{\mbox{\boldmath$ L$}}
\def\bM{\mbox{\boldmath$ M$}}
\def\bN{\mbox{\boldmath$ N$}}
\def\bO{\mbox{\boldmath$ O$}}
\def\bP{\mbox{\boldmath$ P$}}
\def\bQ{\mbox{\boldmath$ Q$}}
\def\bR{\mbox{\boldmath$ R$}}
\def\bS{\mbox{\boldmath$ S$}}
\def\bT{\mbox{\boldmath$ T$}}
\def\bU{\mbox{\boldmath$ U$}}
\def\bV{\mbox{\boldmath$ V$}}
\def\bW{\mbox{\boldmath$ W$}}
\def\bX{\mbox{\boldmath$ X$}}
\def\bY{\mbox{\boldmath$ Y$}}
\def\bZ{\mbox{\boldmath$ Z$}}
\def\ba{\mbox{\boldmath$ a$}}
\def\bb{\mbox{\boldmath$ b$}}
\def\bc{\mbox{\boldmath$ c$}}
\def\bd{\mbox{\boldmath$ d$}}
\def\be{\mbox{\boldmath$ e$}}
\def\bff{\mbox{\boldmath$ f$}}
\def\bg{\mbox{\boldmath$ g$}}
\def\bh{\mbox{\boldmath$ h$}}
\def\bi{\mbox{\boldmath$ i$}}
\def\bj{\mbox{\boldmath$ j$}}
\def\bk{\mbox{\boldmath$ k$}}
\def\bl{\mbox{\boldmath$ l$}}
\def\bm{\mbox{\boldmath$ m$}}
\def\bn{\mbox{\boldmath$ n$}}
\def\bo{\mbox{\boldmath$ o$}}
\def\bp{\mbox{\boldmath$ p$}}
\def\bq{\mbox{\boldmath$ q$}}
\def\br{\mbox{\boldmath$ r$}}
\def\bs{\mbox{\boldmath$ s$}}
\def\bt{\mbox{\boldmath$ t$}}
\def\bu{\mbox{\boldmath$ u$}}
\def\bv{\mbox{\boldmath$ v$}}
\def\bw{\mbox{\boldmath$ w$}}
\def\bx{\mbox{\boldmath$ x$}}
\def\by{\mbox{\boldmath$ y$}}
\def\bz{\mbox{\boldmath$ z$}}
\newcommand*\mycirc[1]{%
  \begin{tikzpicture}
    \node[draw,circle,inner sep=1pt] {#1};
  \end{tikzpicture}
}
\providecommand{\keywords}[1]{\textbf{\textit{Keywords:}} #1}
\newcommand{\upcite}[1]{\textsuperscript{\textsuperscript{\cite{#1}}}}
\newcommand{\kg}[1]{\textcolor{red}{\textbf{(kg)} #1}}
\newcommand{\zh}[1]{\textcolor{ForestGreen}{\textbf{(zh)} #1}}
\newcommand{\xh}[1]{\textcolor{blue}{\textbf{(xh)} #1}}          \makeatletter
\newcommand{\res}[1]{\textcolor{black}{ #1}} 
\def\@biblabel#1{#1.}
\makeatother
\title{Variational system identification of the partial differential equations governing microstructure evolution in materials: Inference over sparse and spatially unrelated data}
\author{Z. Wang\thanks{Department of Mechanical Engineering, University of Michigan}, X. Huan\thanks{Department of Mechanical Engineering, Michigan Institute for Computational Discovery \& Engineering, University of Michigan, University of Michigan} and K. Garikipati\thanks{Departments of Mechanical Engineering, and Mathematics, Michigan Institute for Computational Discovery \& Engineering, University of Michigan, corresponding author, {\tt krishna@umich.edu}}}
\maketitle

\begin{abstract}
Pattern formation is a widely observed phenomenon in diverse fields including materials physics, developmental biology and ecology, among many others. The physics underlying the patterns is specific to the mechanisms, and is encoded by partial differential equations (PDEs). With the aim of discovering hidden physics, we have previously presented a variational approach to identifying such systems of PDEs in the face of noisy data at varying fidelities (\emph{Computer Methods in Applied Mechanics and Engineering}, \textbf{353}:201-216, 2019). Here, we extend our variational system identification methods to address the challenges presented by image data on microstructures in materials physics. PDEs are formally posed as initial and boundary value problems over combinations of time intervals and spatial domains whose evolution is either fixed or can be tracked. However, the vast majority of microscopy techniques for evolving microstructure in a given material system deliver micrographs of pattern evolution over domains  that bear no relation with each other at different time instants. The temporal resolution can rarely capture the fastest time scales that dominate the early dynamics, and noise abounds. Furthermore, data for evolution of the same phenomenon in a material system may well be obtained from different physical specimens. Against this backdrop of spatially unrelated, sparse and multi-source data, we exploit the variational framework to make judicious choices of weighting functions and identify PDE operators from the dynamics. A consistency condition arises for parsimonious inference of a minimal set of the spatial operators at steady state. It is complemented by a confirmation test that provides a sharp condition for acceptance of the inferred operators. The entire framework is demonstrated on synthetic data that reflect the characteristics of the experimental material microscopy images.
\end{abstract}

\begin{keywords}
Inverse problems, system identification, pattern formation, incomplete data 
\end{keywords}

\section{Introduction}
\label{sec:intro}

Pattern-formation is ubiquitous in many branches of the physical sciences. It occurs prominently in material microstructures driven by diffusion, reaction and phase transformations, and is revealed by a range of microscopy techniques that delineate the components or constituent phases. 
In developmental biology, examples include the organization of cells in the early embryo, markings on animal coats, insect wings, plant petals and leaves, as well as the segregation of cell types during the establishment of tissues. In ecology, patterns are formed on larger scales as types of vegetation spreads across forests. For context, we briefly discuss the role of pattern forming systems of equations in these phenomena. Pattern formation during phase transformations in materials physics can happen as the result of instability-induced bifurcations from a uniform composition \cite{Jiangetal2016,Rudrarajuetal2016,Teichertetal2017}, which was the original setting of the Cahn-Hilliard treatment \cite{CahnHilliard1958}. Following Alan Turing's seminal work on reaction-diffusion systems \cite{Turing1952}, a robust literature has developed on the application of nonlinear versions of this class of partial differential equations (PDEs) to model pattern formation in developmental biology \cite{Gierer1972,Murray1981,Dillon1994,Barrio1999,Barrio2009,MainiByrne2012,Spill2015,Korvasova2015,GarikipatiJMPS2017}. The Cahn-Hilliard phase field equation has also been applied to model other biological processes with evolving fronts, such as tumor growth and angiogenesis \cite{Wise2008,Cristini2009,Lowengrub2010,Lowengrub2009,Vilanova2013,Vilanova2014,Oden2010,Xu2016}. Reaction-diffusion equations also appear in ecology, where they are more commonly referred to as activator-inhibitor systems, and are found to underlie large scale patterning \cite{HilleRisLambers2001,Rietkerk2008}. All of these pattern forming systems fall into the class of nonlinear, parabolic PDEs, and have spawned a vast literature in mathematical physics. They can be written as systems of first-order dynamics driven by a number of time-independent terms of algebraic and differential form. The spatio-temporal, differentio-algebraic operators act on either a composition (normalized concentration) or an order parameter. It also is common for the algebraic and differential terms to be coupled across multiple species.

Patterns in physical systems are of interest because, up to a point, human experts in each of the above fields (materials microscopists, developmental biologists and ecologists) are able to identify phenomena solely on the basis of patterns. This success of intuition fed by experience does, however, break down when, for instance, the materials scientist is confronted by the dynamic processes in an unstudied alloy, or the developmental biologist considers a previously neglected aspect of morphogenesis. In such settings the challenge is to discover the operative physics from among a range of mechanisms. As with all of quantitative physics, the only rigorous route to such discovery is the mathematical one. For systems that vary over space and time, this description is in the form of PDEs. Identification of participating PDE operators from spatio-temporal observations can thus uncover the underlying physical mechanisms, and lead to improved understanding, prediction and manipulation of these systems.

Distinct from classical adjoint-based approaches to inverse problems, the system identification problem on PDEs is to balance accuracy and complexity in finding the best model among many possible candidates that could explain the spatio-temporal data on the dynamical system. The proposed models can be parameterized by the coefficients of candidate operators, which serve as a basis. The task of solving this inverse problem can then be posed as
finding the best coefficient values that allow a good agreement between model predictions and data.
Sparsity of these coefficients is further motivated by the principle that parsimony of physical mechanisms is favored for applicability to wider regimes of initial and boundary value problems. The comparison procedure between models and data may be set up via the following two broad approaches. (a) When only sparse and noisy data are available, which might also be of indirect quantities that do not explicitly enter the PDEs, quantifying uncertainty in the identification becomes important and a Bayesian statistical approach is very useful.
(b) When relatively larger volumes of observed data are accessible for quantities that directly participate in the PDEs, regression-based methods seeking to minimize an appropriate loss function can be highly efficient.

The first (Bayesian) approach has been successfully used for inferring a fixed number of coefficients in specified PDE models, leveraging efficient sampling algorithms such as Markov Chain Monte Carlo~\cite{Various2011}. However, they generally require many forward solutions of the PDE models, which is computationally expensive and may quickly become impractical with growth in the number of PDE terms (i.e., dimension of the identification problem).
As a result, the use of Bayesian approaches for identifying the best model from a large candidate set remains challenging. 
If made practical, they can be advantageous in providing uncertainty information, and offering significant flexibility in accommodating the data, which could be sparse and noisy, only available over some subset of the domain, collected at a few time steps, and composed of multiple statistical measures, functionals of the solution, or other indirect {Quantities of Interests} (QoIs). In the second (regression) approach, the PDEs themselves have to be represented by the data, which can be achieved by constructing the operators either in strong form such as finite difference, as in the Sparse Identification of Nonlinear Dynamics (SINDy) approach \cite{KutzPNAS2015}, or in weak form built upon basis functions, as in the Variational System Identification (VSI) approach \cite{WangCMAME2019}. Both representations pose stringent requirements on the data for accurately constructing the operators. The first of these is the need for time series data that can be related to chosen spatial points either directly or by interpolation. This is essential for consistency with a PDE that is written in terms of spatio-temporal operators at defined points in space and instants in time. The second requirement is for data with sufficient spatial resolution to construct spatial differential operators of possibly high order. However, these regression approaches enjoy the advantage over Bayesian methods that repeated forward PDE solutions are not necessary. PDE operators have thus been successfully identified from a comprehensive library of candidates \cite{WangCMAME2019, KutzPNAS2015, KutzIEEE2016, KutzSCIADV2017}. In a different approach to solving inverse problems \cite{Raissi2019}, the strong form of a specified PDE is directly embedded in the loss function while training deep neural network representations of the solution variable. However, this approach depends on  data at high spatial and temporal resolution for successful training of the deep neural network representations of the solution variable. A perspective and comparison between these two approaches can be found in Ref.~\cite{TAML2020}. 
\res{Lastly, there has been growing interest in combining graph theory with data-driven physics discovery. These approaches include genetic programming algorithm for discovering governing equations without pre-specification of the operators~\cite{Ghanem2019}, and geometric learning for reconstructing the normal forms of equations using data collected from systems with unknown initial conditions and parameter values~\cite{YairE7865}.} 

A significant discordance can exist in the form of material microstructure datasets when juxtaposed against the underlying premise of all the above approaches. These experimental datasets, while corresponding to different times are also commonly collected over different spatial subdomains of a physical specimen at each instant. This includes scenarios where the spatial subdomains have no overlap and are unrelated to each other. Furthermore, while subject to the same processing conditions, they may well come from different physical specimens. This is because the experimental techniques for extracting data at a given time after specimen preparation (which can include mechanical, chemical and thermal steps) involve destructive processes including cutting and grinding of the specimen. After this procedure, the entire specimen is removed from further experimentation, and a new specimen needs to be created if data is to be collected at a different time. Both of these conditions essentially negate the foundational notion of a PDE describing the temporal evolution of quantities at chosen spatial points for a single instantiation of that initial and boundary value problem (IBVP). The spatial subdomains on which microscopy is conducted may represent only  small portions of entire specimens. Boundary data, if present, are also prey to the above loss of spatial localization over time. Finally, the effort of processing (specimen preparation, mechanical, chemical and heat treatment) to attain the desired kinetic rates and thermodynamic driving forces, and subsequently to obtain microscopy images, leads to sparsity of data in time; even tens of instants are uncommon.

In this communication, we extend our VSI
methods \cite{WangCMAME2019} to sparse and spatially unrelated data, motivated by the above challenges of experimental microscopy in materials physics. The goal remains to discover the physical mechanisms underlying patterns of diffusion, reaction and phase transformation in materials physics. We present novel advances that exploit the variational framework and a notion of similarity between \emph{snapshots} of data, which we make precise, to 
circumvent the loss of spatial relatedness so that temporally sparse data prove sufficient for inferring the kinetics. These methods are described in Section \ref{sec:incompdata} and demonstrated on synthetic data in Section \ref{sec:Example}. Furthermore, imposition
of consistency on the steady state versions of candidate PDEs opens the door to parsimonious choice of a minimal set of spatial operators. To complement this condition, we also present a confirmation test that is a sharp condition for acceptance of the inferred operators. The confirmation test is described in Section \ref{sec:three} and demonstrated on synthetic data in Section \ref{sec:data at steady state without noise}. Discussions and concluding remarks are presented in Section \ref{sec:concl}.

\section{Variational identification of PDE systems from sparse and spatially unrelated  data}
\label{sec:incompdata}

\subsection{The Galerkin weak form for system identification}
We first provide a brief discussion on the weak form of PDEs as used in this work.
We start with the general strong form for first-order dynamics written as
\begin{align}
   \frac{\partial C}{\partial t}-\Bchi\cdot\Btheta=0, 
   \label{eq:general_strongForm}
\end{align}
where $\Bchi$ is a vector containing all possible linearly independent terms expressed as algebraic and differential operators on the scalar solution $C$:
\begin{align}
\Bchi=[1, C, C^2,...,\nabla^2 C,...],
\end{align}
and $\Btheta$ is the vector of scalar coefficients of these operators. For example, using this nomenclature, the one-field diffusion reaction equation
\begin{align}
\frac{\partial C}{\partial t}-D\nabla^2 C-f=0
\end{align}
with constant diffusivity $D$ and reaction rate $f$ has
\begin{align}
\Bchi=[1, C, C^2,\nabla^2 C]\hspace{2em} \mathrm{and} \hspace{2em}
\Btheta=[f,0,0,D].
\end{align}
Note that the time derivative $\partial C/\partial t$ is treated separately from the other terms in order to highlight the first-order dynamics of the problems that we target in this work. Given some observational data for the dynamical system, our system identification problem entails finding the correct governing equation via the coefficient vector, $\Btheta$, 
which specifies the active operators from among a dictionary of candidates, $\Bchi$.
 
As argued by Wang et al. \cite{WangCMAME2019}, the weak form offers particular advantages for the system identification problem. It allows the use of basis functions that can be drawn from function classes that have higher-order regularity.  Furthermore, the weak form transfers spatial derivative operators from the trial solution, represented by the data on $C$, to the weighting function. These two features significantly mollify the noise, stiffness and general loss of robustness associated with constructing spatial derivatives from data. They are especially relevant for the consideration of high-order gradient operators, which are often found in many types of pattern-forming phenomena in physics. Another advantage of writing the PDEs in weak form is that boundary conditions can be constructed as operators, thus making their identification a natural outcome. This weak form approach then leads to what we refer to as Variational System Identification \cite{WangCMAME2019}, abbreviated here to VSI. We note that, very recently, there has been growing interest in exploiting the weak form for system identification in numerical \cite{Reinbold2020PRE} as well as analytical settings \cite{Bortz2020weakSINDY,Bortz2020weakSINDYPDE}.

For infinite-dimensional problems with Dirichlet boundary conditions on $\Gamma^c$, the weak form is stated as: $\forall ~w \in \mathscr{V}$ where $\mathscr{V}= \{w\vert ~w = ~0 \;\mathrm{on}\;  \Gamma^c\}$, find $C \in \mathscr{S}$ where $\mathscr{S} = \{C\vert C = \overline{C} \text{ on } \Gamma^c\}$ such that
\begin{align}
\int_{\Omega}w\left(\frac{\partial C}{\partial t}-\Bchi\cdot\Btheta\right) \text{d}v=0.
\label{eq:weak_form}
\end{align}
Integration by parts and the application of boundary conditions, which are determined by the operators in $\Bchi$, leads to the final weak form. In the interest of brevity we have not written out the weak form for each instance of $\Bchi$. For finite-dimensional fields, $C^h$ and $w^h$, respectively, replace $C$ and $w$ in the statement of Equation (\ref{eq:weak_form}): $C^h\in \mathscr{S}^h \subset \mathscr{S}$ where $\mathscr{S}^h= \{ C^h \in \mathscr{H}^2(\Omega) ~\vert  ~C^h = ~\bar{C}\; \mathrm{on}\;  \Gamma^u\}$, and $w^h \in \mathscr{V}^h \subset \mathscr{V}$ where $\mathscr{V}^h= \{ w^h \in\mathscr{H}^2(\Omega)~\vert  ~w^h = ~0 \;\mathrm{on}\;  \Gamma^u\}$. The choice of $\mathscr{H}^2(\Omega)$ as the Sobolev space is motivated by the differential operators, which reach the highest order of two in the weak forms we consider (and order four in strong form). The variations $w^h$ and trial solutions $C^h$ are defined component-wise using a finite number of basis functions:
\begin{align}
w^h = \sum_{a=1}^{n_\mathrm{b}} d^a \phi^a \label{eq:basisw} \\
C^h = \sum_{a=1}^{n_\mathrm{b}} c^a \phi^a,
\label{eq:basisu}
\end{align}
\noindent where $n_\mathrm{b}$ is the dimensionality of the function spaces $\mathscr{S}^h$ and $\mathscr{V}^h$, and $\phi^a$ represents the basis functions. Equation (\ref{eq:basisu}) represents interpolation of discrete data to obtain fields with regularity specified by $\mathscr{H}^2$.

\begin{figure}[hbtp]
\centering
\vspace{-1cm}
\includegraphics[scale=0.15]{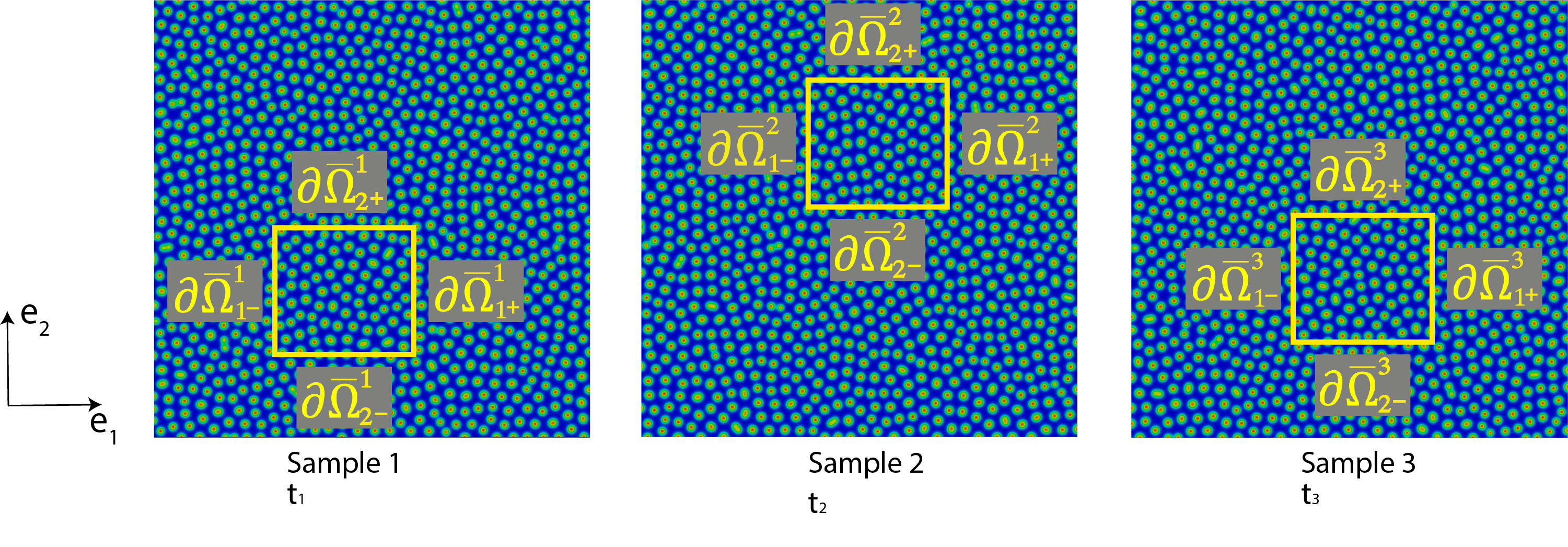}
\caption{The sample data shown within the snapshots (yellow square) corresponding to each time instant are only available over subdomains of the full field.  They are also spatially unrelated over time.}
\label{fig:sample}
\end{figure}

If, at each time instant of interest, $t$, the data available are in the form $C(\bx_I,t)$ such that $\bx_I \in \Omega$ for $I = 1,\dots,n_\text{sp}$, with $n_\text{sp}$ being the number of spatial points---that is, if data is available corresponding to discretization points for the entire spatial domain---then VSI as presented in its original framework
\cite{WangCMAME2019} presents itself as a powerful approach. However, the experimental characterization of microstructure evolution is typically carried out on physical samples that are cut out of larger wafers at chosen times, further processed for microscopy and then discarded. For this reason, the data acquired at each time are only available over subdomains of the full field that 
do not correspond to the same spatial locations---i.e., they are spatially unrelated (Figure \ref{fig:sample}), as explained in the Introduction.  For brevity we use the term \emph{snapshot} to refer to the data field over a chosen subdomain and at a given time. For experimental convenience, these snapshots are typically rectangular--a fact that we leverage for our methods. Furthermore, each such snapshot is usually produced from a different physical specimen due to the destructive and intrusive nature of data acquisition in the experiments. We next present a 
two-stage approach that accommodates these scenarios by exploiting the variational setting.

\subsection{Similarity of snapshots}
\label{sec:statsim}

The snapshots are chosen to be sufficiently distant from boundaries of the wafer so that they are representative of any other region, also similarly removed from the boundaries. If such rectangular snapshots (typically used for experimental characterization), say $\overline{\Omega}^1, \overline{\Omega}^2,\overline{\Omega}^3,\dots$ are identical in geometry  and subject to periodic boundary conditions on $\partial\overline{\Omega}^I_{i}$, where $I = 1,2,3,\dots$, and $i = 1,2$ denotes coordinate planes relative to which the boundaries $\partial\overline{\Omega}^I_{i}$ are perpendicular (see Figure \ref{fig:sample}), it follows that the field $C$ is identical within $\overline{\Omega}^1, \overline{\Omega}^2,\overline{\Omega}^3,\dots$. 
However, this condition requiring identical physical samples is too strong to impose on the experimental techniques. Instead, two weaker conditions are assumed to hold. The first states that for rectangular and identical snapshots, $\overline{\Omega}^I, \overline{\Omega}^J$, the mass influxes, $j_n$, integrated over the corresponding boundary subsets, $\partial\overline{\Omega}^I_{i^+}, \partial\overline{\Omega}^J_{i^+}$ or $\partial\overline{\Omega}^I_{i^-}, \partial\overline{\Omega}^J_{i^-}$, $i = 1,2$  respectively, are equal (see Figure \ref{fig:sample}):
\begin{equation}
    \int\limits_{\partial\overline{\Omega}^I_{i^\pm}}j_n \text{d}S = \int\limits_{\partial\overline{\Omega}^J_{i^\pm}}j_n \text{d}S,\quad i = 1,2.
    \label{eq:statsimboundary0}
\end{equation}
By choosing $\overline{\Omega}^I, \overline{\Omega}^J$ to be adjacent with a common boundary, say  $\partial\overline{\Omega}^I_{i^+}=\partial\overline{\Omega}^J_{i-}$, noting that the influx $j_n = -\boldsymbol{j}\cdot\bn$ for flux $\boldsymbol{j}$ and unit outward normal to a boundary being $\boldsymbol{n}$ we have :
\begin{equation}
   - \int\limits_{\partial\overline{\Omega}^I_{i+}} \boldsymbol{j}\cdot\boldsymbol{n}\text{d}S = -\int\limits_{\partial\overline{\Omega}^J_{i+}}\boldsymbol{j}\cdot\boldsymbol{n} \text{d}S = \int\limits_{\partial\overline{\Omega}^J_{i-}}\boldsymbol{j}\cdot\boldsymbol{n} \text{d}S,\quad i = 1,2.
    \label{eq:statsimboundary1}
\end{equation}
Where, by virtue of being the unit outward normal, $\boldsymbol{n}$ over $\partial\overline{\Omega}^I_{i^+}$ is opposite to $\boldsymbol{n}$ over $\partial\overline{\Omega}^J_{i^-}$. From the last two members of the above equation, we have:
\begin{equation}
   \int\limits_{\partial\overline{\Omega}^J_{i}} \boldsymbol{j}\cdot\boldsymbol{n}\text{d}S = \int\limits_{\partial\overline{\Omega}^J_{i+}}\boldsymbol{j}\cdot\boldsymbol{n} \text{d}S + \int\limits_{\partial\overline{\Omega}^J_{i-}}\boldsymbol{j}\cdot\boldsymbol{n} \text{d}S = 0,\quad i = 1,2.
    \label{eq:statsimboundary12}
\end{equation}
This means that the total mass fluxes integrated over opposite boundaries are equal and opposite, thus ensuring that the total mass flowing in over one boundary exits over the opposite one. This implies the vanishing of mass flux upon integration over the boundary
\begin{equation}
    \int\limits_{\partial\overline{\Omega}^I}\boldsymbol{j}\cdot\boldsymbol{n}\text{d}S = 0
    \label{eq:statsimboundary2}
\end{equation}
for an arbitrary snapshot $\overline{\Omega}^I$.

As the second condition we assume that volume-averaged $k^\text{th}$ powers of the composition field are equal between the rectangular snapshots:
\begin{equation}
    \frac{1}{\overline{V}^1}\int\limits_{\overline{\Omega}^1} C^k \text{d}V =\frac{1}{\overline{V}^2}\int\limits_{\overline{\Omega}^2} C^k \text{d}V = \frac{1}{\overline{V}^3}\int\limits_{\overline{\Omega}^3} C^k \text{d}V = \dots
    \label{eq:statsimregion}
\end{equation}
where $\overline{V}^I = \text{vol}(\overline{\Omega}^I)$.  In practice, Equations (\ref{eq:statsimboundary2}) and (\ref{eq:statsimregion}) are expected to hold approximately, so that
\begin{equation}
    \left\vert\,\,\int\limits_{\overline{\Omega}^I}\boldsymbol{j}\cdot\boldsymbol{n} \text{d}S\,\right\vert \le \varepsilon_\text{b},\quad i = 1,2
    \label{eq:statsimboundaryeps}
\end{equation}
and
\begin{equation}
    \left\vert\frac{1}{\overline{V}^I}\int\limits_{\overline{\Omega}^I} C^k \text{d}V - \frac{1}{\overline{V}^J}\int\limits_{\overline{\Omega}^J} C^k \text{d}V\right\vert \le \varepsilon_\text{r}
    \label{eq:statsimregioneps}
\end{equation}
for $0 < \varepsilon_\text{b}, \varepsilon_\text{r}$ being bounded. Equations (\ref{eq:statsimboundaryeps}) and (\ref{eq:statsimregioneps}) represent the notion of similarity that is assumed between snapshots used for material characterization. Since the methods developed in this communication are concerned with such experiments, we use this notion of similarity in developing variational system identification techniques in the following section. Additionally, Section \ref{sec:statistical similarity} demonstrates the approximate satisfaction of Equations (\ref{eq:statsimboundaryeps}) and (\ref{eq:statsimregioneps}) for the class of first-order PDEs that govern the evolution of microstructure in materials.

\subsection{Two stage variational system identification for dynamics}
\label{sec:Method of moments}
First, we rewrite Equation (\ref{eq:weak_form}) as an integral over $\Omega_n \subseteq\Omega$; a subset representing the snapshot over which data was acquired at time $t_n$. Additionally, we split operators $\Bchi$ and coefficients $\Btheta$ into two parts such that $\Bchi\cdot\Btheta = \Bchi_1\cdot\Btheta_1+\Bchi_2\cdot\Btheta_2$. The entire equation is furthermore normalized by $V_n = \text{vol}(\Omega_n)$:
\begin{align}
\frac{1}{V_n}\int_{\Omega_n}w\left(\frac{\partial C}{\partial t}-\left(\Bchi_1\cdot\Btheta_1+\Bchi_2\cdot\Btheta_2\right)\right) \text{d}v=0.
\label{eq:weak_form_2}
\end{align}
Here, the components of $\Bchi_1$ contain all the algebraic operators and those of $\Bchi_2$ contain all the differential operators:
\begin{align}
\Bchi_1=[1, C, C^2,...],\quad
\Bchi_2=[\nabla^2C, \nabla\cdot C\nabla C,\nabla^4 C,...],
\end{align}
with $\Btheta_1$ and $\Btheta_2$ being the corresponding coefficient vectors. We then use a backward difference approximation of the dynamics, write $\Bchi_2\cdot\Btheta_2$ in divergence form as $\Bchi_2\cdot\Btheta_2 = \nabla\cdot(\widehat{\Bchi}_2\cdot\widehat{\Btheta}_2)$ and apply integration by parts to arrive at:
\begin{align}
\frac{1}{V_n} \int_{\Omega_n} w\frac{C_{n}-C_{n-1}}{\Delta t}\text{d}v=\frac{1}{V_n}\left(\int_{\Omega_n} w\Bchi_{1_n} \text{d}v\cdot \Btheta_1- \int_{\Omega_n}\nabla w\cdot \widehat{\Bchi}_{2_n} \text{d}v\cdot \widehat{\Btheta}_2+ \int_{\partial\Omega_n}w\widehat{\Bchi}_{2_n}\cdot\bn \text{d}s\cdot\widehat{\Btheta}_2\right),
\label{eq:weak_form_3}
\end{align}
where $(\bullet)_n$ denotes the time discrete approximation of a field at time $t_n$. While other time-marching schemes are certainly also possible, we use backward differencing here for its simplicity, stability, and as a representative choice for illustration purposes when the focus of our problem is not on time-marching specifically. While, with noise-free data, the time discrete approximation grows more accurate for smaller time steps, noise could be amplified by the time discretization and overwhelm the true value \cite{WangCMAME2019}. Larger time steps can prevent this dominance of the noise.

We define Stage 1 by choosing $w =1$, yielding:
\begin{align}
\frac{1}{V_n}\int_{\Omega_n}C_{n}\text{d}v= \frac{1}{V_n}\Delta t\int_{\Omega_n}\Bchi_{1_n} \text{d}v\cdot \Btheta_1+\frac{1}{V_n}\int_{\Omega_n}C_{n-1}\text{d}v+\frac{1}{V_n}\Delta t\int_{\partial\Omega_n}\widehat{\Bchi}_{2_n}\cdot\bn \text{d}s\cdot\widehat{\Btheta}_2
\label{eq:moment_1_BC}
\end{align}
The left hand-side is the mean composition field. The $\widehat{\Bchi}_2$ operators vanish in volume integrands since, for a constant weighting function, $\nabla w = 0$. From Equation (\ref{eq:statsimboundaryeps}) the boundary term on the right satisfies
\begin{equation}
 \frac{1}{V_n}\Delta t\int_{\partial\Omega_n}\widehat{\Bchi}_{2_n}\cdot\bn \text{d}s\cdot\widehat{\Btheta}_2 \le \frac{\varepsilon_\text{b}\Delta t}{V_n},
 \label{eq:boundflux}
\end{equation}
and for small enough $\varepsilon_\text{b}$ can be neglected relative to the other terms of order $\Delta t/V_n$. See Section \ref{sec:statistical similarity} for numerical evidence of this result for similarity of snapshots. As a consequence, only algebraic (non-differential) operators remain:
\begin{align}
\frac{1}{V_n}\int_{\Omega_n}C_{n}\text{d}v= \frac{1}{V_n}\Delta t\int_{\Omega_n}\Bchi_{1_n} \text{d}v\cdot \Btheta_1+\frac{1}{V_n}\int_{\Omega_n}C_{n-1}\text{d}v.
\label{eq:moment_1}
\end{align}

For Stage 2, we choose $w=C_n$ over $\Omega_n$, to give:
\begin{align}
\frac{1}{V_n}\int_{\Omega_n}C_n\frac{\partial C_n}{\partial t}\text{d}v&=\frac{1}{V_n}\int_{\Omega_n}C_n\Bchi\cdot\Btheta \text{d}v\\
\Rightarrow \quad \frac{1}{V_n}\int_{\Omega_n}\frac{1}{2}\frac{\partial C_n^2}{\partial t}\text{d}v&=\frac{1}{V_n}\int_{\Omega_n}C_n\Bchi\cdot\Btheta \text{d}v\\
&=\frac{1}{V_n}\int_{\Omega_n}C_n\Bchi_1\cdot\Btheta_1\text{d}v+\frac{1}{V_n}\int_{\Omega_n}C_n\Bchi_2\cdot\Btheta_2 \text{d}v.
\end{align}
Again writing a backward difference approximation of the dynamics, using the divergence form as above and integrating by parts, we have:
\begin{align}
\frac{1}{V_n}\int_{\Omega_n}C^2_{n}\text{d}v&=\frac{1}{V_n}2\Delta t\left(\int_{\Omega_n}C_n\Bchi_{1_n}\text{d}v\cdot\Btheta_1- \int_{\Omega_n}\nabla C_n\cdot \widehat{\Bchi}_{2_n} \text{d}v\cdot \widehat{\Btheta}_2\right)\nonumber\\&\qquad+ \frac{1}{V_n}2\Delta t\int_{\partial\Omega_n}C_n\widehat{\Bchi}_{2_n}\cdot\bn \text{d}s\cdot\widehat{\Btheta}_2+\frac{1}{V_n}\int_{\Omega_n}C^2_{n-1}\text{d}v.
\label{eq:moment_2}
\end{align}
The left hand-side of Equation (\ref{eq:moment_2}) is the second power of the concentration field. Note that the boundary term is different than in Equation (\ref{eq:moment_1_BC}), and cannot be assumed to vanish. Before moving on with the formulation we emphasize that $\Omega_n\subseteq\Omega$ in Equations (\ref{eq:moment_1}) and (\ref{eq:moment_2}) can be the entire physical specimen, $\Omega$, or any subdomain. 

Relation (\ref{eq:statsimregioneps}) guarantees that the volume-normalized first and second powers of $C_n$ over a chosen subdomain can be approximated by the corresponding volume-normalized powers evaluated over different subdomains. We can thus write
%
\begin{align}
 \frac{1}{V_{i_n}}\int_{\Omega_{i_n}}C^k_{n}\text{d}v\approx \frac{1}{V_{j_n}}\int_{\Omega_{j_n}}C^k_{n}\text{d}v,\quad\text{where }\Omega_{i_n},\Omega_{j_n}\subseteq \Omega, \quad i,j = 1,2,\dots.
 \label{eq:assumption1}
\end{align}
%
%
Equation (\ref{eq:assumption1}) allows us to use data collected over different subdomains, $\Omega_{i_n}$, $i = 1,2,\dots$ at time $t_n$. This further allows us to estimate the volume-normalized powers by replacing the integrals of $C_{n-1}$ and $C^2_{n-1}$ over $\Omega_n$ in Equations (\ref{eq:moment_1}) and (\ref{eq:moment_2}), respectively, by integrals of the same integrands, but over $\Omega_{n-1}$:
\begin{align}
\frac{1}{V_{n}}\int_{\Omega_{n}} C_{n}\text{d}v&\approx \frac{1}{V_{n}}\Delta t\int_{\Omega_{n}}\Bchi_1 \text{d}v\cdot \Btheta_1+\frac{1}{V_{n-1}}\int_{\Omega_{n-1}}C_{n-1}\text{d}v
\label{eq:moment_1_asumption_1}\\
\frac{1}{V_{n}}\int_{\Omega_{n}}C^2_{n}\text{d}v&\approx\frac{1}{V_{n}}2\Delta t\left(\int_{\Omega_{n}}C_{n}\Bchi_1\text{d}v\cdot\Btheta_1- \int_{\Omega_{n}}\nabla C_{n}\cdot\widehat{\Bchi}_2 \text{d}v\cdot \widehat{\Btheta}_2\right)+\frac{1}{V_{n-1}}\int_{\Omega_{n-1}}C^2_{n-1}\text{d}v\nonumber\\
&\qquad + \frac{1}{V_n}2\Delta t\int_{\partial\Omega_n}C_n\widehat{\Bchi}_{2_n}\cdot\bn \text{d}s\cdot\widehat{\Btheta}_2
\label{eq:moment_2_asumption_1}
\end{align}
Thus, the data $C_k$ corresponding to a specific time instant, $t_k$, need only be considered over the snapshot, $\Omega_k$, observed at that time in an experiment.

Relation (\ref{eq:statsimregioneps}) also guarantees that the volume-normalized first and second powers of $C_n$ over sub-domain $\Omega_n$ at time $t_n$ can be approximated by the corresponding volume-normalized powers evaluated using data, $\widetilde{C}_n$, collected from sub-domains $\widetilde{\Omega}$ on \emph{different physical specimens} at time $t_n$. Here, the tilde denotes data from a different specimen. This reflects experimental practice in which data from different specimens are measured under the same conditions and physical time of the experiment, and similarity of snapshots is assumed. The mathematical and numerical equivalence to this procedure is modeled by employing the same initial condition with different spatially randomized perturbations. This is explained in detail in Section \ref{sec:Data_preparaiton}. We can then write
\begin{align}
\frac{1}{V_{n}}\int_{\Omega_n}C^k_{n}\text{d}v \approx\frac{1}{\widetilde{V}_{n}}\int_{\widetilde{\Omega}_{n}}\widetilde{C}^k_{n}\text{d}v.
 \label{eq:assumption2}
\end{align}
%
%

Relation (\ref{eq:statsimregioneps}) is a statement on the similarity of fields over different subdomains and physical specimens at the same time. The data collected from pattern forming phenomena better satisfies this approximation of similarity for larger subdomains because localized features are averaged out. The extent to which the  similarity implied in relations (\ref{eq:statsimboundaryeps}) and (\ref{eq:statsimregioneps}) is realized in the data is discussed in Section \ref{sec:statistical similarity}.
Replacing the approximations by equalities in our VSI algorithm, Equations (\ref{eq:moment_1_asumption_1}) and (\ref{eq:moment_2_asumption_1}) can be rewritten as: 
\begin{align}
&\text{Stage 1:}\nonumber\\
&\qquad\frac{1}{\Delta t }\left(\frac{1}{V_{n}}\int_{\Omega_n} C_{n}\text{d}v-\frac{1}{V_{n-1}}\int_{\Omega_{n-1}}C_{n-1}\text{d}v\right)= \frac{1}{V_{n}}\int_{\Omega_n}\Bchi_1 \text{d}v\cdot \Btheta_1 
\label{eq:linearsystem_1}\\
&\text{Stage 2:}\nonumber\\
&\qquad\frac{1}{2\Delta t }\left(\frac{1}{V_{n}}\int_{\Omega_n}C^2_{n}\text{d}v-\frac{1}{V_{n-1}}\int_{\Omega_{n-1}}C^2_{n-1}\text{d}v\right)-\frac{1}{V_{n}}\int_{\Omega_n}C_{n}\Bchi_1\text{d}v\cdot\Btheta_1=- \frac{1}{V_{n}}\int_{\Omega_n}\nabla C_{n}\cdot\widehat{\Bchi}_2 \text{d}v\cdot \widehat{\Btheta}_2\nonumber\\
&\qquad\qquad+ \frac{1}{V_n}\int_{\partial\Omega_n}C_n\widehat{\Bchi}_{2_n}\cdot\bn \text{d}s\cdot\widehat{\Btheta}_2
\label{eq:linearsystem_2}
\end{align}
where, following Equation (\ref{eq:assumption2}), the fields $C_n,C_{n-1}$ and sub-domains $\Omega_n,\Omega_{n-1}$ could come from any specimen, and we have dispensed with the tildes. Given data on $C$ at suitable times, we define the left hand-sides of Equations (\ref{eq:linearsystem_1}) and (\ref{eq:linearsystem_2}) to be the labels. In Section \ref{sec:ID-2Step} we show how they lead to target vectors in VSI, and the right hand-sides lead to matrices of candidate basis operators.

We recall that, in the weak form, the choice of the weighting function $w^h \in \mathscr{V}^h$ is arbitrary. However, the judicious choice of $w^h$ is critical for the success of VSI. A spatially constant $w^h$ eliminates volume integrals of differential operators in the weak form leaving only algebraic (non-differential) operators as volume terms in Stage 1. The choice of $w^h=C^h$ in Stage 2 turns differential operators, such as the Laplace and biharmonic operators into quadratic forms in the weak form (see Section \ref{sec:candidate basis operators} below). This is suitable for pattern forming dynamics, because it ensures that the integrand of, e.g. $\int_\Omega\nabla C\cdot\nabla C\text{d}v$ is non-negative, and that the corresponding weak operators are identifiable. However certain choices of $w^h$ could render some operators unidentifiable. For example, choosing $w^h=x_1$ (spatial coordinate), the integral  $\int_\Omega\nabla x_1\cdot\nabla C \text{d}v$ proves insignificant when evaluated over a (quasi-)periodic pattern. Contributions to this term are nearly cancelled out by the $x_1$-symmetry of the (quasi-)periodicity that is seen, for instance, in Figure \ref{fig:sample}.

\subsection{Candidate basis operators}
\label{sec:candidate basis operators}
Suppose we have data on certain degrees of freedom (DOFs) of a discretization, $C^h$, for a series of time steps. These DOFs are the nodal values in finite element methods, and the control variables if iso-geometric analytic methods are used. Using these DOFs, a Galerkin representation of the field could be constructed over the domain. Below we explain the procedure of generating candidate basis operators in weak form with three examples.

\subsubsection{Weak forms of polynomial approximations of algebraic operators at time $t_n$ with $w^h=1$}

\begin{align}
\chi^{C^k,1}_{1_n} \colon=\frac{1}{V_n}\int_{\Omega_n}(C_n^h)^k\text{d}v
\end{align}
The above equation represents all algebraic operators without loss of generality, since in principle they can all be expanded via polynomial expansions (e.g., Taylor series).  If specific forms are desired, for instance exponential terms that are common in reaction terms, they would be used directly. While $w^h=C^h$ (and higher powers) does not eliminate algebraic operators in Stage 2, $w^h = 1$ does eliminate differential operators from the volume integrals allowing robust identification of the algebraic operators in Stage 1.

\subsubsection{Weak form of the Laplacian operator at time $t_n$ with $w^h=C^h$}
Multiplying by the weighting function and integrating by parts, we have
\begin{equation}
\int_{\Omega_n}C^h_n\nabla^2 C^h_{n}\text{d}v=-\int_{\Omega_n}\nabla C^h_n\cdot\nabla C^h_{n}\text{d}v+\frac{1}{2}\int_{\partial\Omega_n}\nabla (C^h_n)^2\cdot\bn\text{d}s.
\end{equation}
We define 
\begin{align}
    \chi^{\nabla^2 C,C}_{2_n}=-\int_{\Omega_n}\nabla C^h_n\cdot\nabla C^h_{n}\text{d}v+\frac{1}{2}\int_{\partial\Omega_n}\nabla (C^h_n)^2\cdot\bn\text{d}s 
\end{align}
as the Laplacian in weak form with $w^h = C^h$.

\subsubsection{Weak form of the biharmonic operator $\nabla^4 C_n$ at time $t_n$ with $w^h=C^h$}
Multiplying by the weighting function and integrating by parts twice, we obtain
\begin{equation}
\int_{\Omega_n}w^h\nabla^4 C^h_{n}\text{d}v=\int_{\Omega_n} \nabla^2 C^h_{n} \nabla^2 C^h_{n}\text{d}v-\int_{\partial\Omega_n}\nabla C^h_{n}\cdot\bn \nabla^2 C^h_{n}\text{d}s +\int_{\partial\Omega_n}C^h_{n} \nabla(\nabla^2 C^h_{n})\cdot\bn\text{d}s.  \label{eq:weakBiharmonic3}
\end{equation}
We then define
\begin{align}
\chi^{\nabla^4 C,C}_{2_n}=\int_{\Omega_n} \nabla^2 C^h_{n} \nabla^2 C^h_{n}\text{d}v -\int_{\partial\Omega_n}\nabla C^h_{n}\cdot\bn \nabla^2 C^h_{n}\text{d}s +\int_{\partial\Omega_n}C^h_{n} \nabla(\nabla^2 C^h_{n})\cdot\bn\text{d}s
\label{eq:basis_Biharmonic}
\end{align}
as the biharmonic operator in weak form with $w^h = C^h$.

The second-order gradients in Equation (\ref{eq:basis_Biharmonic}) require the solutions and basis functions to lie in $\mathscr{H}^2(\Omega)$, while the Lagrange polynomial basis functions traditionally used in finite element analysis only lie in $\mathscr{H}^1(\Omega)$. We therefore draw the basis functions, $\phi^{a}$ in Equations (\ref{eq:basisw}) and (\ref{eq:basisu}), from the family of Non-Uniform Rational B-Splines (NURBS), and adopt Isogeomeric Analysis (IGA) in our simulations to find the solutions in  $\mathscr{H}^2(\Omega)$. A discussion of the NURBS basis and IGA is beyond the scope of this communication; interested readers are directed to the original works on this topic, such as Ref. \cite{CottrellHughesBazilevs2009} and references therein.


\subsection{Identification of basis operators via two stage stepwise regression}
\label{sec:ID-2Step}
To identify a dynamical system, we need to generate all possible basis operators, $\Bchi_1$ and $\Bchi_2$, acting on the solution, compute the non-zero coefficient for each basis operator that is in the model (active bases), while also attaining coefficients of zero for the basis operators that are not in the model (inactive bases).
We begin by putting together the right hand-sides of Equations (\ref{eq:linearsystem_1}) and (\ref{eq:linearsystem_2}) at each time $\{ \dots, t_{n-1}, t_n, t_{n+1},\dots\}$. Let 
\begin{align}
\by_1=\left[
\begin{array}{c}
\vdots\\
\frac{1}{\Delta t }\left(\frac{1}{V_{n-1}}\int_{\Omega_{n-1}} C_{n-1}\text{d}v-\frac{1}{V_{n-2}}\int_{\Omega_{n-2}}C_{n-2}\text{d}v\right)\\
\frac{1}{\Delta t }\left(\frac{1}{V_{n}}\int_{\Omega_n} C_{n}\text{d}v-\frac{1}{V_{n-1}}\int_{\Omega_{n-1}}C_{n-1}\text{d}v\right)\\
\frac{1}{\Delta t }\left(\frac{1}{V_{n+1}}\int_{\Omega_{n+1}} C_{n+1}\text{d}v-\frac{1}{V_{n}}\int_{\Omega_n}C_{n}\text{d}v\right)\\
\vdots\\
\end{array}
\right]
\label{eq:targety_1}
\end{align}
and 
\begin{align}
\by_2=\left[
\begin{array}{c}
\vdots\\
\frac{1}{2\Delta t }\left(\frac{1}{V_{n-1}}\int_{\Omega_{n-1}}C^2_{n-1}\text{d}v-\frac{1}{V_{n-2}}\int_{\Omega_{n-2}}C^2_{n-2}\text{d}v\right)-\frac{1}{V_{n-1}}\int_{\Omega_{n-1}}C\Bchi_1\text{d}v\cdot\Btheta_1\\
\frac{1}{2\Delta t }\left(\frac{1}{V_{n}}\int_{\Omega_n}C^2_{n}\text{d}v-\frac{1}{V_{n-1}}\int_{\Omega_{n-1}}C^2_{n-1}\text{d}v\right)-\frac{1}{V_{n}}\int_{\Omega_n}C\Bchi_1\text{d}v\cdot\Btheta_1\\
\frac{1}{2\Delta t }\left(\frac{1}{V_{n+1}}\int_{\Omega_{n+1}}C^2_{n+1}\text{d}v-\frac{1}{V_{n}}\int_{\Omega_n}C^2_{n}\text{d}v\right)-\frac{1}{V_{n+1}}\int_{\Omega_{n+1}}C\Bchi_1\text{d}v\cdot\Btheta_1\\
\vdots\\
\end{array}
\right]
\label{eq:targety_2}
\end{align}
be the target vectors formed as labels from the data at the two stages, respectively. Likewise, we form the respective matrices, $\BXi_1$ and $\BXi_2$, containing all possible operator bases:
\begin{align}
\BXi_1=\left[
\begin{array}{cccc}
\vdots&\vdots&\vdots&\vdots\\
\frac{1}{V_{n-1}}\int_{\Omega_{n-1}}1 \text{d}v&\frac{1}{V_{n-1}}\int_{\Omega_{n-1}}C_{n-1} \text{d}v&\frac{1}{V_{n-1}}\int_{\Omega_{n-1}}C^2_{n-1} \text{d}v &...\\
\frac{1}{V_{n}}\int_{\Omega_{n-1}} \text{d}v&\frac{1}{V_{n}}\int_{\Omega_n}C_{n} \text{d}v&\frac{1}{V_{n}}\int_{\Omega_n}C^2_{n} \text{d}v &...\\
\frac{1}{V_{n+1}}\int_{\Omega_{n+1}}1 \text{d}v&\frac{1}{V_{n+1}}\int_{\Omega_{n+1}}C_{n+1} \text{d}v&\frac{1}{V_{n+1}}\int_{\Omega_{n+1}}C^2_{n+1} \text{d}v &...\\
\vdots&\vdots&\vdots&\vdots
\end{array}
\right]
\label{eq:xi_1}
\end{align}


\begin{align}
\BXi_2=\left[\footnotesize
\begin{array}{ccc}
\vdots&\vdots&\vdots\\
\chi^{\nabla^2 C,C}_{2_{n-1}}&\chi^{\nabla^4 C,C}_{2_{n-1}} &...\\
\chi^{\nabla^2 C,C}_{2_{n}}&\chi^{\nabla^4 C,C}_{2_{n}} &...\\
\chi^{\nabla^2 C,C}_{2_{n+1}}&\chi^{\nabla^4 C,C}_{2_{n+1}} &...\\
\vdots&\vdots&\vdots
\end{array}
\right]
\label{eq:xi_2}
\end{align}
Then Equations (\ref{eq:linearsystem_1}) and (\ref{eq:linearsystem_2}) can be expressed as
\begin{align}
 \text{Stage 1:}\qquad   \by_1=\BXi_1\Btheta_1
    \label{eq:least-square_1}
\end{align}
and 
\begin{align}
 \text{Stage 2:}\qquad   \by_2(\Btheta_1)=\BXi_2\Btheta_2.
    \label{eq:least-square_2}
\end{align}
Equations (\ref{eq:least-square_1}) and (\ref{eq:least-square_2}) immediately deliver the two-stage algorithm of VSI, first for the coefficient vector $\Btheta_1$, and then for $\Btheta_2$ given the estimate of $\Btheta_1$. Thus splitting the process into two stages also alleviates the curse of dimensionality as a smaller number of operators needs to be considered at each stage for parsimonious identification.

The coefficient vectors $\Btheta_1$ and $\Btheta_2$ in Equations (\ref{eq:least-square_1}) and (\ref{eq:least-square_2}) can be solved by minimizing the squared-norm loss functions $l_1$ and $l_2$:
\begin{align}
\Btheta_1 &=\text{arg }\underset{\widetilde{\Btheta}_1}\min\; \left\{ l_1 := \left\|\, \by_1-\BXi_1\widetilde{\Btheta}_1 \,\right\|_{2}^2\right\}
\label{eq:OP1}\\
\Btheta_2 &=\text{arg }\underset{\widetilde{\Btheta}_2}\min\; \left\{ l_2 :=\left\|\, \by_2-\BXi_2\widetilde{\Btheta}_2 \,\right\|_{2}^2\right\},
\label{eq:OP2}
\end{align}
%
%
which have analytical solutions expressible via the pseudo-inverses:
\begin{align}
\Btheta_1&=(\BXi_1^T\BXi_1)^{-1}\BXi_1^T\by_1\\
\Btheta_2&=(\BXi_2^T\BXi_2)^{-1}\BXi_2^T\by_2.
\label{eq:sol_least-square}
\end{align}
For robustness in the presence  of noise and outliers, we use ridge regression to inject regularization. The optimization problems in Equations (\ref{eq:OP1}) and Equation (\ref{eq:OP2}) are then updated with additional penalty terms:
\begin{align}
\Btheta_1 &=\text{arg }\underset{\widetilde{\Btheta}_1}\min \text{ } \left\{l_1(\widetilde{\Btheta}_1) +\lambda_1 \left\|\, \widetilde{\Btheta}_1 \,\right\|_{2}^2 \right\} \label{eq:ridge1}\\
\Btheta_2 &=\text{arg }\underset{\widetilde{\Btheta}_2}\min \text{ } \left\{l_2(\widetilde{\Btheta}_2) +\lambda_2 \left\|\, \widetilde{\Btheta}_2 \,\right\|_{2}^2 \right\} \label{eq:ridge2}
\end{align}
where $\lambda_1$ and $\lambda_2$ are regularization hyperparameters, with solutions
\begin{align}
\Btheta_1&=(\BXi_1^T\BXi_1+\lambda_1\mathbf{I})^{-1}\BXi_1^T\by_1\\
\Btheta_2&=(\BXi_2^T\BXi_2+\lambda_2\mathbf{I})^{-1}\BXi_2^T\by_2.
\label{eq:sol_ridge}
\end{align}

However, performing a standard, one-shot ridge regression for the full $\Btheta$ vector will lead to a solution with nonzero contributions in all components of the vector. Such a solution of the system identification will be overfit to the data (and potentially to noise in the data), and lose parsimony by failing to sharply delineate the relevant bases. Another option is to use compressive sensing techniques \cite{Candes2006a,Donoho2006a} that employ $\ell_1$-regularization, instead. However, we found their performance for system identification to be highly sensitive to the selection of regularization hyperparameters, leading to VSI results that are similar to one-shot ridge regression for $\Btheta$. These observations then motivate us to take a different approach to iteratively eliminate the operators that have been identified to be inactive.

\subsubsection{Stepwise regression}
In this work, we use backward model selection by stepwise regression \cite{ISL}, which, as we have demonstrated, delivers parsimonious results with VSI \cite{WangCMAME2019}. The algorithm is summarized below.
\\

\begin{figure}[hbtp]
  \centering
\includegraphics[width=0.65\textwidth]{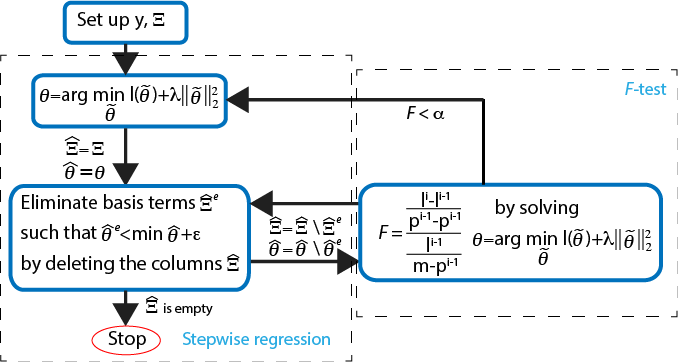}
\caption{Schematic of the algorithms for stepwise regression and the $F$-test. }
\label{fig:system_algorithm_flowchart}
\end{figure}

\noindent\fbox{%
\parbox{\textwidth}{%
\textbf{Algorithm 1: Model selection by Stepwise regression:}
\\
\\
\texttt{Step 0: Establish target vector $\by$ and matrix of bases $\BXi$.}
\vspace{0.25cm}

\texttt{Step 1: Solve $\Btheta^i$ by the linear regression problems, Equations (\ref{eq:least-square_1}) and (\ref{eq:least-square_2}), using ridge regression. Calculate the loss function at this iteration, $l^i$.}
\vspace{0.25cm}

\texttt{Step 2: 
Eliminate basis operators in matrix $\BXi$  by deleting their columns, using the $F$-test introduced below. Set to zero the corresponding components of $\Btheta^i$. GOTO Step 1. Note that at this stage the loss function remains small (\text{$l^{i}\sim l^{i-1}$}), and the solution may be overfit. }
\vspace{0.25cm}

\texttt{Step 3: The algorithm stops if the $F$-test does not allow elimination of any more basis operators. Beyond this, the loss function increases dramatically for any further reduction. }
}
}
\\
\\
 \res{Starting from a dictionary containing all potential relevant operators and while the residual remains small, we eliminate the inactive operators iteratively until the model becomes underfit as indicated by a drastic increase of residual norm. Since the model at iteration $i$ contains fewer bases than at iteration $i-1$ and is therefore more restrictive, the loss function, in general, is higher at iteration $i$ than at iteration $i-1$. In this sense, model always becomes ``worse'' after eliminating any basis. We want to determine whether the model at iteration $i$ is significantly worse than at iteration $i-1$. If so, the basis is not eliminated.} There are several choices for the criterion for eliminating basis terms. Here, we adopt a widely used statistical criterion called the $F$-test, also used by us previously \cite{WangCMAME2019}. The significance of the change between the model at iterations $i$ and $i-1$ is evaluated by:
\begin{align}
F=\frac{ \frac{l^i-l^{i-1}}{p^{i-1}-p^{i}}}{\frac{l^{i-1}}{m-p^{i-1}}}
\end{align}
where $p^i$ is the number of bases at iteration $i$ and $m$ is the total number of operator bases. Model selection is achieved through the following algorithm:
\\
\\
\noindent\fbox{%
\parbox{\textwidth}{%
\textbf{Algorithm 2: Model selection by the $F$-test:}
\\
\texttt{Step 1:}\\
\texttt{Tentatively eliminate the basis corresponding to coefficients in $\Btheta^i$ which are smaller than the pre-defined threshold. Evaluate the $F$ value followed by ridge regression on the reduced basis set.}
\vspace{0.25cm}

\texttt{Step 2:}\\
\texttt{IF $F<\alpha$}\\
\texttt{THEN formally eliminate these bases in matrix $\BXi$, by deleting the corresponding columns. GOTO Step 1.}\\
\texttt{ELSE GOTO Step 1, and choose another basis as a candidate for elimination.}
}
}
\\
\\
In this work we adopt the threshold:
\begin{align}
\text{threshold}^i&=\widehat{\theta}^i+\epsilon\label{eq:threshold}\\
\widehat{\theta}^i &= \text{arg }\underset{\{\theta^i_\alpha\}_{\alpha=1}^m}\min\; \vert\theta^i_\alpha\vert
\label{eq:thresholdomega}
\end{align}
where, as defined in Equation (\ref{eq:thresholdomega}) $\widehat{\theta}^i$ is the component with the smallest magnitude in ${\{\theta^i_\alpha\}_{\alpha=1}^m}$, and $\epsilon$ is a small tolerance.\footnote{\res{The $F$ test serves a similar purpose as the Pareto
analysis presented in~\cite{KutzPRS2017} to select parsimonious governing equations from a large set of potential models.}} 
The penalty coefficients in ridge regression, $\lambda_1$ and $\lambda_2$, are chosen to lie in the range $[10^{-10},10^{-1}]$ by leave-one-out cross-validation at each iteration. The hyperparameter $\alpha$ is chosen to lie in the range $[1,10]$, by five-fold cross validation.
We summarize the stepwise regression algorithm and $F$-test in Figure \ref{fig:system_algorithm_flowchart}.

\subsection{Variational System Identification with dynamic data}

\label{sec:Example}
We now use the framework detailed in the preceding section to identify the parabolic PDEs that govern pattern formation. To test our methods, we deploy them here to identify PDEs using data synthesized through high-fidelity, direct numerical simulations (DNS) from known ``true-solution" systems detailed in Section \ref{sec:Data_preparaiton}.
%
We consider test cases with the following two pattern formation models for data generation, with their true parameter values summarized in Table \ref{ta:parameters}: 

\noindent \textbf{Model 1:}
\begin{align}
\frac{\partial C_1}{\partial t}&=D_1\nabla^2C_1+R_{10}+R_{11}C_1+R_{13}C_1^2C_2\\
\frac{\partial C_2}{\partial t}&=D_2\nabla^2C_2+R_{20}+R_{21}C_1^2C_2\\
\text{with}& \quad \nabla C_1\cdot\bn=0; \quad \nabla C_2\cdot\bn=0 \text{ on }\Gamma 
\end{align}
where $\Gamma$ is the domain boundary.
Model 1 represents the coupled diffusion-reaction equations for two species following Schnakenberg kinetics \cite{Schnakenberg1976}, but with different boundary conditions. For an activator-inhibitor species pair, these equations use auto-inhibition with cross-activation of a short range species, and auto-activation with cross-inhibition of a long range species to form so-called Turing patterns \cite{Turing1952}.

\begin{figure}[hbtp]
\centering
\includegraphics[scale=0.2]{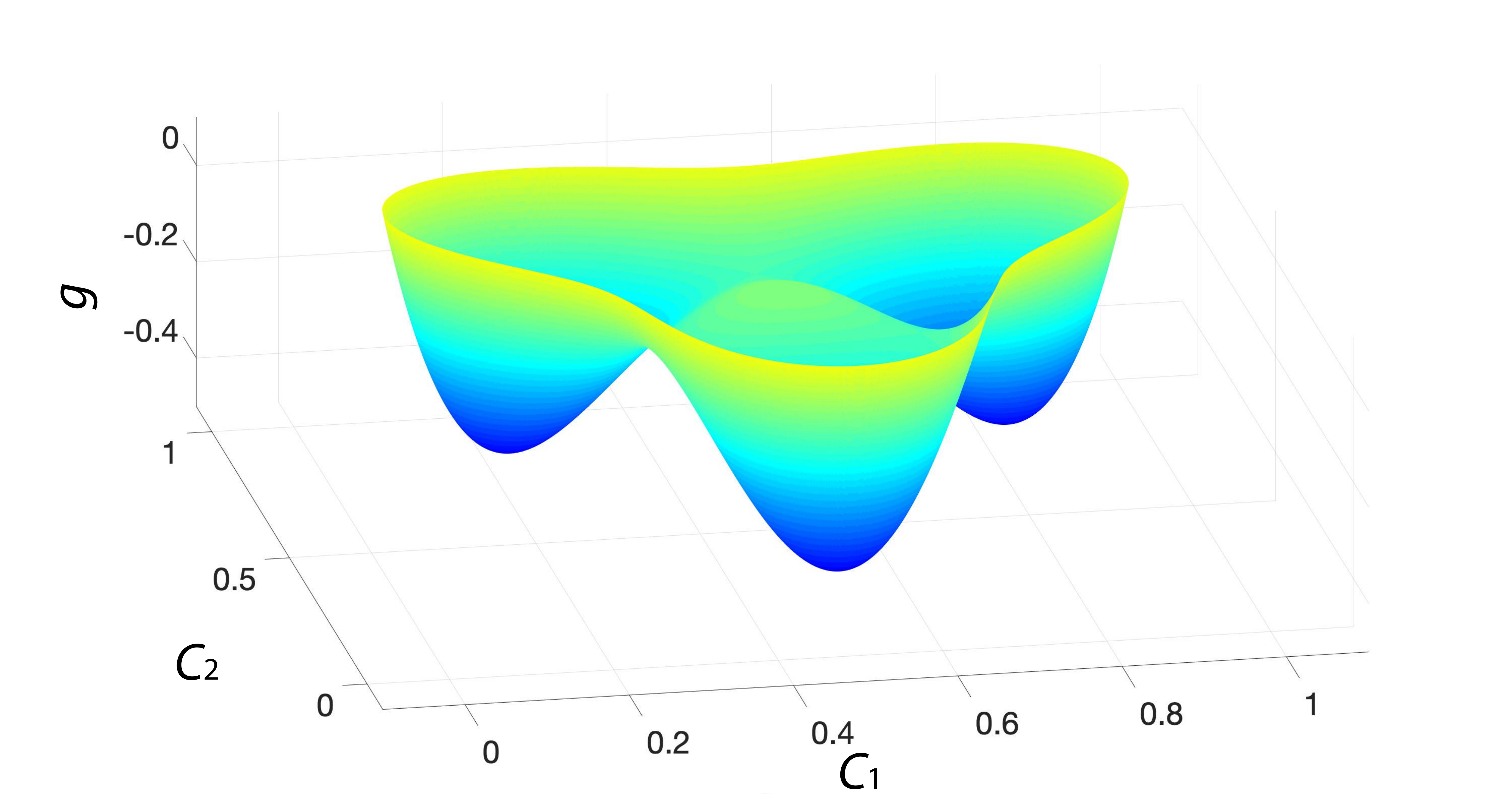}
\caption{The three-well non-convex tissue energy density function. }
\label{fig:free_energy}
\end{figure}

\noindent \textbf{Model 2:}
\begin{align}
  \frac{\partial C_1}{\partial t}&=\nabla \cdot (M_1\nabla\mu_1)\label{eq:Cahn-C1}\\
    \frac{\partial C_2}{\partial t}&=\nabla \cdot (M_2\nabla\mu_2)\label{eq:Cahn-C2}\\
\mu_1&=\frac{\partial g}{\partial C_1}-k_1\nabla^2C_1 \label{eq:Cahn-mu_1}\\
\mu_2&=\frac{\partial g}{\partial C_2}-k_2\nabla^2C_2 \label{eq:Cahn-mu_2}\\
 \text{with}& \quad \nabla \mu_1\cdot\bn=0; \quad \nabla C_1\cdot\bn = 0 \text{ on }\Gamma  \label{eq:CahnHillDirBC}\\
 & \quad \nabla \mu_2\cdot\bn=0; \quad \nabla C_2\cdot\bn = 0  \text{ on }\Gamma
\end{align}
where $g$ is a non-convex, ``homogeneous'' free energy density function, whose form has been chosen from Ref. \cite{GarikipatiJMPS2017}:
\begin{align}
   g(C_1,C_2)&=\frac{3d}{2s^4}\left((2C_1-1)^2+(2C_2-1)^2\right)^2+\frac{d}{s^3}(2C_2-1)\left((2C_2-1)^2-3(2C_1-1)^2\right)\nonumber\\
      &\quad-\frac{3d}{2s^2}\left((2C_1-1)^2+(2C_2-1)^2\right).
   \label{eq:freeEnergy_g}
 \end{align}
Model 2 is a two-field Cahn-Hilliard system with fourth-order terms in the concentrations, $C_1$ and $C_2$, which become apparent on substituting Equations (\ref{eq:Cahn-mu_1}) and (\ref{eq:Cahn-mu_2}) into (\ref{eq:Cahn-C1}) and (\ref{eq:Cahn-C2}), respectively. The three-well non-convex free energy density function (see Figure \ref{fig:free_energy}), $g(C_1,C_2)$, drives segregation of the system into two distinct phases. We have previously used this system to make connections with cell segregation in developmental biology \cite{GarikipatiJMPS2017}.
The diffusion-reaction equations occur widely in the context of nucleation and growth phenomena in materials physics. The Cahn-Hilliard equation \cite{CahnHilliard1958} also occupies a central role in the materials physics literature for modelling phase transformations developing from a uniform concentration field in the presence of a concentration instability.

\begin{table}[h]
\centering
 \begin{tabular}{|c|c|c|c|c|c|c|c|c|c|c|c|c|c|}
 \hline
$D_1$&$D_2$ & $R_{10}$ & $R_{11}$ & $R_{13}$ & $R_{20}$ &$R_{21}$ & $M_1$& $M_2$& $k_1$&$k_2$ &$d$ &$s$\\\hline
1&40&0.1&$-1$&1&0.9&$-1$& 0.1& 0.1& 10&10& 0.4& 0.7\\ \hline
\end{tabular}
\caption{True parameter values used in DNS for synthesizing test data.}
\label{ta:parameters}
\end{table}
Substituting the parameter values from Table \ref{ta:parameters}, we present the weak form of each model:

\noindent \textbf{Model 1:}
\begin{align}
\int_{\Omega}w_1\frac{\partial C_1}{\partial t}\text{d}v&=\int_{\Omega}-1\nabla w_1 \cdot\nabla C_1\text{d}v
+\int_{\Omega}w_1 (0.1-C_1+1C_1^2C_2)\text{d}v 
\label{eq:weak_value_model1-1}\\    
\int_{\Omega}w_2\frac{\partial C_2}{\partial t}\text{d}v&=\int_{\Omega}-40\nabla w_2\cdot\nabla C_2\text{d}v+\int_{\Omega}w_2(0.9-1C_1^2C_2)\text{d}v
\label{eq:weak_value_model1-2}
\end{align}

\noindent \textbf{Model 2:}
\begin{align}
\int_{\Omega}w_1\frac{\partial C_1}{\partial t}\text{d}v=&\int_{\Omega}\nabla w_1\cdot\left(-17.8126+47.98C_1+21.591C_2-47.98C_1^2-15.9933C_2^2 \right)\nabla C_1\text{d}v\nonumber\\
&+ \int_{\Omega}\nabla w_1\cdot\left(-10.7955+21.591C_1+15.9933C_2-31.9867C_1C_2 \right)\nabla C_2\text{d}v\nonumber\\
&+\int_{\Omega}-1\nabla^2w_1\nabla^2C_1\text{d}v
\label{eq:weak_value_model2-1}
\\
\int_{\Omega}w_2\frac{\partial C_2}{\partial t}\text{d}v=&\int_{\Omega}\nabla w_2\cdot\left(-10.7955+21.591C_1+15.9933C_2-31.9867C_1C_2 \right)\nabla C_1\text{d}v\nonumber\\
&+ \int_{\Omega}\nabla w_2\cdot\left(-12.2149+15.9933C_1+42.3823C_2-15.9933C_1^2-47.98C_2^2 \right)\nabla C_2\text{d}v\nonumber\\
&+\int_{\Omega}-1\nabla^2w_2\nabla^2C_2\text{d}v
\label{eq:weak_value_model2-2}
\end{align}

\subsubsection{Data preparation}
\label{sec:Data_preparaiton}
All computations have been implemented in the \texttt{mechanoChemIGA} code framework, a library for modeling mechano-chemical problems using isogeometric analytics, available at \url{https://github.com/mechanoChem/mechanoChem}. The IBVPs presented here are two-dimensional. Data generation by direct numerical simullation was carried out on uniformly discretized $400\times400$ meshes. Adaptive time stepping is used\footnote{The time step size lies in the range $\Delta t=[0.25,2]$ depending on the convergence of the solver. } with a direct solver. 
The initial conditions for all simulations are a constant value corrupted with random noise at each grid point:
\begin{align}
C_{1,i}=0.5+\delta_{1,i} \label{eq:ini_1}\\
C_{2,i}=0.5+\delta_{2,i} 
\label{eq:ini_2}
\end{align}
where $\delta_{1,i},\delta_{2,i}\sim \mathcal{U}(-0.01,0.01)$ are independent (across $i$), uniform, random variables at each nodal mesh point $i$.

\begin{figure}[hbtp]
\centering
\includegraphics[scale=0.53]{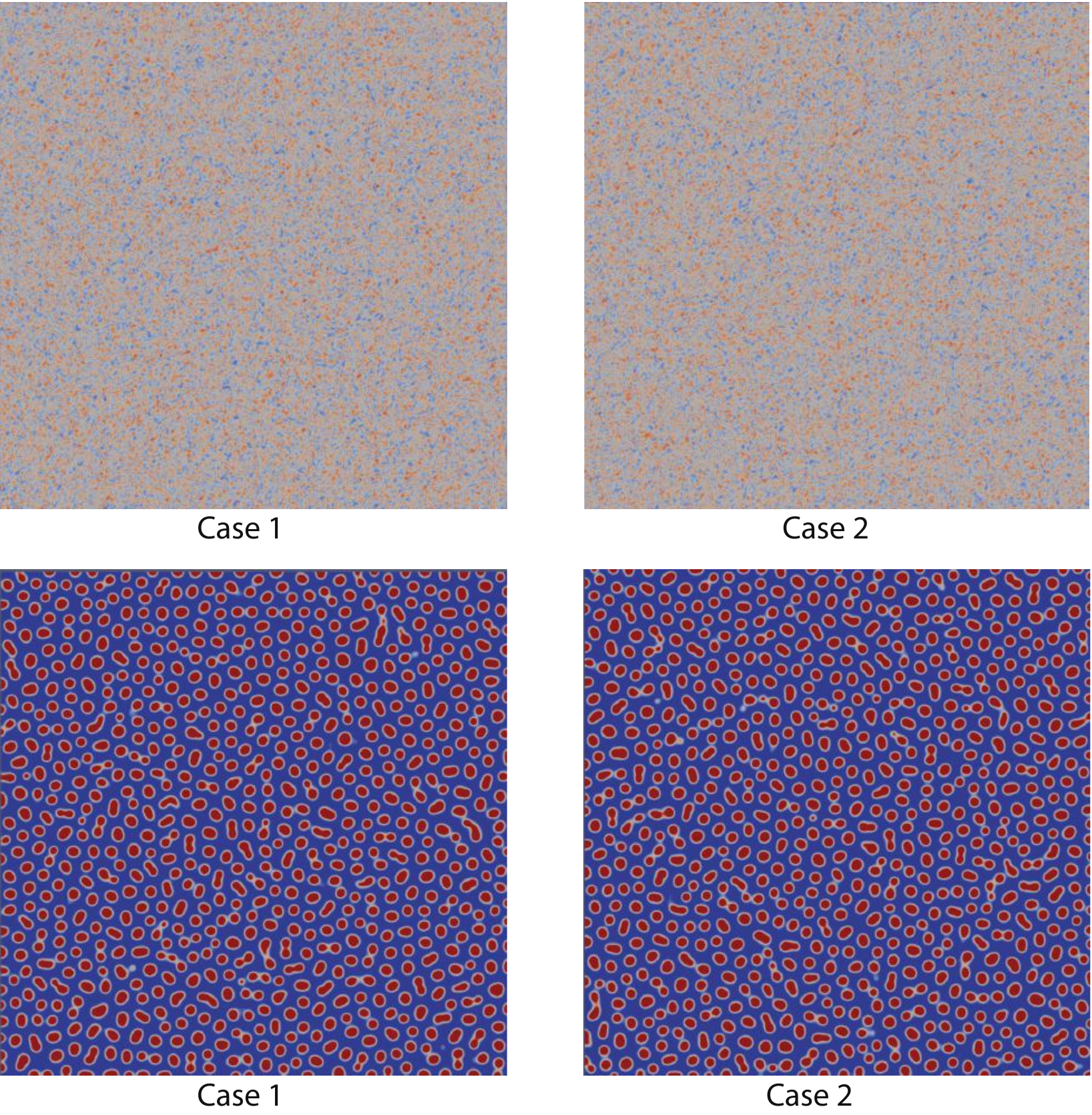}
\caption{Upper plots: Concentration at $t=0$ of two cases representing distinct experimental specimens. The perturbation is re-generated for every simulation. Lower plots: Concentration at $t=20$. The Turing patterns generated from different initial conditions appear statistically similar.}
\label{fig:diffusion_ini_pattern}
\end{figure}

To mimic the experimental data that would be necessarily collected from multiple physical specimens due to the destructive nature of the acquisition procedure, we run each model for 30 different simulations, each
with an independently and identically distributed (i.i.d.) realization of the initial condition in Equations (\ref{eq:ini_1}) and (\ref{eq:ini_2}) (e.g., see upper plots in Figure \ref{fig:diffusion_ini_pattern}). The patterns generated from separate simulations, with initial conditions randomized as above, are non-identical but statistically similar (e.g., see lower plots in Figure \ref{fig:diffusion_ini_pattern}). In this study we collected the data at 30 time steps, each over a spatially unrelated snapshot from a different simulation. Therefore, the total dataset of 30 snapshots does not correspond to a common initial condition or spatial subdomain.

The experimental data may also be noisy and of varying sparsity, often available only at low spatial resolution due to cost limitations or accessibility constraints of historical data. To mimic noisy data,
we superpose the synthetic solutions $C_1$ and $C_2$ with an i.i.d. Gaussian noise $\mathcal{N}(0,\sigma^2)$ at each grid point. To simulate scenarios with different data sparsity, we use the full $400\times 400$ domain as well as proportionally smaller
subdomains of $300\times 300$, $200\times 200$, $100\times 100$ and $50\times 50$ meshes. These proper subsets represent incomplete data and, as discussed, could possibly come from spatially unrelated and even non-overlapping subdomains at different times. Note that while each snapshot consists of
values from multiple grid points, a single snapshot yields only one ``data point''  due to the integration forms at the two stage of identification (See Equation \eqref{eq:linearsystem_1} and \eqref{eq:linearsystem_2}). 
%
Having clean and noisy snapshots at varying sparsity, from different simulations representing distinct specimens and unrelated, non-overlapping subdomains, we generate 34 candidate basis operators in addition to the time derivative terms, summarized in Table \ref{ta:basis}. \res{The construction of operator basis in weak form is quite inexpensive, equivalent to the finite element assembly process (even when considering a large set of candidate operators). Furthermore, the bases only need to be constructed once in the entire VSI procedure, and the subsequent regression steps do not ever need them to be reconstructed; this is because the bases are constructed from data which are given and fixed. The computational expense of Stepwise regression is equal to $\sum^N_i t_i$ where $N$ is the total iteration number and $t_i$ is the expense of regression problem at the $i$th iteration. $N$ increases linearly and $t_i$ increases roughly quadratically, for iterative solvers, relative to the size of the basis that was considered. As a result, the total computational expense increases roughly cubically. For the problems considered in this communication, VSI generally takes only minutes on a personal laptop. }

\begin{table}[h]
\centering
\caption{Candidate basis operators for model selection. Asterisks, ($\ast$) in the left column represent algebraic operators on $C_1$ and $C_2$. }
 \begin{tabular}{c|c}
 \hline
Type of basis& Basis in weak form\\\hline
\multirow{3}{*}{$\nabla(*\nabla C_1)$}& $\int_{\Omega}\nabla w\nabla C_1\text{d}v \quad \int_{\Omega}\nabla wC_1\nabla C_1\text{d}v \quad \int_{\Omega}\nabla w\nabla C_2\text{d}v$\\
     &$\int_{\Omega}\nabla wC_1^2\nabla C_1\text{d}v \quad \int_{\Omega}\nabla wC_1C_2\nabla C_1\text{d}v \quad \int_{\Omega}\nabla wC_2^2\nabla C_1\text{d}v$ \\
     &$\int_{\Omega}\nabla wC_1^3\nabla C_1\text{d}v \quad \int_{\Omega}\nabla wC_1^2C_2\nabla C_1\text{d}v \quad \int_{\Omega}\nabla wC_1C_2^2\nabla C_1\text{d}v \quad \int_{\Omega}\nabla wC_2^3\nabla C_1\text{d}v$\\
     \hline
\multirow{3}{*}{$\nabla(*\nabla C_2)$}& $\int_{\Omega}\nabla w\nabla C_2\text{d}v \quad \int_{\Omega}\nabla wC_1\nabla C_2\text{d}v \quad \int_{\Omega}\nabla w\nabla C_2\text{d}v$\\
     &$\int_{\Omega}\nabla wC_1^2\nabla C_2\text{d}v \quad \int_{\Omega}\nabla wC_1C_2\nabla C_2\text{d}v \quad \int_{\Omega}\nabla wC_2^2\nabla C_2\text{d}v$ \\
     &$\int_{\Omega}\nabla wC_1^3\nabla C_2\text{d}v \quad \int_{\Omega}\nabla wC_1^2C_2\nabla C_2\text{d}v \quad \int_{\Omega}\nabla wC_1C_2^2\nabla C_2\text{d}v \quad \int_{\Omega}\nabla wC_2^3\nabla C_2\text{d}v$\\
     \hline  
$\nabla^2(*\nabla^2 C)$& $\int_{\Omega}\nabla^2 w\nabla^2 C_1\text{d}v \quad \int_{\Omega}\nabla^2 wC_1\nabla^2 C_1\text{d}v \quad \int_{\Omega}\nabla^2 w\nabla^2 C_2\text{d}v \quad \int_{\Omega}\nabla^2 wC_2\nabla^2 C_2\text{d}v $\\
\hline
\multirow{3}{*}{non-gradient}& $-\int_{\Omega} w1\text{d}v \quad -\int_{\Omega} wC_1 \text{d}v \quad -\int_{\Omega} w C_2\text{d}v$\\
     &$-\int_{\Omega} wC_1^2\text{d}v \quad -\int_{\Omega} wC_1C_2\text{d}v \quad -\int_{\Omega} wC_2^2\text{d}v$ \\
     &$-\int_{\Omega} wC_1^3\text{d}v \quad -\int_{\Omega} wC_1^2C_2\text{d}v \quad -\int_{\Omega} wC_1C_2^2\text{d}v \quad -\int_{\Omega} wC_2^3\text{d}v$\\
     \hline
\end{tabular}
\label{ta:basis}
\end{table}

\subsubsection{Similarity between snapshots of data in pattern forming physics}
\label{sec:statistical similarity}
\begin{figure}[hbtp]
\centering
\subfigure[The total flux for $C_1$]{\includegraphics[scale=0.5]{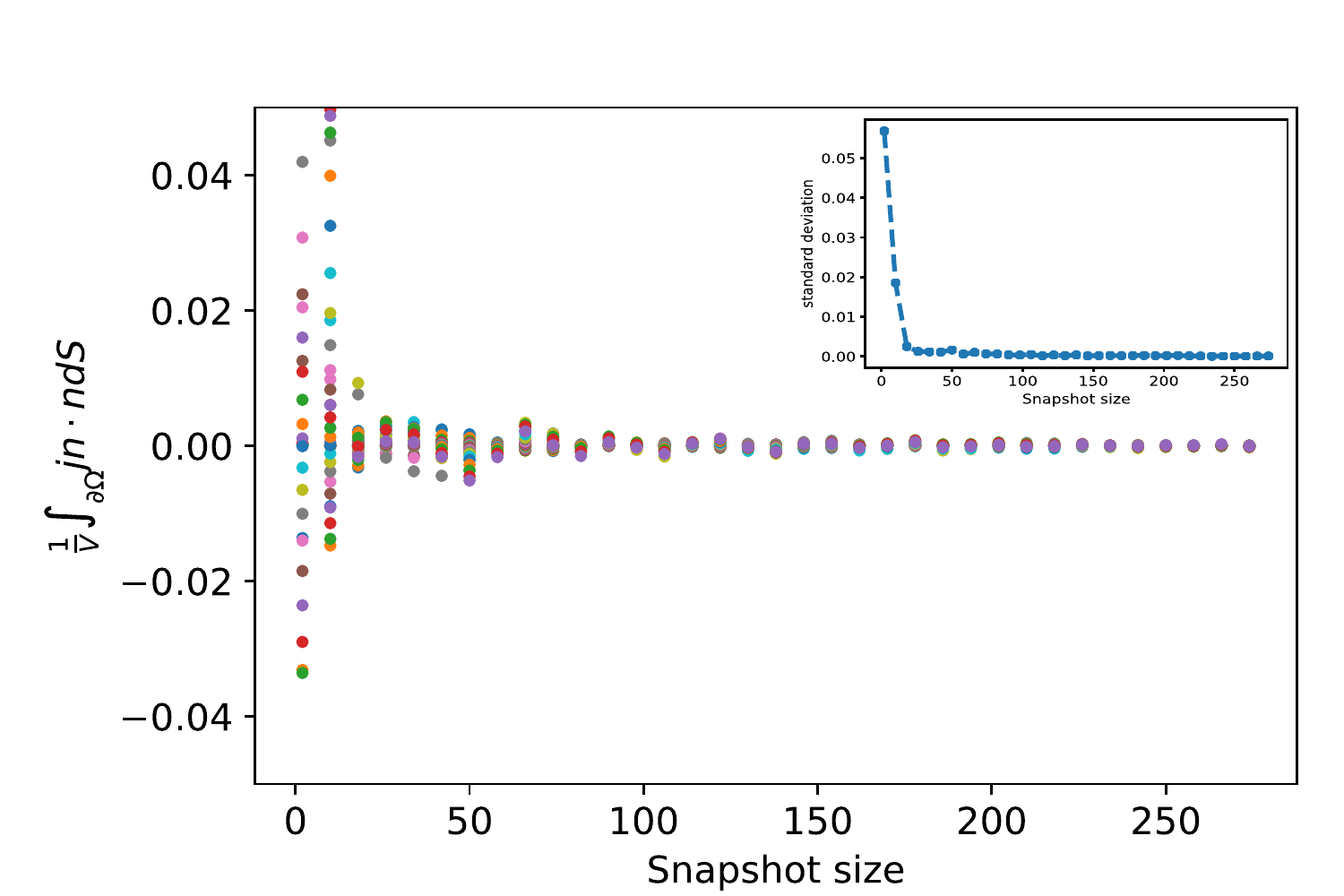} } 
\subfigure[The total flux for $C_2$]{\includegraphics[scale=0.5]{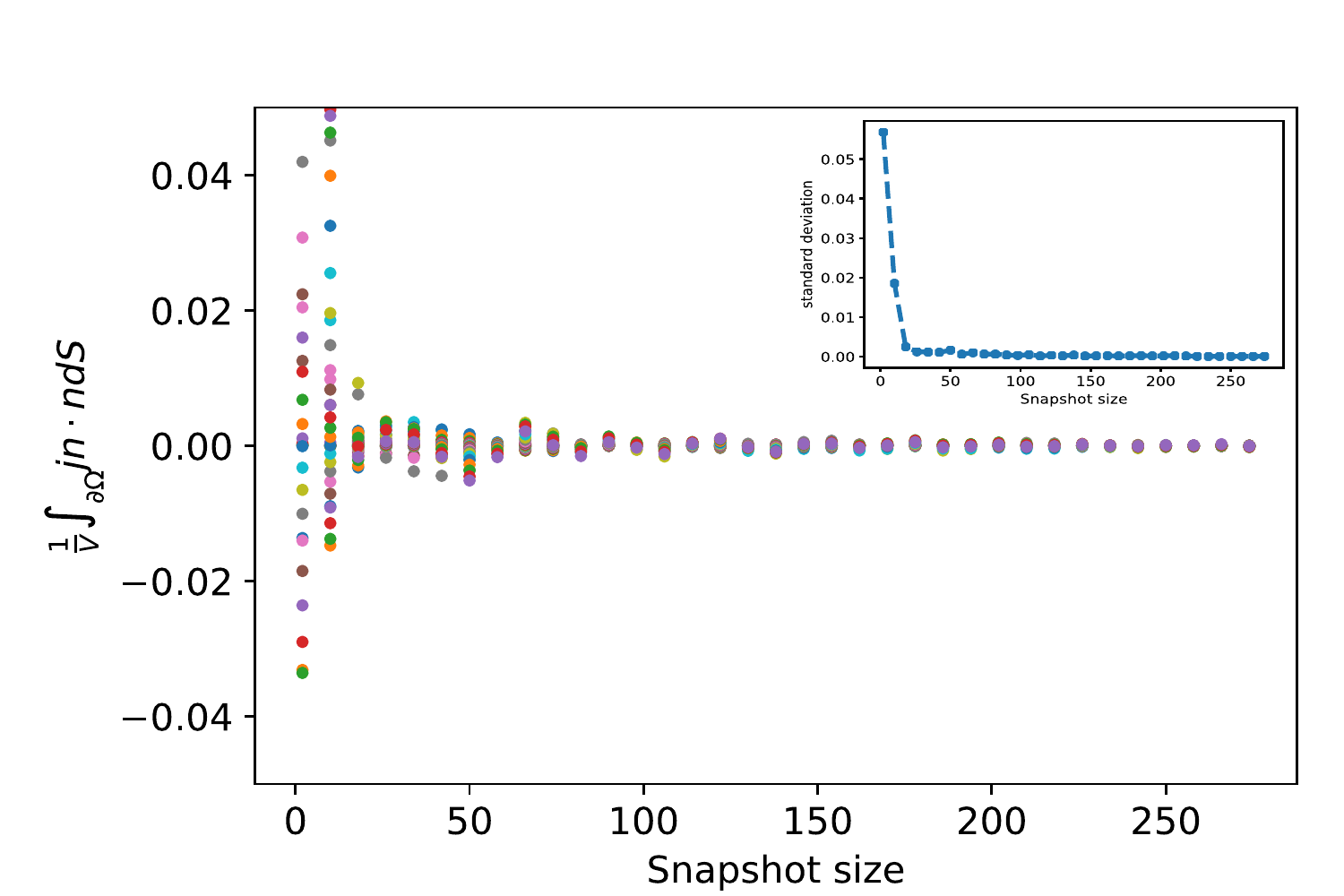} } 
\caption{The total flux over snapshots of different sizes 
marked by different colors. The total flux converges to zero with increasing snapshot size. The embedded subplot shows the decreasing standard deviation. }
\label{fig:zero_flux_DR}
\end{figure}

The similarity between snapshots of data arises due to the random distribution of features/particles in the patterns. Before identifying the governing PDEs, we examine the validity of Equations (\ref{eq:statsimboundaryeps}) and (\ref{eq:statsimregioneps}) arrived at from assumptions  in Section \ref{sec:statsim}. We first evaluate the total flux over the boundaries of snapshots with different sizes as shown in Figure \ref{fig:zero_flux_DR}. As snapshot size increases, the total flux, scaled by its volume, vanishes. This validates the relation (\ref{eq:statsimboundaryeps}) in the limit of large snapshots.

\begin{figure}[hbtp]
\centering
\subfigure[First power of $C_1$ ]{\includegraphics[scale=0.53]{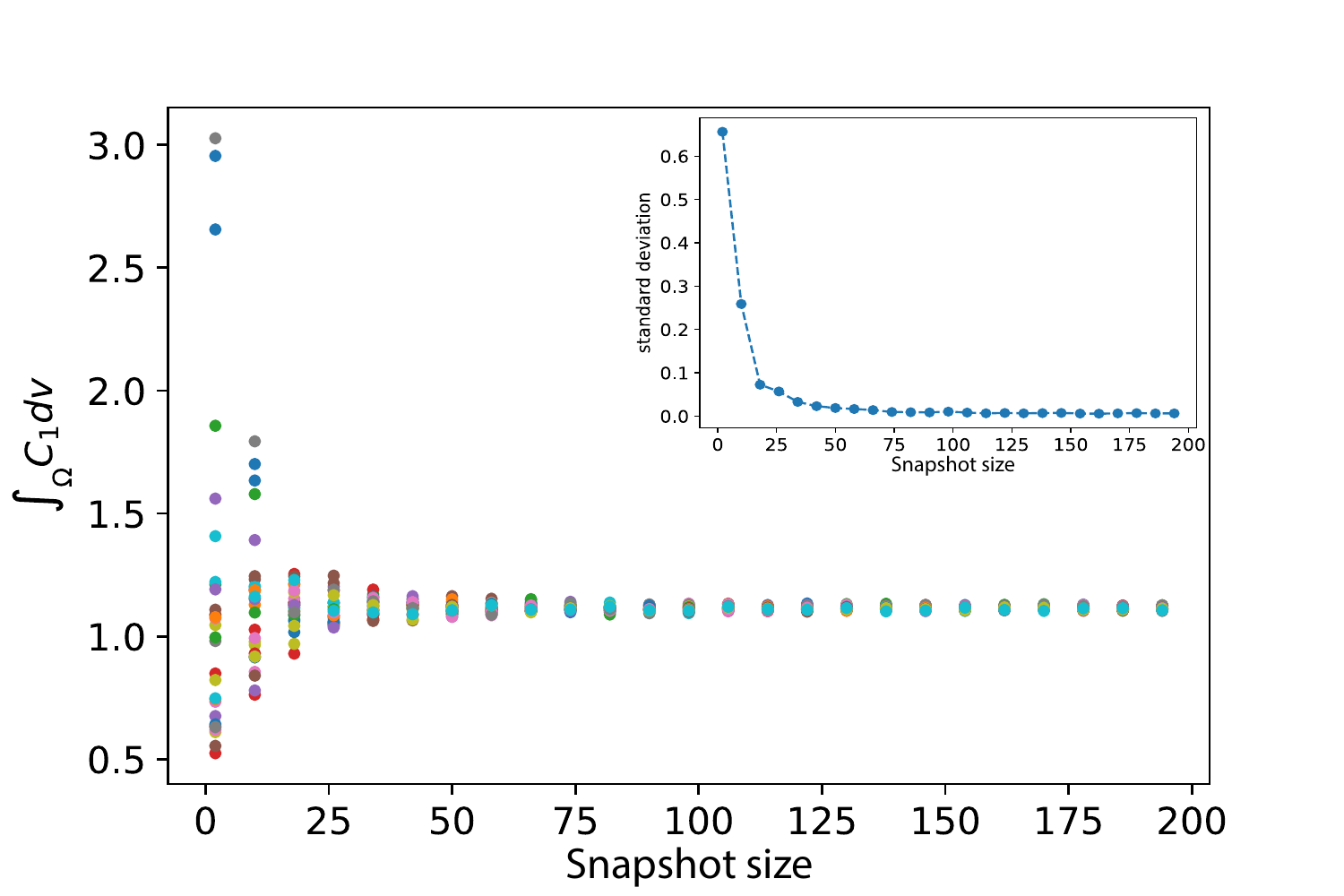}}
\subfigure[Second power of $C_1$ ]{\includegraphics[scale=0.53]{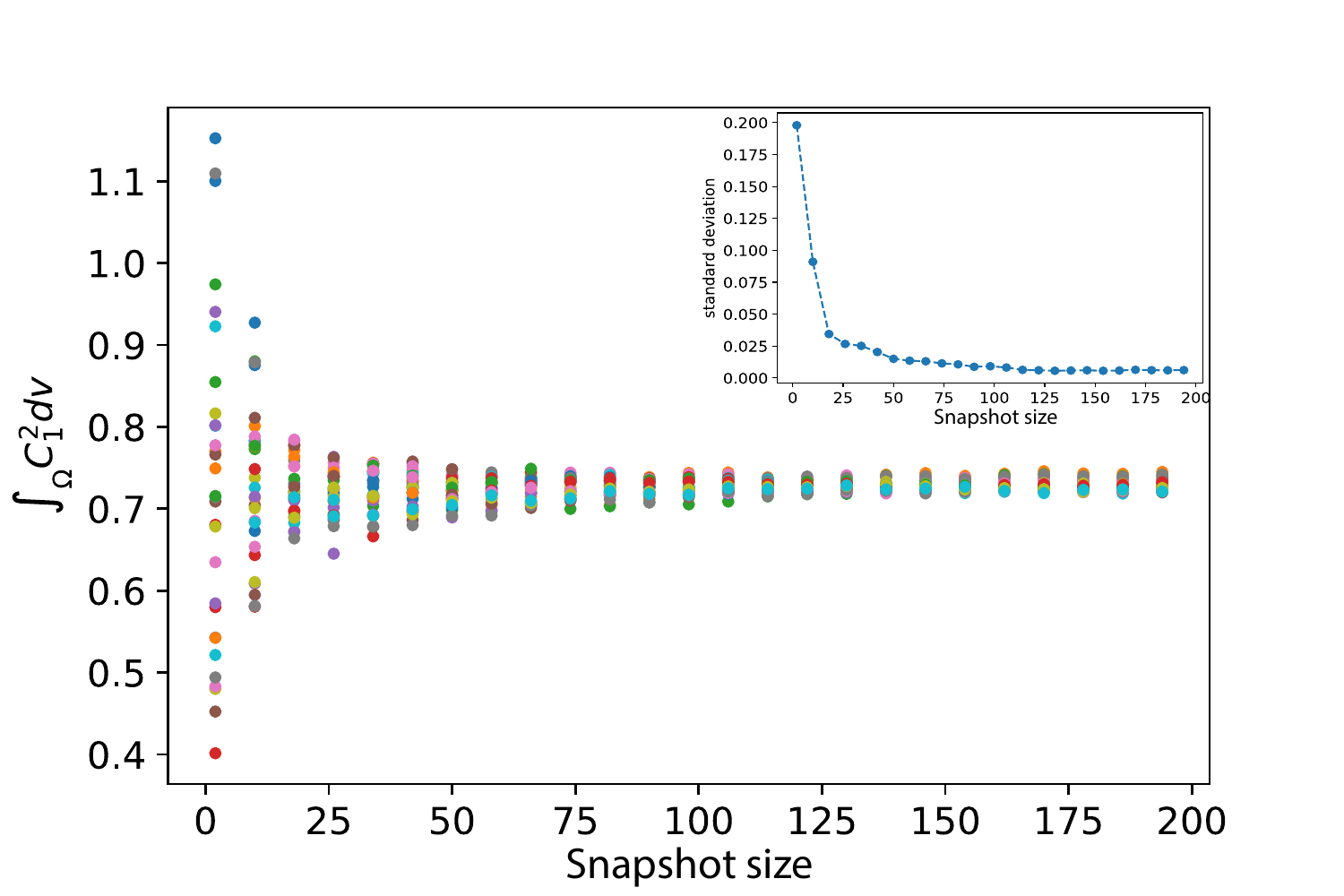}}
\subfigure[First power of $C_2$ ]{\includegraphics[scale=0.53]{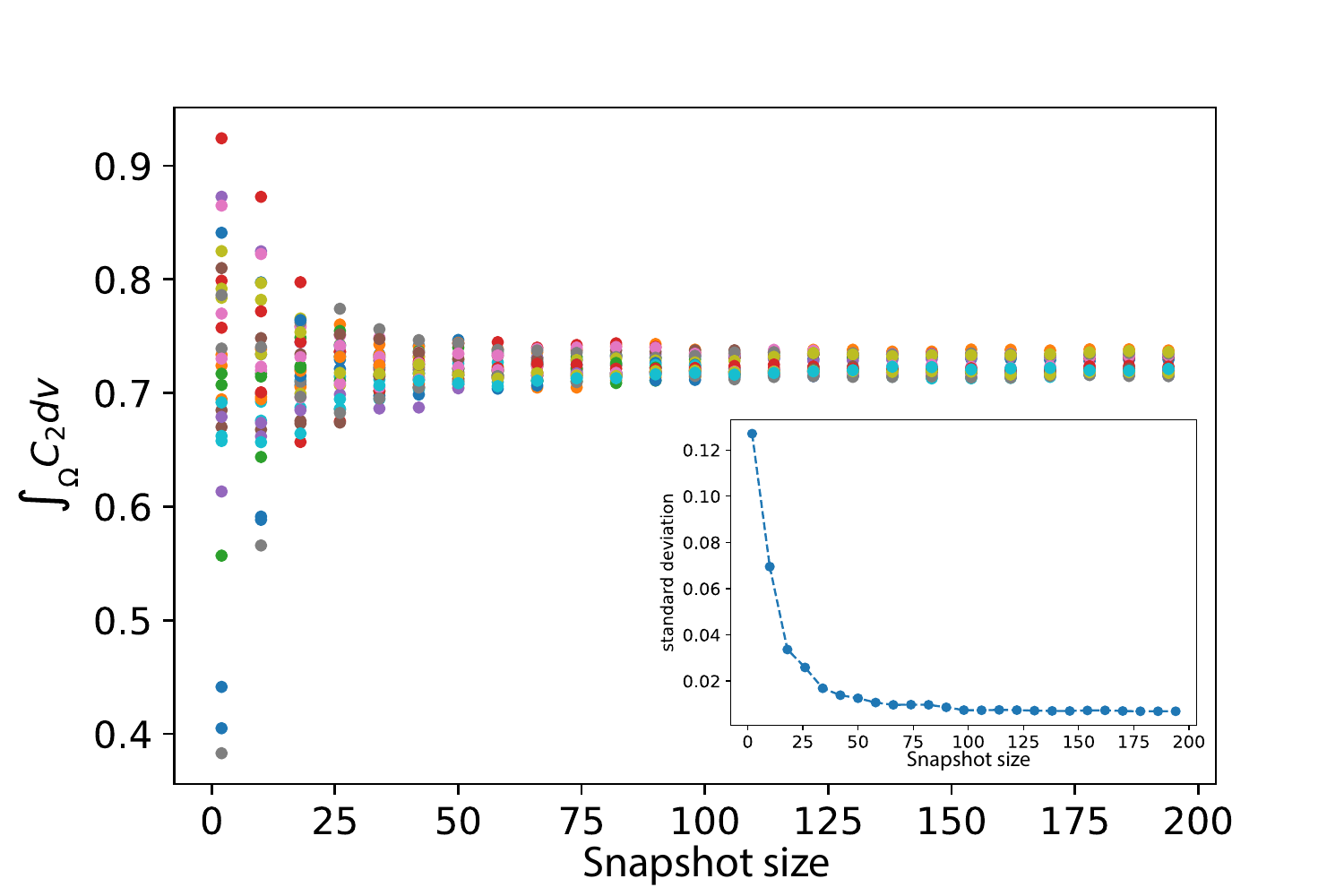}}
\subfigure[Second power of $C_2$]{\includegraphics[scale=0.53]{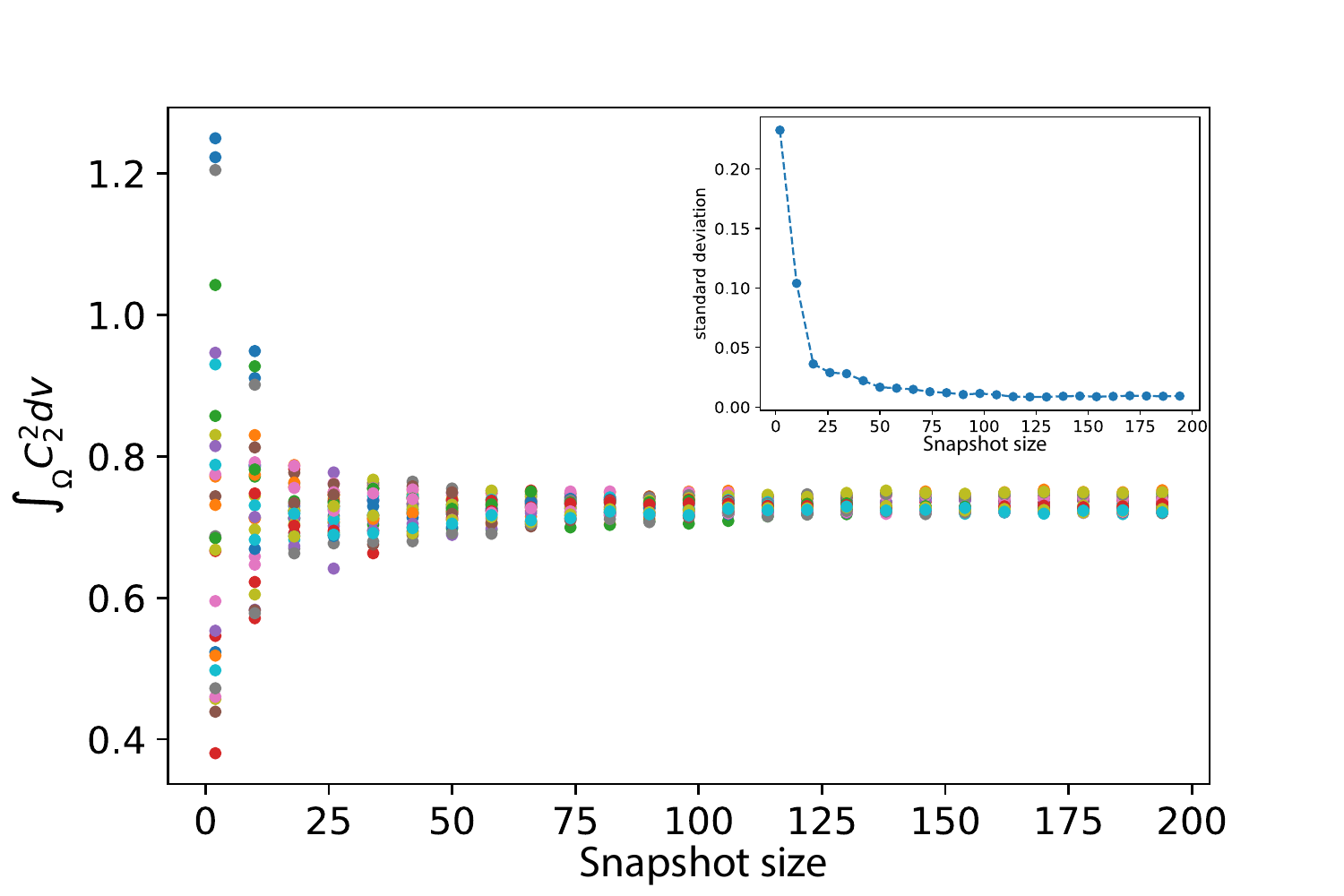}}
\caption{The powers evaluated using data generated by Model 1 from 30 simulations, marked by different colors, with different randomized initial conditions modeling data from different specimens. The embedded subplot shows the decreasing standard deviation of moments. Note the convergence with increasing snapshot size in the main and sub-plot.  }
\label{fig:moments_multiple_sample_diffu-react}
\end{figure}

The patterns formed by the diffusion-reaction system are shown in the lower plot in Figure \ref{fig:diffusion_ini_pattern}. The ``particles'' representing high concentrations of one species, attain random distributions over the entire domain. 
 Figure \ref{fig:moments_multiple_sample_diffu-react} shows the first and second powers of $C_1$ and $C_2$ evaluated using data with different sizes of snapshots from separate simulations. 
The random perturbation affects the local features of $C_1$ and $C_2$; however, with increasing size of snapshots their similarity becomes evident. Consequently powers of the composition evaluated using data from snapshots (30 different simulations are shown) converge to their values using full field data on the entire domain. 
Together they validate Relation (\ref{eq:statsimregioneps}). The statistics start to converge when snapshots are bigger than $50\times50$, and show very low variance relative to the entire domain when the snapshot size exceeds $200\times 200$. The similarity between snapshots becomes more evident as the ratio of particle spacing to the snapshots' linear dimension decreases (0.1 for the $50\times50$ snapshot to 0.025 for the $200\times 200$ snapshot). The numerical validation of  similarity between snapshots using Cahn-Hilliard and Allen-Cahn equations that govern the evolution of microstructure in materials are presented in the Appendix (see Section \ref{sec:Appendix:statistical_similarity} and Figures \ref{fig:patterns_CH_AC}-\ref{fig:moments_multiple_sample_AC}).

\subsection{Numerical examples}
\label{sec:results in dynamic regime}
\begin{figure}[hbtp]
\centering
\includegraphics[scale=0.53]{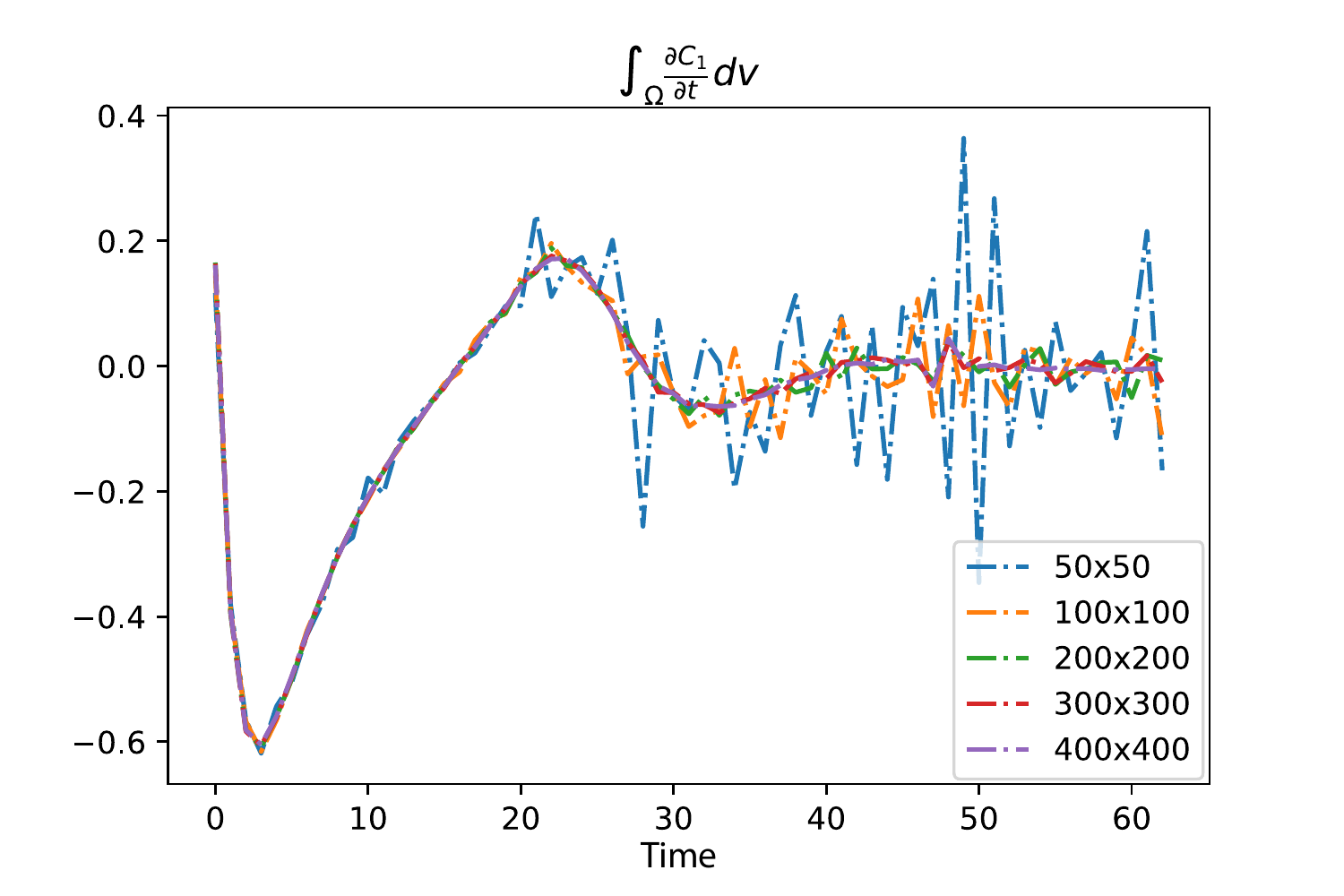}
\includegraphics[scale=0.53]{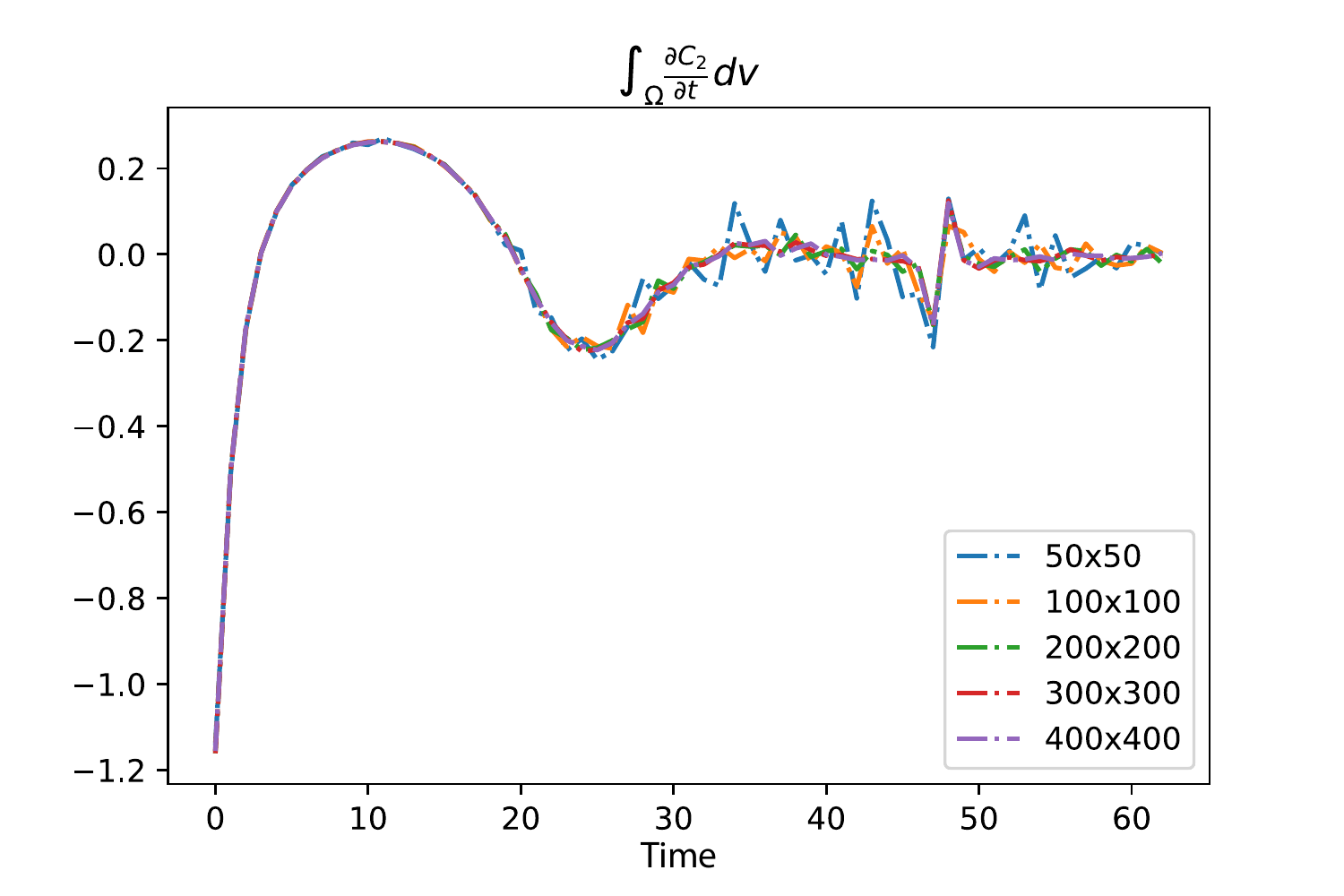}
\caption{Time evolution of $\int_{\Omega}\frac{\partial C_1}{\partial t}\text{d}v$ and $\int_{\Omega}\frac{\partial C_2}{\partial t}\text{d}v$. Time derivatives constructed from small snapshots show higher variance due to initial conditions close to steady state.}
\label{fig:C_dot}
\end{figure}

The approximation of time derivatives needs data from at least two time steps, which are collected from different subdomains of the same simulation, or from simulations with different initial conditions, as discussed previously. Our first example uses noise-free data. Note however that since similarity is better satisfied between larger snapshots, an approximation error remains. Consequently the linear systems (Equations (\ref{eq:least-square_1}) and (\ref{eq:least-square_2})) constructed by these noise-free data do not hold exactly. For instance Figure \ref{fig:C_dot} shows the time derivatives constructed using data collected from different sizes of snapshots at each time step. Smaller snapshots yield high variance resulting from the initial condition's randomness (the only source of stochasticity, since we are utilizing noise-free data in this example), especially at later times. This is because variations with time are small when the dynamical system is close to steady state, and are easily dominated by the variance induced by initial conditions. Their poor representation challenges system identification, as we will discuss below.

\begin{figure}[hbtp]
\centering
\includegraphics[scale=0.31]{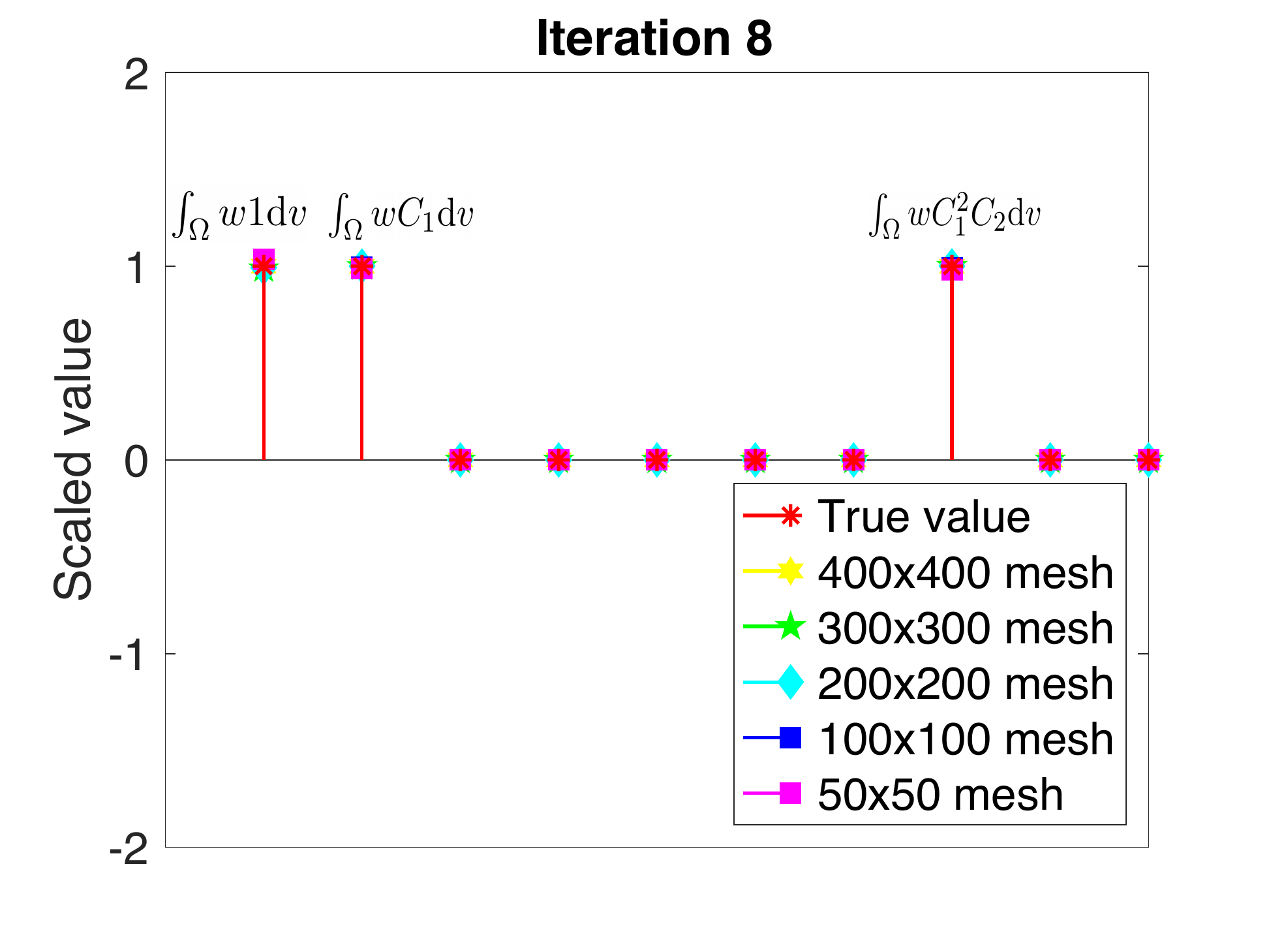}
\includegraphics[scale=0.18]{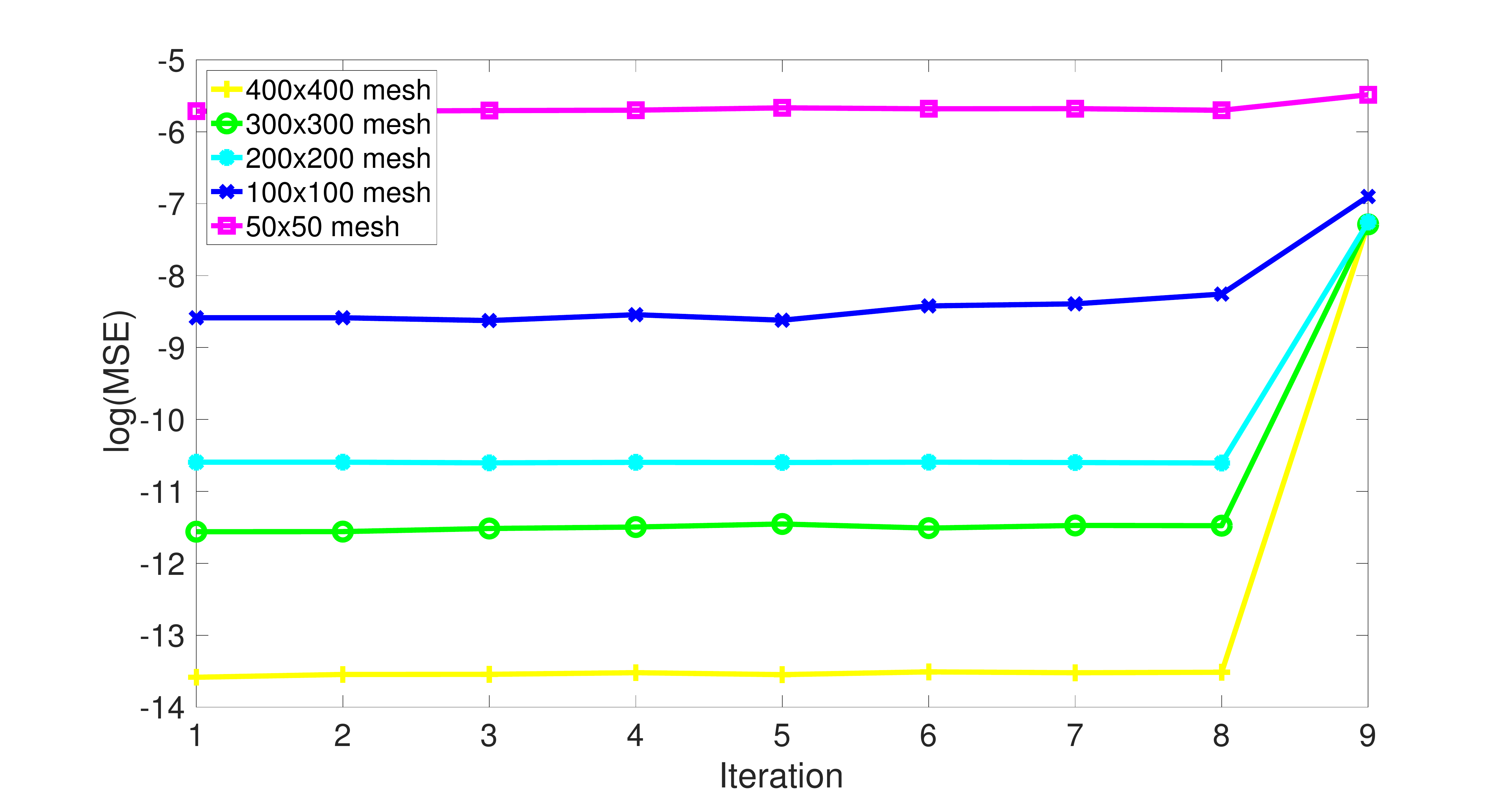}
\caption{Inferred operators for $C_1$ from Stage 1 (left panel), and the loss (right panel) at each iteration using data generated from Model 1. The identified coefficients of active operators are scaled by their true values. The algorithm converges at iteration 8, beyond which the loss increases dramatically if any more operators are eliminated.}
\label{fig:results_diffu_C1_stage1}
\end{figure}

\begin{figure}[hbtp]
\centering
\includegraphics[scale=0.31]{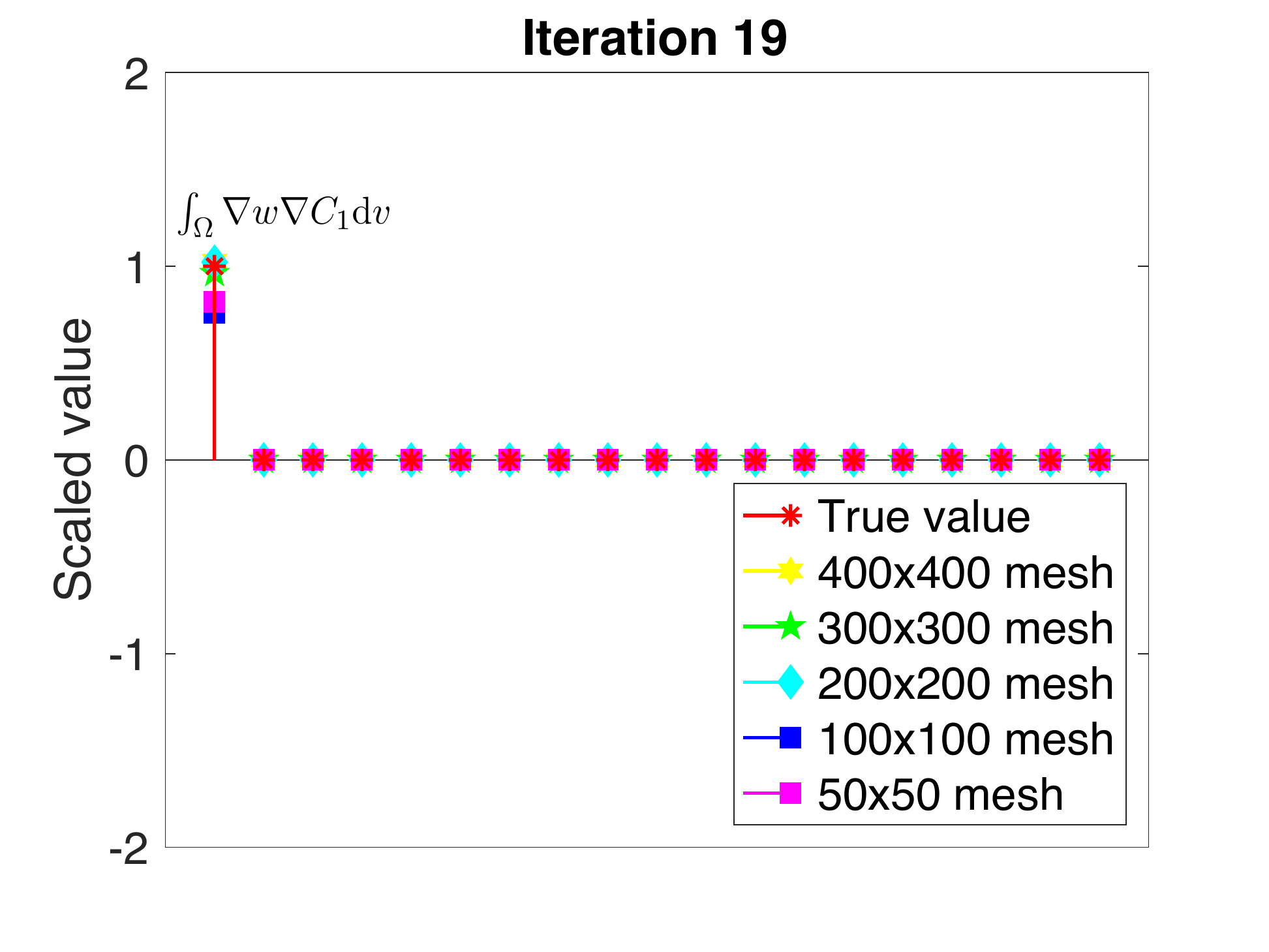}
\includegraphics[scale=0.18]{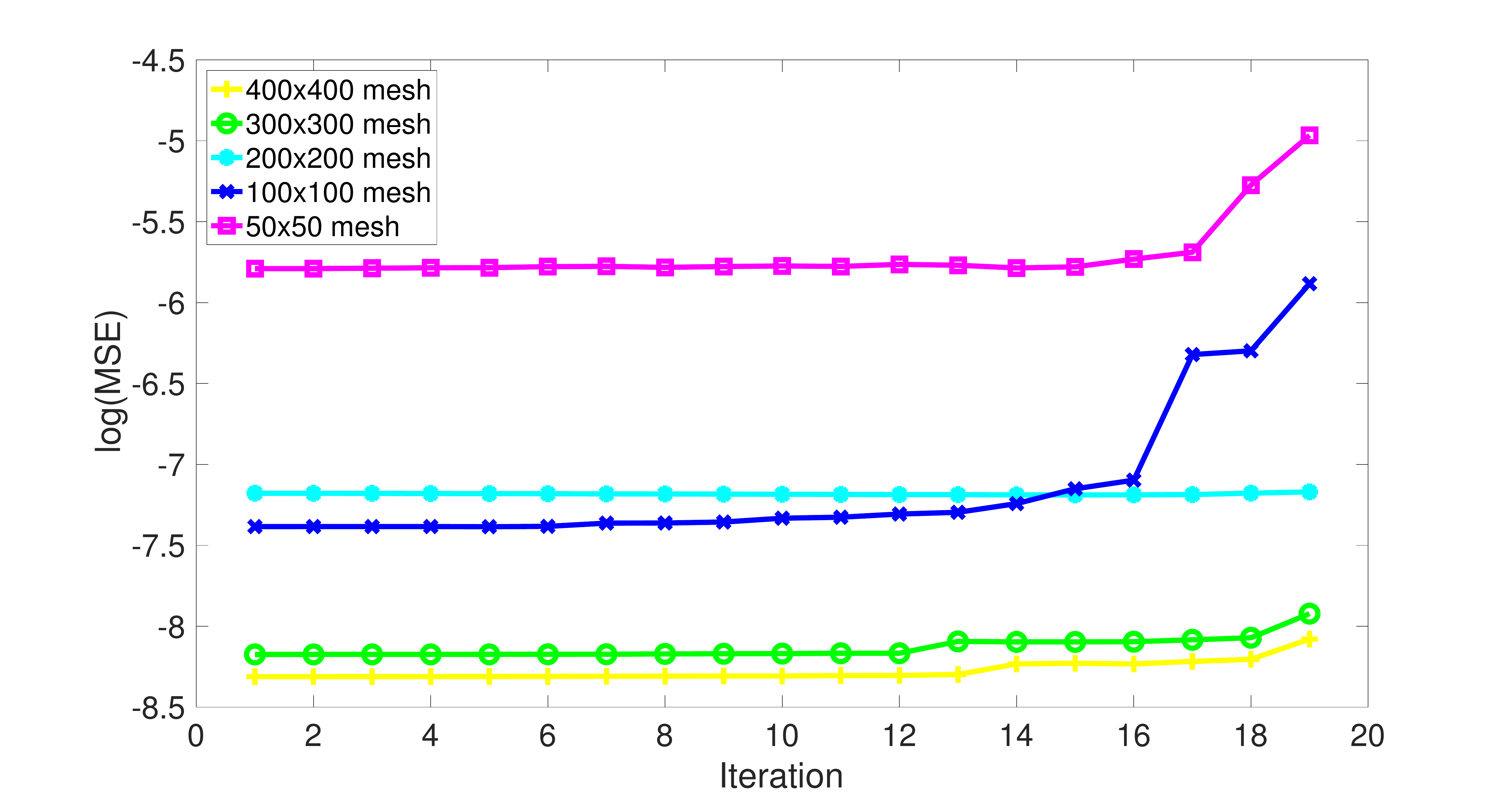}
\caption{Inferred operators for $C_1$ from Stage 2 (left panel), and the loss (right panel) at each iteration using data generated from Model 1. The identified coefficients of active operators are scaled by their true values. The algorithm converges at iteration 19 with only one operator remaining.}
\label{fig:results_diffu_C1_stage2}
\end{figure}

We successfully identified the algebraic operators in the governing equation for $C_1$ in Stage 1, as shown in Figure \ref{fig:results_diffu_C1_stage1}. The stem and leaf plots (left panel) show the coefficient of each operator scaled by its true value. The loss remains small until it converges (right panel) \res{(i.e., when the loss significantly increases over an iteration)}. VSI using data collected from larger snapshots yields lower losses, which also increase more dramatically compared to results using smaller snapshots, when an active operator is eliminated by trial in the stepwise regression process. In Stage 2 we successfully identified the one gradient dependent operator in the governing equation shown in Figure \ref{fig:results_diffu_C1_stage2}. We notice that in Stage 1 the gradually increasing loss function using data collected from small snapshots ($50 \times 50$ mesh) allows a relatively small threshold in the $F$-test $(\alpha \sim 1$) to distinguish active and inactive operators. On the other hand, in Stage 2 the loss functions increase by half an order of magnitude in the last few iterations 
for the $100\times 100$ and $50 \times 50$ cases, thus needing a higher $F$-test threshold $(\alpha \sim 10$) to distinguish inactive operators. These changes in loss function trajectories present a challenge in selecting $\alpha$ for the $F$-test, which must be tested via cross validation to balance model complexity and accuracy.

We also successfully identified the algebraic operators in the governing equation for $C_2$. However using the $50 \times 50$ and $100\times 100$ dataset, the sparsity of data leads to the active Laplacian operator being wrongly eliminated. Note that the only surviving operators with these smaller snapshots are different than those identified from the larger snapshot datasets. With the $200\times 200$ and larger snapshots, the Laplacian operator is correctly identified as the sole active operator. The full set of VSI results using noise-free data generated from Model 1 is summarized in Table~\ref{ta:tab:results_model1_dynamic}.
\begin{table}[h]
\centering
\footnotesize
 \begin{tabular}{c|c}
 \hline
 mesh size& results\\
 \hline
\multirow{2}{*}{$400\times 400$} &$\int_{\Omega}w_1\frac{\partial C_1}{\partial t}\text{d}v=\int_{\Omega}-1.0252\nabla w_1 \cdot\nabla C_1\text{d}v
+\int_{\Omega}w_1 (0.1-0.9996C_1+0.9994C_1^2C_2)\text{d}v$\\
 & $\int_{\Omega}w_2\frac{\partial C_2}{\partial t}\text{d}v=\int_{\Omega}-42.0655\nabla w_2\cdot\nabla C_2\text{d}v+\int_{\Omega}w_2(0.8998-0.9997C_1^2C_2)\text{d}v$ \\
\hline
\multirow{2}{*}{$300\times 300$} &$\int_{\Omega}w_1\frac{\partial C_1}{\partial t}\text{d}v=\int_{\Omega}-0.9644\nabla w_1 \cdot\nabla C_1\text{d}v
+\int_{\Omega}w_1 (0.0987-0.9007C_1+1.0009C_1^2C_2)\text{d}v$ \\ 
  & $\int_{\Omega}w_2\frac{\partial C_2}{\partial t}\text{d}v=\int_{\Omega}-42.2344\nabla w_2\cdot\nabla C_2\text{d}v+\int_{\Omega}w_2(0.89982-0.9998C_1^2C_2)\text{d}v$  \\ \hline
\multirow{2}{*}{$200\times 200$} &$\int_{\Omega}w_1\frac{\partial C_1}{\partial t}\text{d}v=\int_{\Omega}-1.0212\nabla w_1 \cdot\nabla C_1\text{d}v
+\int_{\Omega}w_1 (0.0993-1.0007C_1+1.0017C_1^2C_2)\text{d}v$ \\ 
  & $\int_{\Omega}w_2\frac{\partial C_2}{\partial t}\text{d}v=\int_{\Omega}-42.1083\nabla w_2\cdot\nabla C_2+\int_{\Omega}w_2(0.9001-0.9993C_1^2C_2)\text{d}v$\\ \hline
\multirow{2}{*}{$100\times 100$} & $\int_{\Omega}w_1\frac{\partial C_1}{\partial t}\text{d}v=\int_{\Omega}-0.7576\nabla w_1 \cdot\nabla C_1\text{d}v
+\int_{\Omega}w_1 (0.1035-0.9954C_1+0.9908C_1^2C_2)\text{d}v$  \\ 
&$\int_{\Omega}w_2\frac{\partial C_2}{\partial t}\text{d}v=\int_{\Omega}2.77\nabla w_2\cdot C_1^2\nabla C_1\text{d}v+\int_{\Omega}w_2(0.8982-0.9975C_1^2C_2)\text{d}v$ \\
\hline
\multirow{2}{*}{$50\times 50$}  & $\int_{\Omega}w_1\frac{\partial C_1}{\partial t}\text{d}v=\int_{\Omega}-0.8154\nabla w_1 \cdot\nabla C_1\text{d}v
+\int_{\Omega}w_1 (0.1035-0.9889C_1+0.9804C_1^2C_2)\text{d}v$  \\ 
&$\int_{\Omega}w_2\frac{\partial C_2}{\partial t}\text{d}v=\int_{\Omega}3.91\nabla w_2\cdot C_1^2C_2\nabla C_1\text{d}v+\int_{\Omega}w_2(0.8896-0.9886C_1^2C_2)\text{d}v$ \\ \hline
\end{tabular}
\caption{Results using noise-free dynamic data generated from Model 1.}
\label{ta:tab:results_model1_dynamic}
\end{table}

\begin{figure}[hbtp]
\centering
\includegraphics[scale=0.18]{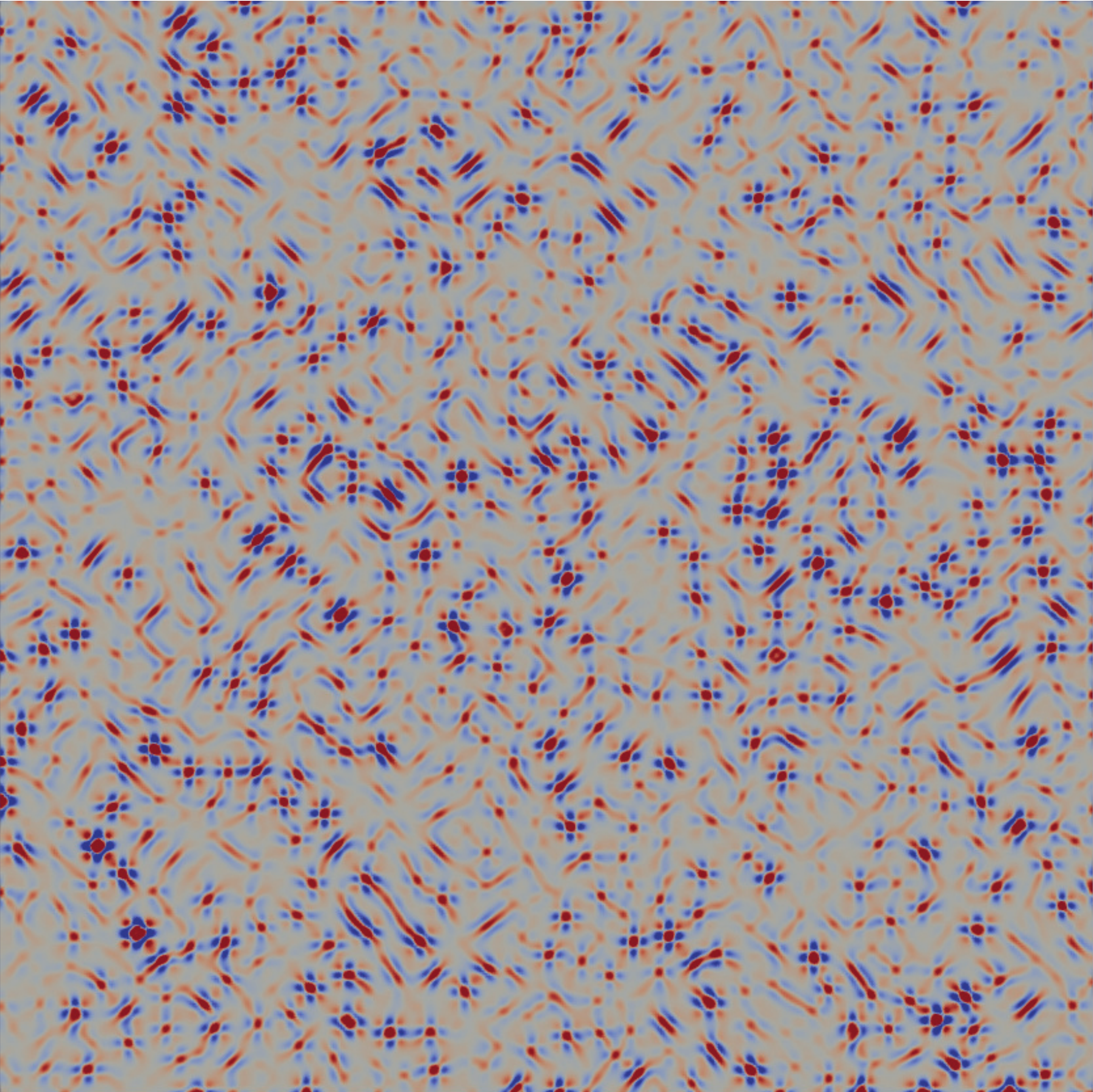}\quad
\includegraphics[scale=0.18]{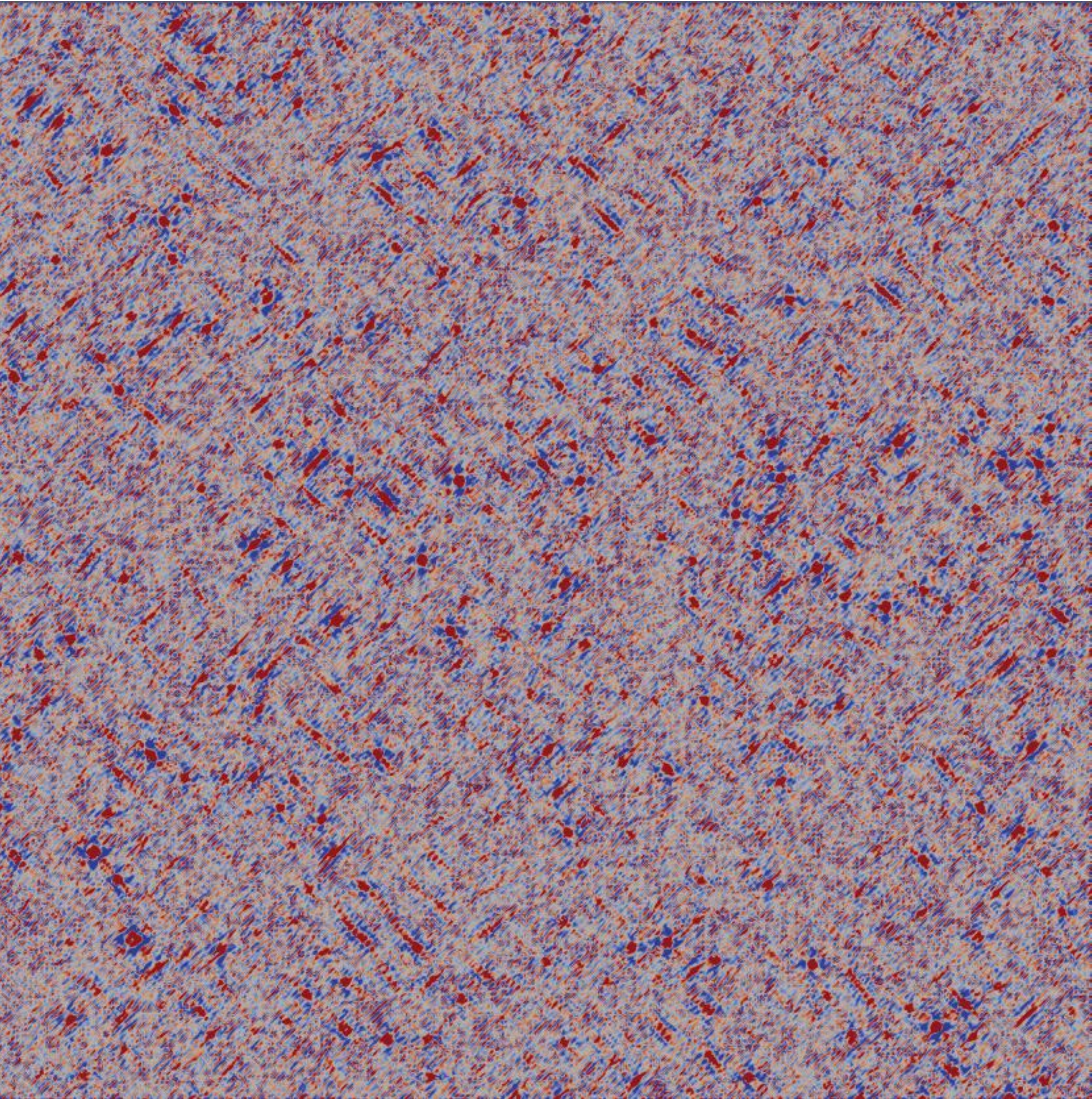}
\caption{Noise-free (left) and noisy (right) fields of $\nabla^2 C_1$.}
\label{fig:diffu-react-laplacian}
\end{figure}

Next, we superimpose a Gaussian noise with zero mean and standard deviation $\sigma=0.01$ over the data generated from Model 1. The noise on $C_1$  and $C_2$ gets amplified in the time derivative and spatial gradients. As shown in Figure \ref{fig:diffu-react-laplacian}, the noise dominates the true value of the Laplacian $\nabla^2 C_1$. The amplification of noise in the differential operators can cause failure of VSI that works with full-field data, as shown by Wang et al. \cite{WangCMAME2019}. However due to the integral nature of the first and second powers in the two stage VSI presented in this work, the noise is smoothed out. The zero mean ensures that the operators constructed with specific weighting functions are insensitive to noise. 

The full set of VSI results using noisy data generated from Model 1 are summarized in Table \ref{ta:tab:results_model1_dynamic_noisy}.  We successfully identified the governing equations of $C_1$. However there remain errors in the coefficients of the identified operators. For the governing equation of $C_2$, all algebraic operators are successfully identified using noisy data. The single active differential operator, the Laplacian, is wrongly identified using smaller snapshots ($200\times 200$, $100\times 100$ and $50 \times 50$ datasets), while the larger snapshots allow its correct identification. 
\begin{table}[h]
\centering
\footnotesize
 \begin{tabular}{c|c}
 \hline
 mesh size& results\\
 \hline
\multirow{2}{*}{$400\times 400$} &$\int_{\Omega}w_1\frac{\partial C_1}{\partial t}\text{d}v=\int_{\Omega}-1.0675\nabla w_1 \cdot\nabla C_1\text{d}v
+\int_{\Omega}w_1 (0.0937-0.9876C_1+0.9912C_1^2C_2)\text{d}v$ \\
 & $\int_{\Omega}w_2\frac{\partial C_2}{\partial t}\text{d}v=\int_{\Omega}-42.2038\nabla w_2\cdot\nabla C_2\text{d}v+\int_{\Omega}w_2(0.8662-0.9661C_1^2C_2)\text{d}v$  \\
\hline
\multirow{2}{*}{$300\times 300$} &$\int_{\Omega}w_1\frac{\partial C_1}{\partial t}\text{d}v=\int_{\Omega}-1.1105\nabla w_1 \cdot\nabla C_1\text{d}v
+\int_{\Omega}w_1 (0.0941-0.9876C_1+0.991C_1^2C_2)\text{d}v$  \\ 
  & $\int_{\Omega}w_2\frac{\partial C_2}{\partial t}\text{d}v=\int_{\Omega}-42.0852\nabla w_2\cdot\nabla C_2\text{d}v+\int_{\Omega}w_2(0.8668-0.9663C_1^2C_2)\text{d}v$ \\ \hline
\multirow{2}{*}{$200\times 200$} &$\int_{\Omega}w_1\frac{\partial C_1}{\partial t}\text{d}v=\int_{\Omega}-1.2102\nabla w_1 \cdot\nabla C_1\text{d}v
+\int_{\Omega}w_1 (0.0944-0.9882C_1+0.9912C_1^2C_2)\text{d}v$ \\ 
  & $\int_{\Omega}w_2\frac{\partial C_2}{\partial t}\text{d}v=\int_{\Omega}-55.257\nabla w_2C_2\cdot\nabla C_2+\int_{\Omega}w_2(0.8662-0.9651C_1^2C_2)\text{d}v$ \\ \hline
\multirow{2}{*}{$100\times 100$} & $\int_{\Omega}w_1\frac{\partial C_1}{\partial t}\text{d}v=\int_{\Omega}-1.1579\nabla w_1 \cdot\nabla C_1\text{d}v
+\int_{\Omega}w_1 (0.0895-0.9897C_1+0.9992C_1^2C_2)\text{d}v$ \\ 
&$\int_{\Omega}w_2\frac{\partial C_2}{\partial t}\text{d}v=\int_{\Omega}-54.989\nabla w_2\cdot C_2\nabla C_2\text{d}v+\int_{\Omega}w_2(-0.8682-0.967C_1^2C_2)\text{d}v$ \\
\hline
\multirow{2}{*}{$50\times 50$}  & $\int_{\Omega}w_1\frac{\partial C_1}{\partial t}\text{d}v=\int_{\Omega}-0.8997\nabla w_1 \cdot\nabla C_1\text{d}v
+\int_{\Omega}w_1 (0.0836-0.9826C_1+0.9973C_1^2C_2)\text{d}v$ \\ 
&$\int_{\Omega}w_2\frac{\partial C_2}{\partial t}\text{d}v=\int_{\Omega}7.93\nabla w_2\cdot C_1\nabla C_1\text{d}v+\int_{\Omega}w_2(-0.8733-0.9726C_1^2C_2)\text{d}v$  \\ \hline
\end{tabular}
\caption{Results using noisy dynamic data generated from Model 1.}
\label{ta:tab:results_model1_dynamic_noisy}
\end{table}

\begin{figure}
    \centering
    \subfigure[$C_1$ concentration: identified Model 1 ]{\includegraphics[scale=0.1]{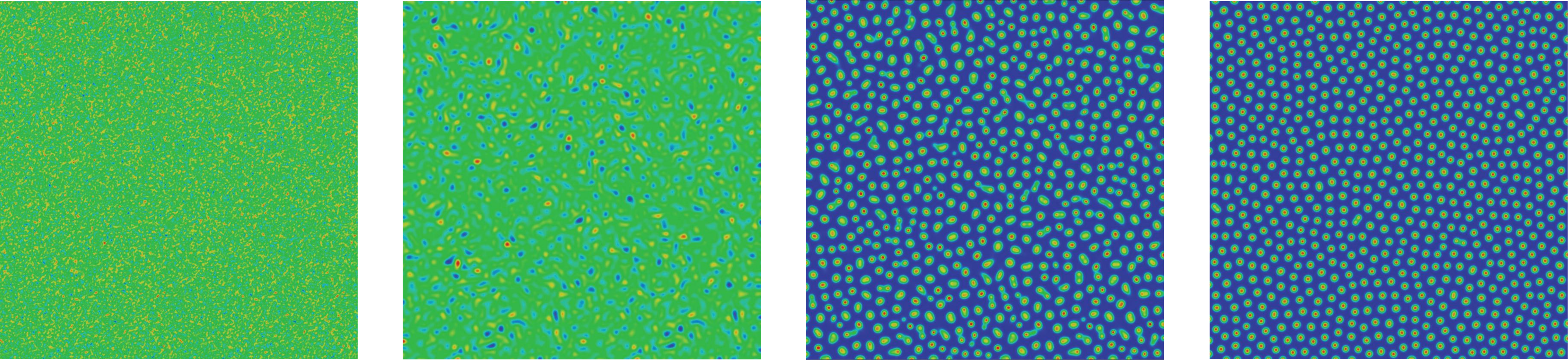} }
    \subfigure[$C_1$ concentration: true Model 1 ]{\includegraphics[scale=0.1]{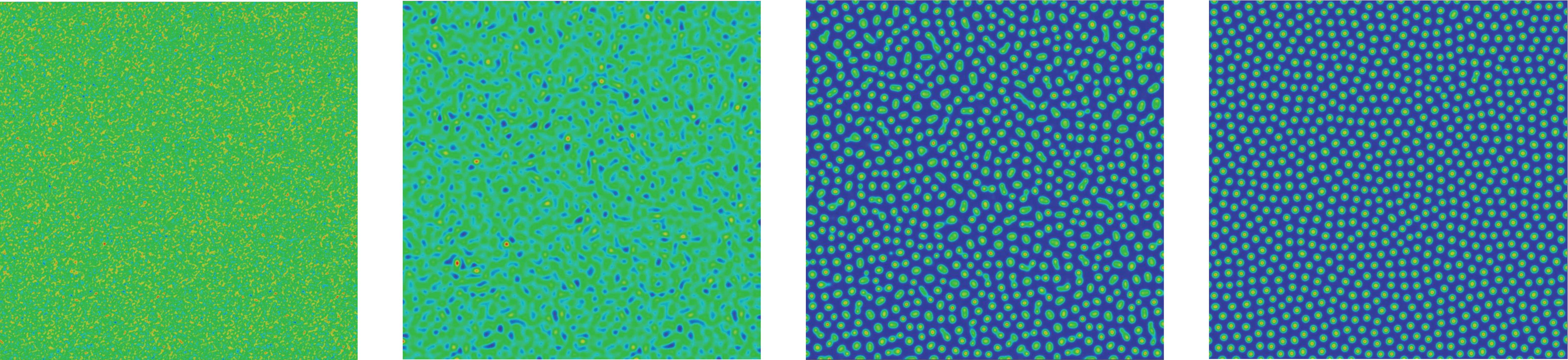} }
    \subfigure[First power ]{\includegraphics[scale=0.5]{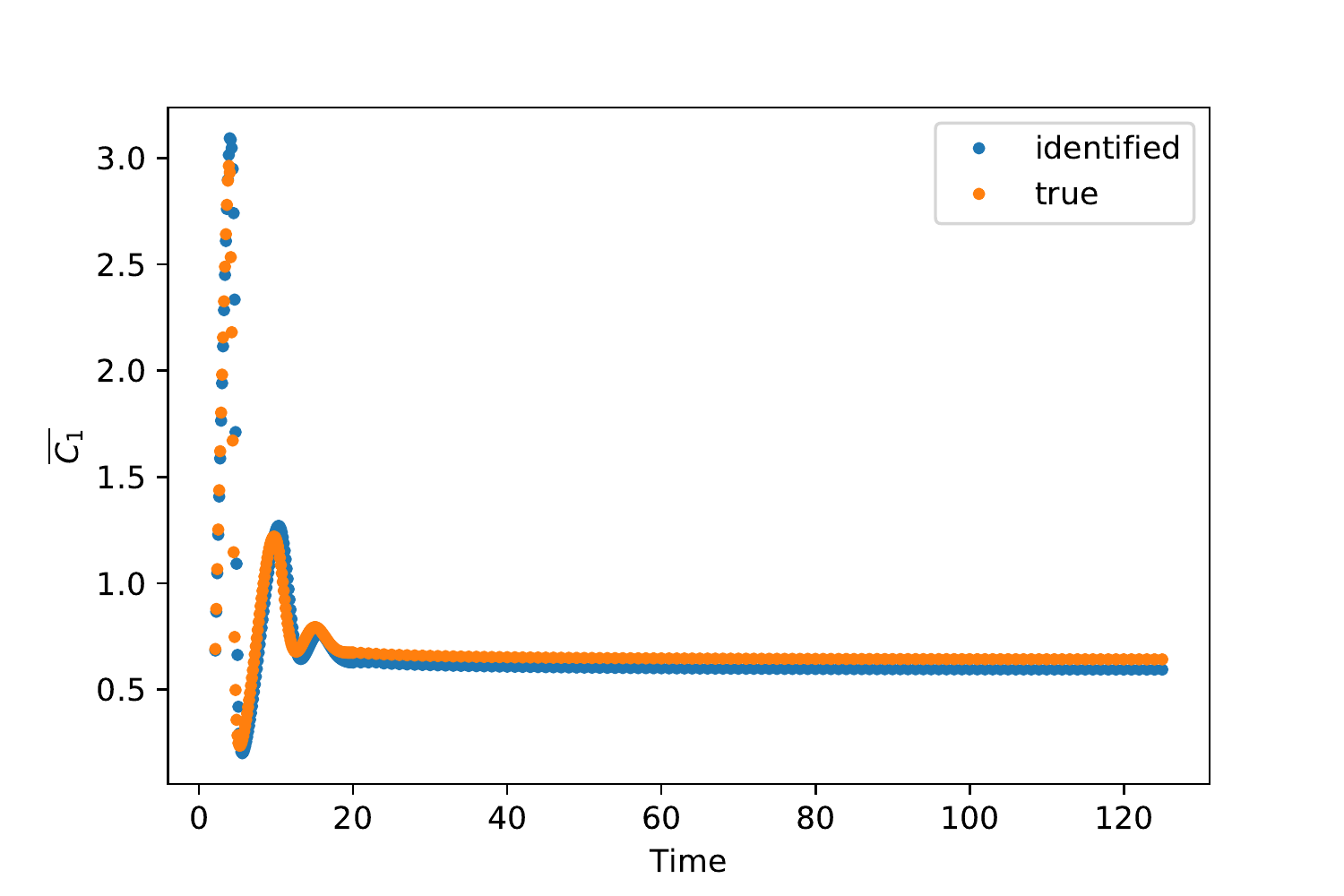} }
    \subfigure[Second power ]{\includegraphics[scale=0.5]{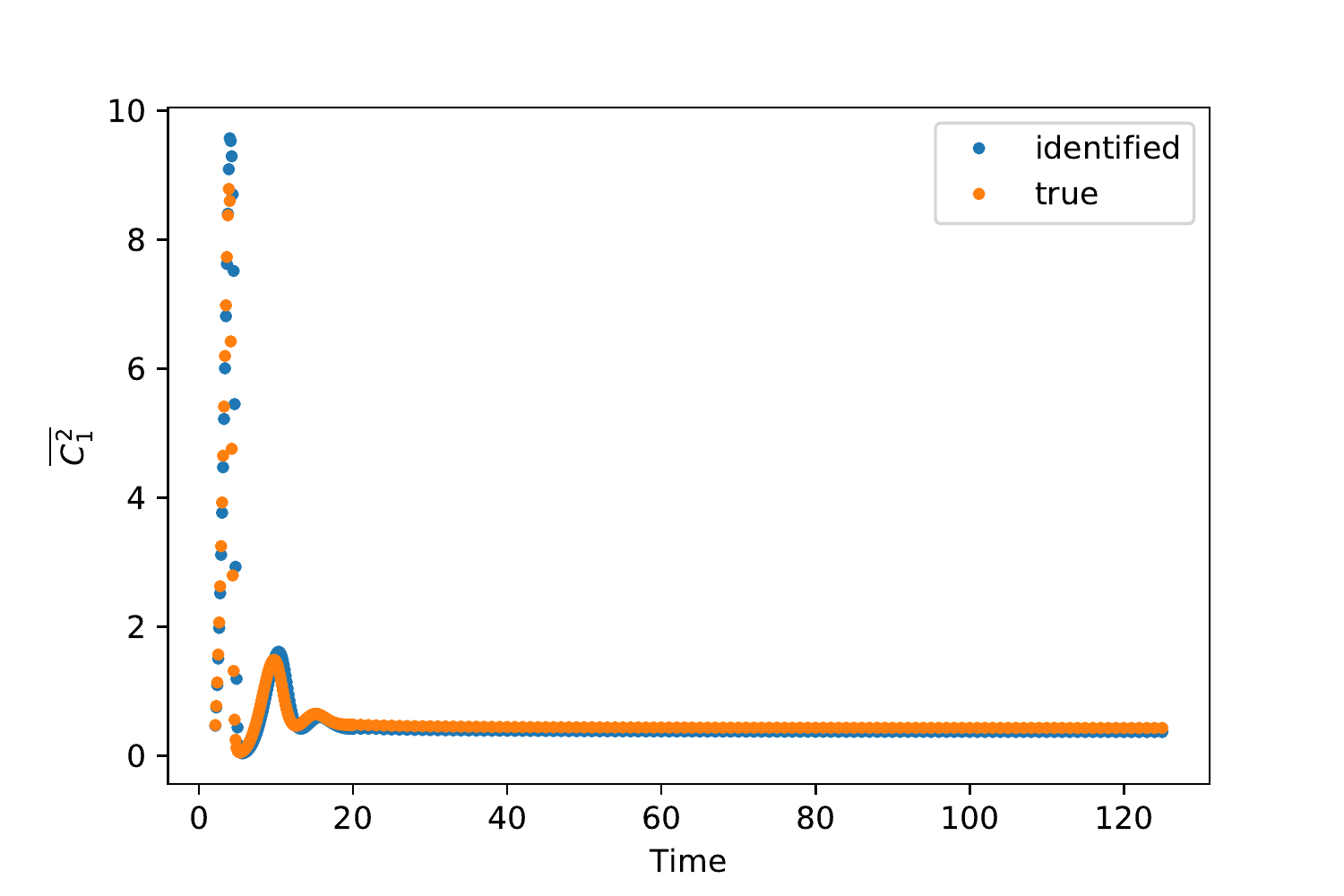} }
    \caption{Comparison of the $C_1$ concentration field simulated by  Model 1 inferred by VSI (a), against the true Model 1 (b). The results are shown at $t=0, 5, 20, 100$. The good agreement for the first and second powers indicates the similarity between the identified model and the true model.}
    \label{fig:comparison_diffu-react}
\end{figure}

We chose the identified model using noisy data collected at $100\times 100$ snapshot for comparison with the true model by solving with the same initial condition. The chosen identified model has the correctly identified operators but with the worst inference of the coefficients over the snapshots considered. Though the identified model yields a slightly sparser distribution of the particles due the the higher diffusivities (upper plots in Figure \ref{fig:comparison_diffu-react}), it produces patterns that are similar to those in the true model in term of first and second powers (lower plots in Figure \ref{fig:comparison_diffu-react}).

Unlike the diffusion-reaction system, Model 2, represented by Cahn-Hilliard equations, is conservative. Consequently the time derivative of the first power vanishes: $\frac{\text{d}}{\text{dt}}\int_{\Omega}C_{n}\text{d}v = 0$. The second moment, as well as other weak forms of algebraic and differential operators with $w^h = 1$ and $w^h = C^h$ also evolve very slowly after the initial regime of spinodal decomposition. With limited dynamic data, identifying the Cahn-Hilliard equations is challenging using the two stage VSI. We next present an alternate approach that leverages the data at steady state to successfully identify both diffusion-reaction and Cahn-Hilliard systems. 

\section{Variational System Identification with dynamic and steady state data}
\label{sec:three}
One challenge of the two-stage procedure introduced in Section~\ref{sec:incompdata} is data availability, where experimental measurements from the dynamic regime (non-steady state) of the system can be especially difficult and expensive to obtain. 
While the regression problems in Equations (\ref{eq:least-square_1}) and (\ref{eq:least-square_2}) generally require over-determined systems 
to arrive at good  solutions, this consequently demands a large dataset to delineate the relevant bases,
since the number of rows in Equations (\ref{eq:least-square_1}) and (\ref{eq:least-square_2}) is the number of time instants minus one. The challenge posed by spatially unrelated data and sparse time instants led us to the two-stage formulation of the problem with integral forms as presented above in Section \ref{sec:Method of moments}. This treatment, however, sacrifices any high spatial resolution available in the experimental snapshots. As a remedy, we have developed an approach that leverages data with high spatial resolution at steady state (or near-steady state). With this new approach, it is possible to first identify all the operators up to a scaling factor, which can subsequently be determined by the dynamics. Inference of this scaling factor, controlled by the dynamics, can then achieved using only one of the two stages described previously for the algebraic and differential operators. The essence of the VSI approach, namely, the treatment of equations in weak form, remains unchanged.

\subsection{Variational System Identification of basis operators with steady state data}
\label{sec:steady state}

Data with high spatial resolution at steady state, or near-steady state, is typically obtained from modern microscopy methods. 
In fact the data satisfying the steady state equation:
\begin{align}
   \Bchi\cdot\Btheta=0, 
   \label{eq:steady-state_strongForm}
\end{align}
already provide rich information about the spatial operators (that is, other than time derivatives) in the system. By choosing a nontrivial $\chi_1$ to be the target operator, the equation can be rewritten as:
\begin{align}
  -\chi_1= \frac{1}{\theta_1}\overline{\Bchi}\cdot\overline{\Btheta}, 
   \label{eq:steady-state_strongForm_2}
\end{align}
where $\overline{\Bchi}=\Bchi\backslash\{\chi_1\}$ and $\overline{\Btheta}=\Btheta\backslash\{\theta_1\}$. We are able to identify the coefficients, $\overline{\Btheta}$, up to a scaling constant, $\theta_1$, using the data at steady state by VSI \cite{WangCMAME2019}. However, in the absence of  prior knowledge about the system, it may be the case that the selected target operator, $\chi_1$, is actually inactive (i.e., its prefactor $\theta_1=0$). Then Equation (\ref{eq:steady-state_strongForm_2}) is not valid for the data. 
Under such conditions, VSI will fail at parsimonious representation over the operators included in $\overline{\Bchi}$. ``False'' outcomes can then be detected by inconsistencies in the system identification, as we elaborate via examples below. 
 
On the other hand, if in two (or more) trials distinct operators are chosen as targets, and if these chosen targets are all indeed active in the true system, then the same set $\overline{\Bchi}$ will be identified in these different trials with coefficients $\overline{\Btheta}$ differing only by a scaling factor. This consistency, up to a scaling factor, provides a confirmation test for VSI of the spatial operators with (near-)steady state data without the need for prior knowledge.

Nevertheless, when confronted with data on systems that are governed by multiple equations, the above approach to identify consistent sets of active operators via the confirmation test can break down. In particular, this can happen if one of the equations has only one operator that is distinct from those appearing in other equations governing the system, and its remaining operators number fewer than in any other equation of the system. In such a situation, the stepwise regression algorithm will detect the steady state equation with the greater number of operators as the only consistent set passing the confirmation test. Even the choice of the one distinct operator in the remaining equation as target operator will not pass the confirmation test for consistency (since upon looking for consistency, the equation with greater number of operators will be identified as it provides a feasible solution with lower loss). Our approach to this detection of only the lower loss solution is to deliberately suppress an operator in a consistent set that passes the confirmation test. This forces the stepwise regression algorithm to detect the equation with fewer operators, provided the suppressed operator is indeed inactive in this set. Subsequent consistency checks with the correctly suppressed operator will pass the confirmation test. The combined confirmation test for consistency with operator suppression appears as Algorithm 3.
 \\
\\
\noindent\fbox{%
\parbox{\textwidth}{%
\textbf{Algorithm 3: Confirmation Test of Consistency with Operator Suppression:}\\

\texttt{Let $N_\chi := \#\Bchi$ (where $\#$ denotes  cardinality) and $N_\text{field} :=$ number of unknown fields in data (e.g., for data on $C_1,C_2$, $N_\text{field} = 2$)}\\

\texttt{Function: Test of Consistency}\\
\text{\qquad}\texttt{FOR $i = 1,\ldots,N_\chi$}\\
\text{\qquad\quad}\texttt{With $\chi_i$ as target vector, do stepwise regression to identify coefficients $\boldsymbol{\vartheta}_i\in \mathbb{R}^{N_\chi-1}$.}\\
\text{\qquad\quad}\texttt{IF $i \ge 2$ \&\& $\boldsymbol{\vartheta}_i = \alpha\boldsymbol{\vartheta}_j$ for $j<i$ THEN}\\
\text{\qquad\qquad}\texttt{Confirmation Test: $\boldsymbol{\vartheta}_i$ and $\boldsymbol{\vartheta}_j$ represent the same consistent set} \\
\text{\qquad\quad}\texttt{ELSE}\\
\text{\qquad\qquad}\texttt{$\boldsymbol{\vartheta}_i$ is inconsistent}\\
\text{\qquad}\texttt{Collect all linearly independent consistent coefficient sets expressible as $\theta_\alpha\boldsymbol{\vartheta}_\alpha, \theta_\beta\boldsymbol{\vartheta}_\beta,\ldots$ for $\alpha,\beta,\ldots \in \{1,\ldots,N_\chi \}$}.\\

\text{\quad}\texttt{While (True):}\\
\text{\qquad} \texttt{Call Test of Consistency.}\\
\text{\qquad} \texttt{IF $\#\{\alpha,\beta,\ldots\} < N_\text{field}$} \\
\text{\qquad\quad}\texttt{FOR $i \in \{\alpha,\beta,\ldots \}$}\\
\text{\qquad\qquad} \texttt{
Define $\widehat{\chi}_k$ to be the operator with coefficient $\theta_{i_k}$ in $\boldsymbol{\vartheta}_i = \{\ldots,\theta_{i_k},\ldots\}$. Call Test of Consistency with $\Bchi\backslash\widehat{\chi}_{k}$.}\\
\text{\qquad} \texttt{ELSE}\\
\text{\qquad\quad}\texttt{EXIT}
\vspace{0.25cm}
}
}
\\
\\

Thereafter, there remains only one unknown in the governing equations, which is the scaling factor, $\theta_1$:
\begin{align}
\frac{\partial C}{\partial t}-\theta_1(\Bchi\cdot\Btheta)=0, 
\end{align}
which then can be identified using dynamic data and one of the two stages of VSI detailed in Sections \ref{sec:Method of moments}--\ref{sec:ID-2Step}. If an algebraic/differential operator was identified via Algorithm 3 for the (near-) steady state data, Stage 1/Stage 2  can be used respectively (Section \ref{sec:Method of moments}). 
 This is elaborated with examples in Section \ref{sec:data at steady state without noise}, and proves to be a powerful feature in favor of robustness and consistency of VSI for steady and dynamic data.

\subsection{Numerical examples with steady state and dynamic data}
\label{sec:data at steady state without noise}
\begin{figure}[hbtp]
\centering
\includegraphics[scale=0.35]{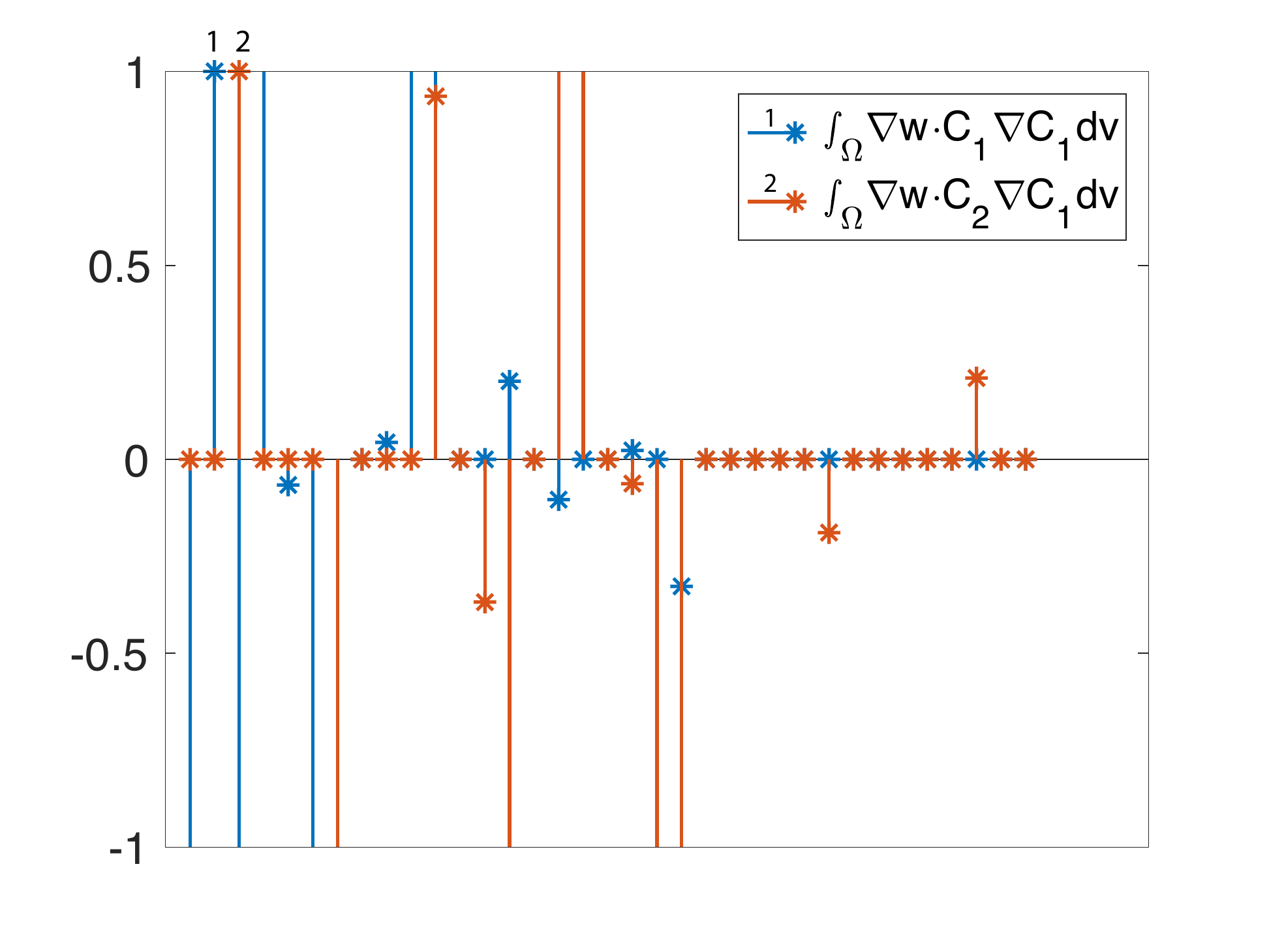}
\caption{One example of inconsistently inferred operators with different targets using data at steady state. Choosing $\int_\Omega \nabla w\cdot C_1\nabla C_1\text{d}v$, labelled by 1, as target operator leads to a set of operators being identified, shown by the blue stems and leaves. This includes the operator $\int_\Omega \nabla w\cdot C_2\nabla C_1\text{d}v$  labelled by 2. However, choosing the latter as the target operator yields a different active set, shown as the orange stems and leaves. Both target operators are actually inactive, leading to inconsistency between the respective sets. }
\label{fig:stem_steady_wrong}
\end{figure}

The steady state form of Model 1 in strong form is:
\begin{align}
D_1\nabla^2C_1+R_{10}+R_{11}C_1+R_{13}C_1^2C_2=0\label{eq:diffu_reac_steady1}\\    
D_2\nabla^2C_2+R_{20}+R_{21}C_1^2C_2=0.
\label{eq:diffu_reac_steady2}
\end{align}
Given steady state, or near-steady state data, such as for full field or snapshots of diffusion-reaction systems, VSI can identify the steady state governing equations using the Confirmation Test of Consistency with Operator Suppression in Algorithm 3. 
As discussed in Section \ref{sec:steady state}, choosing inactive operators to be targets yields inconsistent results. Figure \ref{fig:stem_steady_wrong} shows that if two different (inactive) operators, $\int_\Omega \nabla w\cdot C_1\nabla C_1\text{d}v$ and $\int_\Omega \nabla w\cdot C_2\nabla C_1\text{d}v$ (in weak form), are chosen as targets, the identified results are completely different. The set of identified operators with $\int_\Omega \nabla w\cdot C_1\nabla C_1\text{d}v$ as target contains $\int_\Omega \nabla w\cdot C_2\nabla C_1\text{d}v$ but not \emph{vice versa}. This is because choosing an inactive operator as the target removes it from the stepwise regression and model selection procedures in Algorithms 1 and 2, meaning that it has to be included in the identified set. Consequently, the identified sets of operators do not form a group of governing PDEs that models meaningful physics. This is reflected in the inconsistency between the operator sets. 

\begin{figure}[hbtp]
\centering
\includegraphics[scale=0.35]{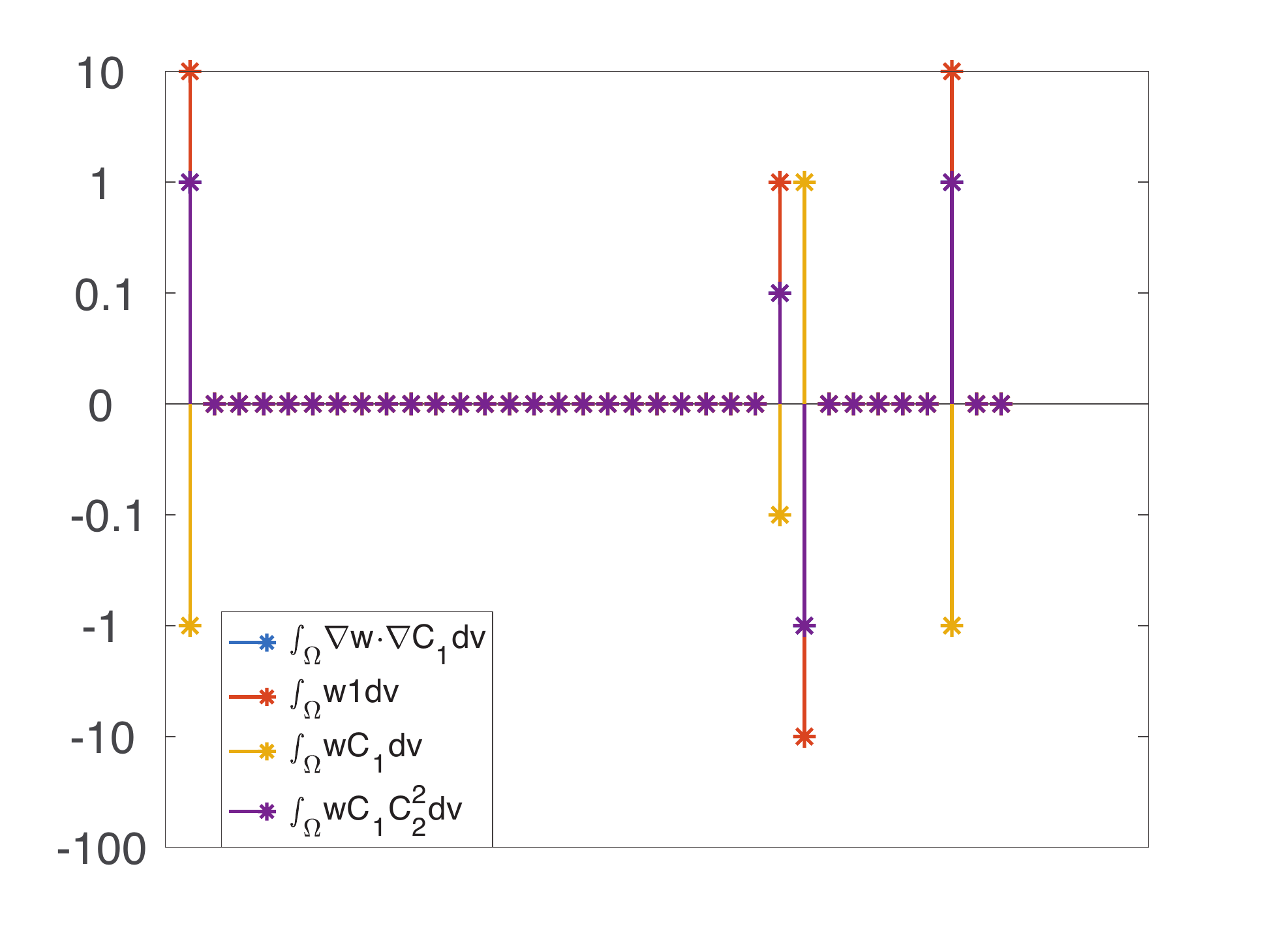}
\includegraphics[scale=0.35]{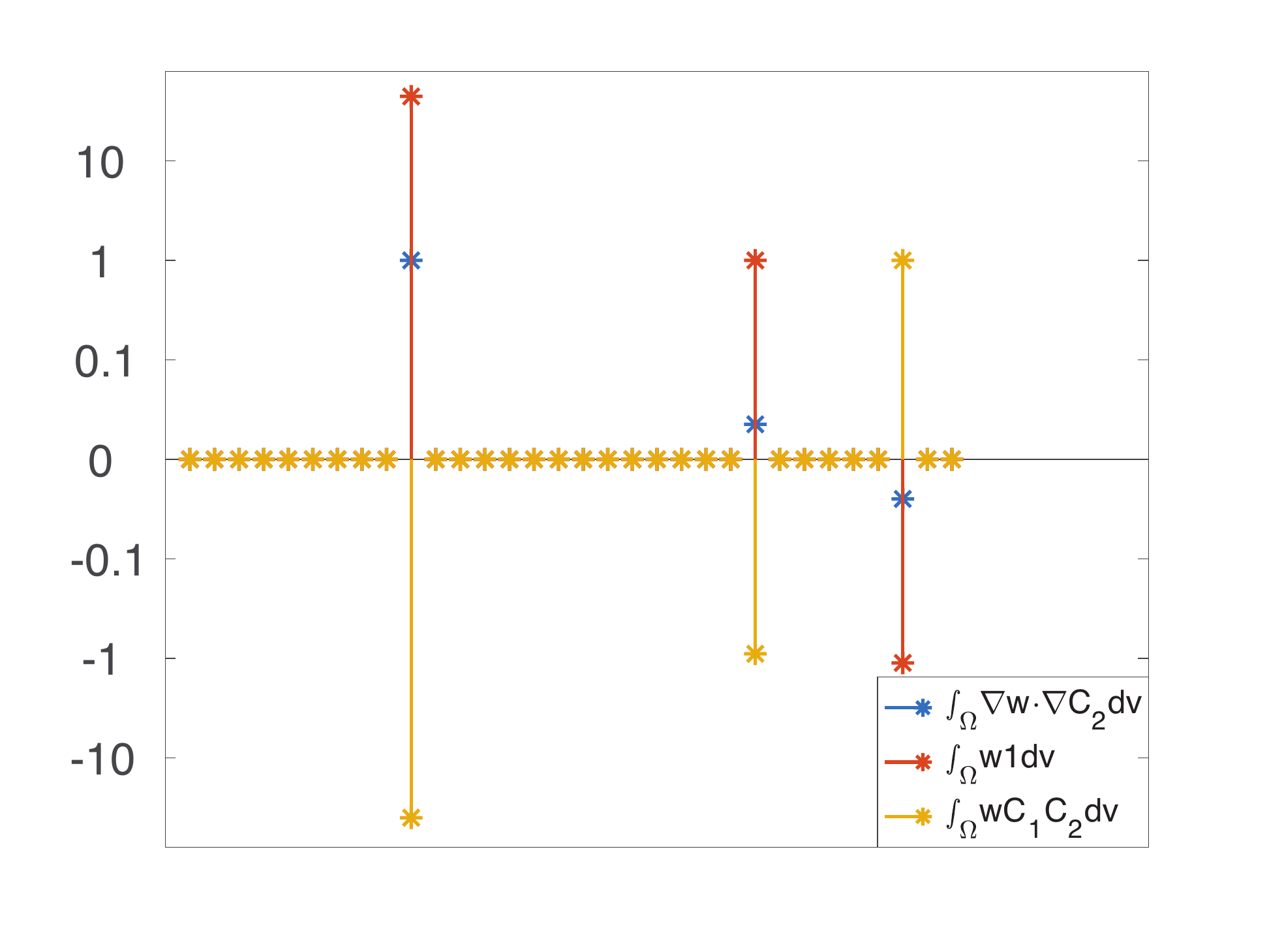}
\caption{Two sets (left and right) of consistently inferred operators with different targets using data generated from Model 1 at steady state. The coefficient values have been plotted on the logarithmic scale in order to make them discernible across orders of magnitude. Each set consists of the operators in the legend of the corresponding sub-plot and is consistent up to a scaling factor. Note that the second set of consistently inferred operators is obtained on a reduced library of candidate operators without $\int_\Omega \nabla w\cdot \nabla C_1\text{d}v$ or $\int_\Omega wC_1\text{d}v$, the two operators in the first set of consistent inferred operators. The two sets, left and right, are the steady state equations for $C_1$ and $C_2$ in Equations (\ref{eq:diffu_reac_steady1}) and (\ref{eq:diffu_reac_steady2}), respectively.}
\label{fig:stem_steady_correct}
\end{figure}

On the other hand, choosing the active operators as the targets does yield consistent identification. The plot on the left in Figure \ref{fig:stem_steady_correct} shows that the identified set is consistent up to a scaling factor, with four different operators as, shown in the legend, as targets. This set identifies the steady state equation (\ref{eq:diffu_reac_steady1}). Note that the weak form operators $\int_\Omega w\text{d}v$ and $\int_\Omega wC_1C_2^2\text{d}v$ are common to Equations (\ref{eq:diffu_reac_steady1}) and (\ref{eq:diffu_reac_steady2}), but choosing either as target only yields the consistent set in Equation (\ref{eq:diffu_reac_steady1}); i.e., the one with more operators, because this set leads to a lower loss. Suppression of either of the identified active operators $\int_\Omega \nabla w\cdot \nabla C_1\text{d}v$ or $\int_\Omega wC_1\text{d}v$, yields another set of consistent inferred operators, shown in the legend on the right in Figure \ref{fig:stem_steady_correct}. This set identifies the steady state equation for $C_2$. This is a manifestation of the Confirmation Test of Consistency with Operator Suppression in Algorithm 3.

In the following we choose the two Laplacian operators, $\nabla^2C_1$ and $\nabla^2C_2$, as  targets and infer the remaining operators, $\overline{\Bchi}$:
\begin{align}
    \nabla^2C_1&=\frac{1}{\theta_{D_1}}\overline{\Bchi}\cdot\overline{\Btheta} \label{eq:steady_C1}\\
    \nabla^2C_2&=\frac{1}{\theta_{D_2}}\overline{\Bchi}\cdot\overline{\Btheta}.\label{eq:steady_C2}
\end{align}
Without noise, very accurate results are obtained even with small snapshots. Next, using Stage 2 of the VSI approach, i.e. by choosing the weighting function to be $w^h = C_1^h$ and $w^h = C_2^h$, the diffusion coefficients $\theta_{D_1}$ and $\theta_{D_2}$ in Equations (\ref{eq:steady_C1}) and (\ref{eq:steady_C2})  are correctly identified. The final identified system using both steady state and dynamic data generated from Model 1 is shown in Table \ref{ta:tab:results_model1_steady_dynamic}. Recall that the diffusion coefficients are not identifiable from early time data collected with small snapshots as discussed before (see Table \ref{ta:tab:results_model1_dynamic}, where this failure was cast in terms of the Laplacian operator). However, they are identified using the Confirmation Test of Consistency with Operator Suppression (Algorithm 3) in combination with Stage 2 of the VSI approach (Algorithms 1 and 2). 
\begin{table}[h]
\centering
\footnotesize
 \begin{tabular}{c|c}
 \hline
 mesh size& results\\
 \hline
\multirow{2}{*}{$400\times 400$} &$\int_{\Omega}w_1\frac{\partial C_1}{\partial t}\text{d}v=\int_{\Omega}-0.9998\nabla w_1 \cdot\nabla C_1\text{d}v
+\int_{\Omega}w_1 (0.1-0.9998C_1+0.9998C_1^2C_2)\text{d}v$ \\
 & $\int_{\Omega}w_2\frac{\partial C_2}{\partial t}\text{d}v=\int_{\Omega}-39.95\nabla w_2\cdot\nabla C_2+\int_{\Omega}w_2(0.8989-0.9987C_1^2C_2)\text{d}v$ \\
\hline
\multirow{2}{*}{$300\times 300$} &$\int_{\Omega}w_1\frac{\partial C_1}{\partial t}\text{d}v=\int_{\Omega}-1.001\nabla w_1 \cdot\nabla C_1\text{d}v
+\int_{\Omega}w_1 (0.1-0.9998C_1+0.9998C_1^2C_2)\text{d}v$ \\
 & $\int_{\Omega}w_2\frac{\partial C_2}{\partial t}\text{d}v=\int_{\Omega}-40.01\nabla w_2\cdot\nabla C_2+\int_{\Omega}w_2(0.9002-1.0002C_1^2C_2)\text{d}v$ \\
\hline
\multirow{2}{*}{$200\times 200$} &$\int_{\Omega}w_1\frac{\partial C_1}{\partial t}\text{d}v=\int_{\Omega}-0.995\nabla w_1 \cdot\nabla C_1\text{d}v
+\int_{\Omega}w_1 (0.1001-1.001C_1+1.001C_1^2C_2)\text{d}v$  \\ 
  & $\int_{\Omega}w_2\frac{\partial C_2}{\partial t}\text{d}v=\int_{\Omega}-40.02\nabla w_2\cdot\nabla C_2+\int_{\Omega}w_2(0.9005-1.005C_1^2C_2)\text{d}v$\\ \hline
\multirow{2}{*}{$100\times 100$} & $\int_{\Omega}w_1\frac{\partial C_1}{\partial t}\text{d}v=\int_{\Omega}-0.0994\nabla w_1 \cdot\nabla C_1\text{d}v
+\int_{\Omega}w_1 (0.0995-0.995C_1+0.995C_1^2C_2)\text{d}v$  \\ 
&$\int_{\Omega}w_2\frac{\partial C_2}{\partial t}\text{d}v=\int_{\Omega}-39.72\nabla w_2\cdot\nabla C_2+\int_{\Omega}w_2(-0.8937-0.993C_1^2C_2)\text{d}v$ \\
\hline
\multirow{2}{*}{$50\times 50$}  & $\int_{\Omega}w_1\frac{\partial C_1}{\partial t}\text{d}v=\int_{\Omega}-0.0994\nabla w_1 \cdot\nabla C_1\text{d}v
+\int_{\Omega}w_1 (0.0994-0.994C_1+0.994C_1^2C_2)\text{d}v$  \\ 
&$\int_{\Omega}w_2\frac{\partial C_2}{\partial t}\text{d}v=\int_{\Omega}-39.23\nabla w_2\cdot\nabla C_2+\int_{\Omega}w_2(-0.8827 -0.9807C_1^2C_2)\text{d}v$   \\ \hline
\end{tabular}
\caption{Results using steady state and dynamic data generated from Model 1.}
\label{ta:tab:results_model1_steady_dynamic}
\end{table}

\begin{figure}[hbtp]
\centering
\includegraphics[scale=0.53]{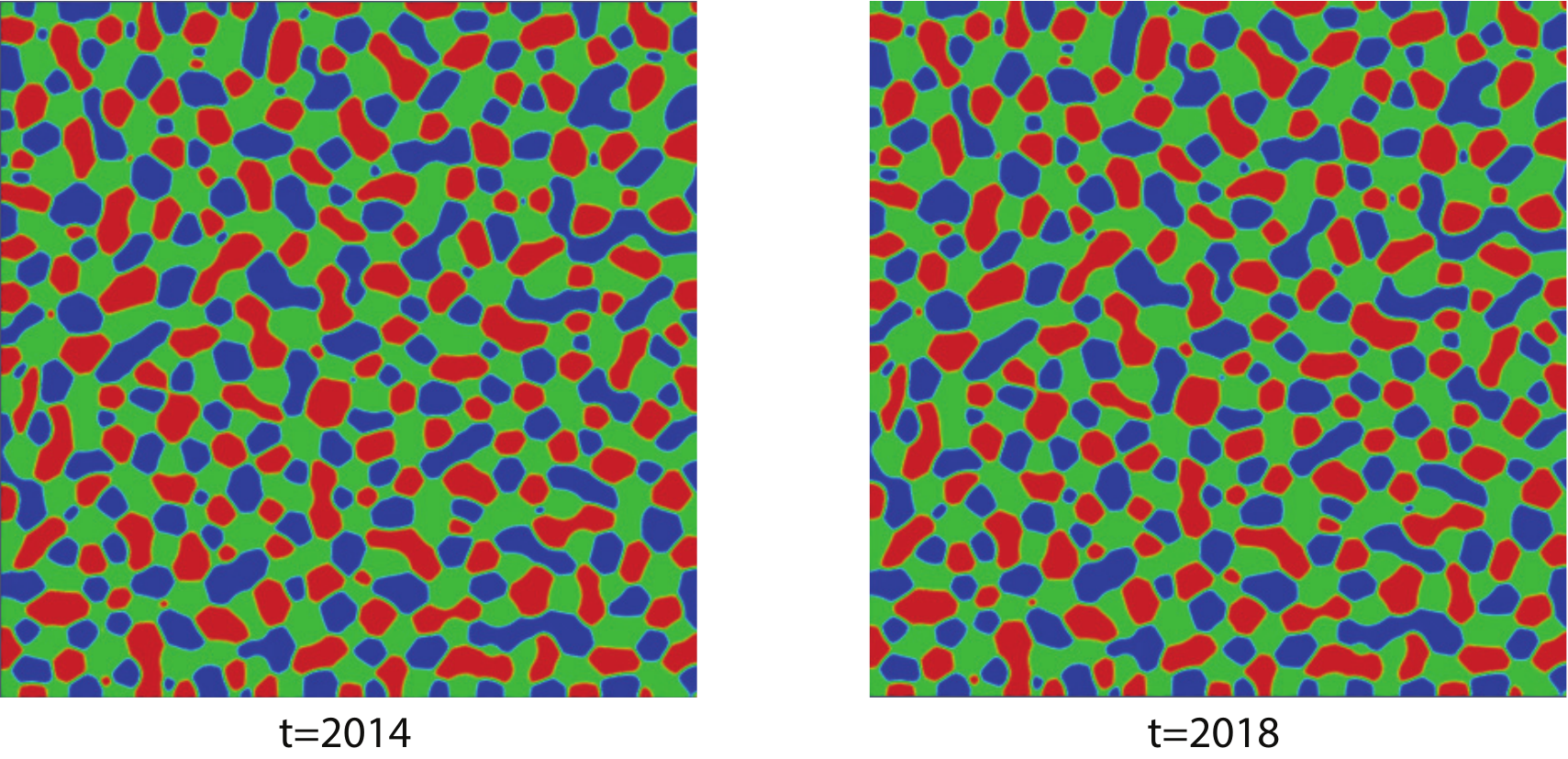}
\caption{$C_1$ concentrations at closely spaced time instants with Model 2. The Cahn-Hilliard equations evolve very slowly in the Ostwald ripening stage following the initial, fast spinodal decomposition regime. The system is at near-steady state around $t=2018$.}
\label{fig:Cahn_steady}
\end{figure}

\begin{figure}[hbtp]
\centering
\includegraphics[scale=0.35]{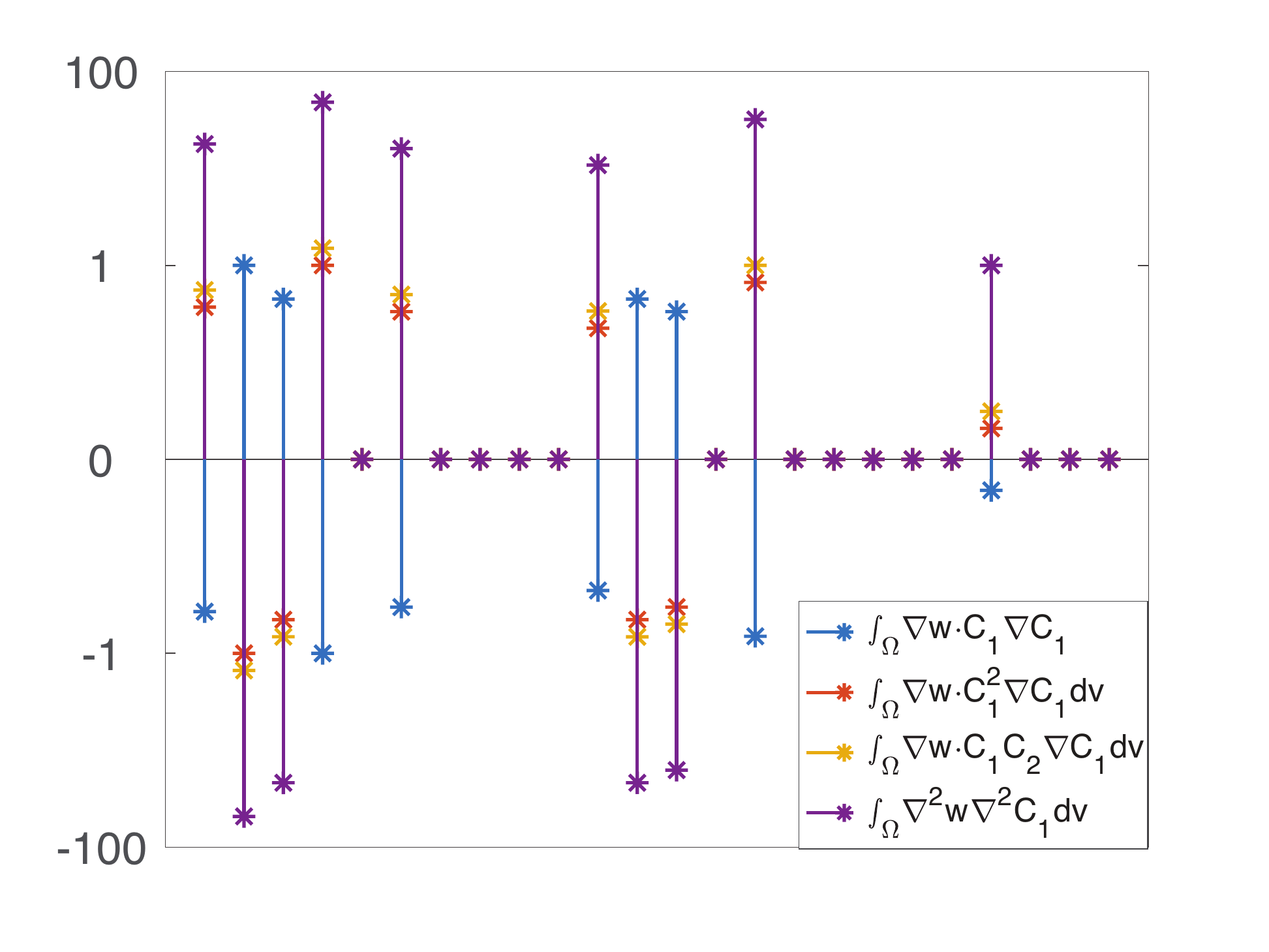}
\includegraphics[scale=0.35]{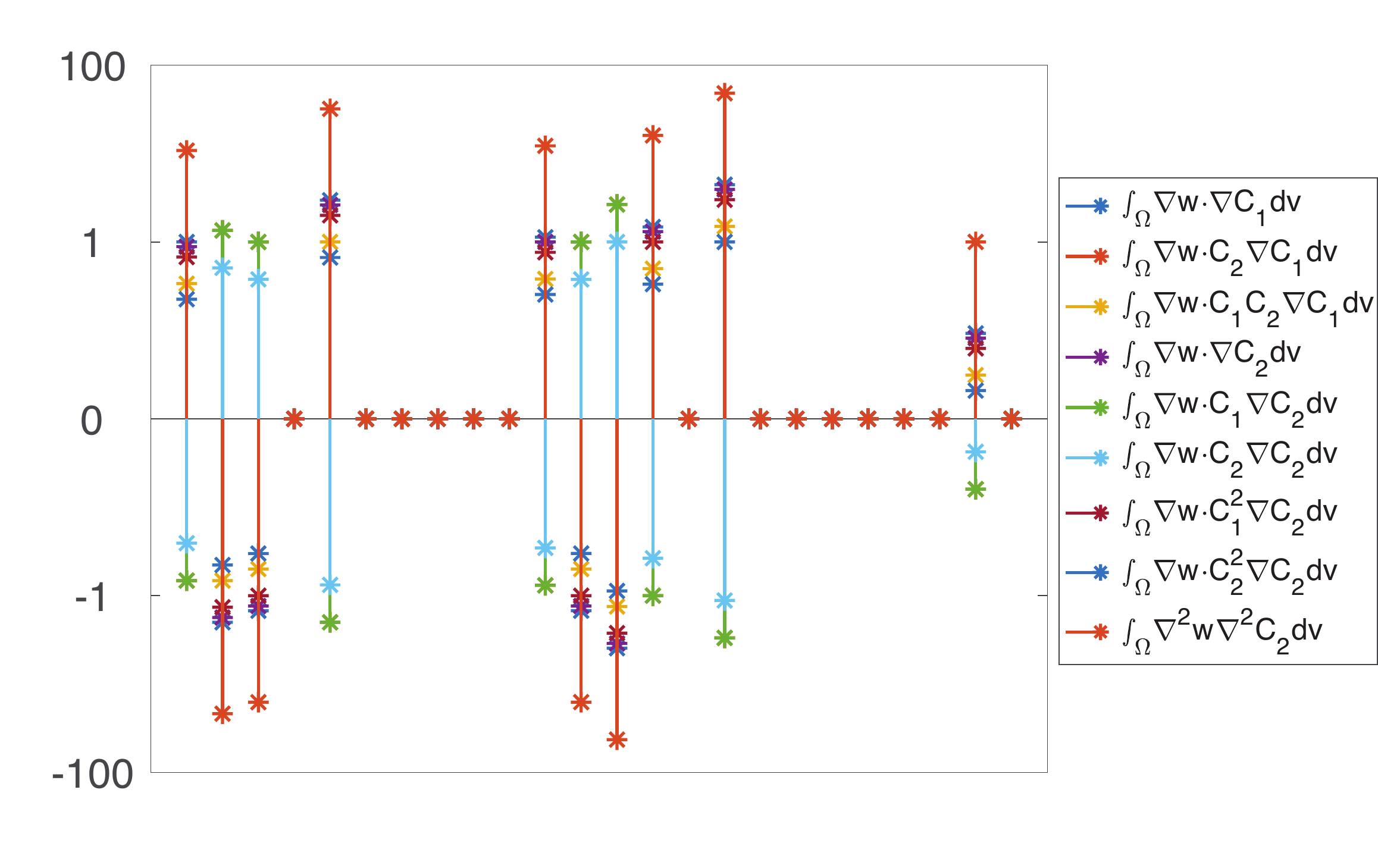}
\caption{Two sets of consistently inferred operators starting with different targets and using data generated from Model 2 at steady state. The coefficient values have been plotted on the logarithmic scale in order to make them discernible across orders of magnitude. The two consistent sets consist of the operators, from left to right, appearing on the right hand-side of Equations (\ref{eq:weak_value_model2-1}) and (\ref{eq:weak_value_model2-2}), and differ only by a scaling factor. The two consistent sets are labelled by the target operator shown in the legends of the two sub-plots. }
\label{fig:results_cahn_roundDrop_consistent}
\end{figure}

Cahn-Hilliard equations behave differently from the diffusion-reaction equations in terms of how they attain steady state. The initial, fast, spinodal decomposition regime is followed by the slower Ostwald ripening mchanism, in which the larger particles grow slowly at the expense of smaller ones. As a result, following the initial spinodal decomposition, the evolution of concentrations is extremely slow (see Figure \ref{fig:Cahn_steady}), and thus the system is at near-steady state:
\begin{align}
\nabla \cdot (M_1\nabla\mu_1)\approx0 \label{eq:CH1-SS}\\
\nabla \cdot (M_2\nabla\mu_2)\approx0
\label{eq:CH2-SS}
\end{align}
We again find the sets of consistent operators using Algorithm 3, and they appear in Figure \ref{fig:results_cahn_roundDrop_consistent}. Unlike the manner in which Algorithm 3 played out for the steady state diffusion-reaction equations, the choice of target operators that are common to both governing equations in the system does not result in convergence to a single set of operators. This is because the two equations in Model 2 have more than a single operator that is unique to each of them. When these are chosen as targets, Algorithm 3 correctly identifies two distinct sets of operators, each being self-consistent. These sets represent the two Cahn-Hilliard equations (\ref{eq:CH1-SS}) and (\ref{eq:CH2-SS}).

In the following we choose the two biharmonic operators, $\nabla^4C_1$ and $\nabla^4C_2$, as the targets and infer the remaining operators, $\overline{\Bchi}$:
\begin{align}
    \nabla^4C_1&=\frac{1}{\theta_{k_1}}\overline{\Bchi}\cdot\overline{\Btheta} \label{eq:steady_Cahn_C1}\\
    \nabla^4C_2&=\frac{1}{\theta_{k_2}}\overline{\Bchi}\cdot\overline{\Btheta}.\label{eq:steady_Cahn_C2}
\end{align} 
Very accurate results are obtained from data without noise, even if the snapshots are small.
 Following identification of the two consistent sets of operators at steady state by Algorithm 3, the scaling factors, $\theta_{k_1}$ and $\theta_{k_2}$ in Equations (\ref{eq:steady_Cahn_C1}) and (\ref{eq:steady_Cahn_C2}), i.e. the coefficients for biharmonic operators, are correctly identified using Stage 2 of VSI with dynamic data and weighting function $w^h=C_1^h$ and $w^h = C_2^h$. 
The final VSI results using both steady state and dynamic data generated from Model 2 are shown in Table \ref{ta:tab:results_model2_steady_dynamic}. 

\begin{table}[h]
\centering
\scriptsize
 \begin{tabular}{c|c}
 \hline
 mesh size& results\\
 \hline
\multirow{4}{*}{$400\times 400$} &$\int_{\Omega}w_1\frac{\partial C_1}{\partial t}\text{d}v=\int_{\Omega}\nabla w_1\cdot\left(-17.417+46.9146C_1+21.114C_2-46.91C_1^2-15.64C_2^2 \right)\nabla C_1\text{d}v$ \\
& $+ \int_{\Omega}\nabla w_2\nabla\cdot\left(-10.55+21.11C_1+15.64C_2-31.177C_1C_2 \right)\nabla C_2\text{d}v+\int_{\Omega}-0.978\nabla^2w_1\nabla^2C_1\text{d}v $\\
& $\int_{\Omega}w_2\frac{\partial C_2}{\partial t}\text{d}v=\int_{\Omega}\nabla w_2\cdot\left(-10.73+21.46C_1+15.9C_2-31.79C_1C_2 \right)\nabla C_1\text{d}v$  \\
& $+ \int_{\Omega}\nabla w_2\nabla\cdot\left(-12.13+15.88C_1+42.12C_2-15.88C_1^2-47.68C_2^2 \right)\nabla^2C_2\text{d}v+\int_{\Omega}-0.994\nabla^2w_2\nabla^2C_2\text{d}v$  \\
\hline
\multirow{4}{*}{$300\times 300$} &$\int_{\Omega}w_1\frac{\partial C_1}{\partial t}\text{d}v=\int_{\Omega}\nabla w_1\cdot\left(-15.992+43.08C_1+19.38C_2-43.08C_1^2-14.364C_2^2 \right)\nabla C_1\text{d}v$ \\
& $+ \int_{\Omega}\nabla w_2\nabla\cdot\left(-9.69+19.38C_1+14.36C_2-28.71C_1C_2 \right)\nabla C_2\text{d}v+\int_{\Omega}-0.898\nabla^2w_1\nabla^2C_1\text{d}v $ \\
& $\int_{\Omega}w_2\frac{\partial C_2}{\partial t}\text{d}v=\int_{\Omega}\nabla w_2\cdot\left(-9.82+19.65C_1+14.55C_2-29.11C_1C_2 \right)\nabla C_1\text{d}v$  \\
& $+ \int_{\Omega}\nabla w_2\nabla\cdot\left(-11.11+14.54C_1+38.56C_2-14.54C_1^2-43.66C_2^2 \right)\nabla^2C_2\text{d}v+\int_{\Omega}-0.91\nabla^2w_2\nabla^2C_2\text{d}v$  \\
\hline
\multirow{4}{*}{$200\times 200$} &$\int_{\Omega}w_1\frac{\partial C_1}{\partial t}\text{d}v=\int_{\Omega}\nabla w_1\cdot\left(-16.42+44.23C_1+19.9C_2-44.23C_1^2-14.74C_2^2 \right)\nabla C_1\text{d}v$  \\
& $+ \int_{\Omega}\nabla w_2\nabla\cdot\left(-9.95+19.9C_1+14.74C_2-29.48C_1C_2 \right)\nabla C_2\text{d}v+\int_{\Omega}-0.86\nabla^2w_1\nabla^2C_1\text{d}v $ \\
& $\int_{\Omega}w_2\frac{\partial C_2}{\partial t}\text{d}v=\int_{\Omega}\nabla w_2\cdot\left(-9.82+19.65C_1+14.55C_2-29.11C_1C_2 \right)\nabla C_1\text{d}v$  \\
& $+ \int_{\Omega}\nabla w_2\nabla\cdot\left(-11.1+14.54C_1+38.56C_2-14.53C_1^2-43.66C_2^2 \right)\nabla^2C_2\text{d}v+\int_{\Omega}-0.91\nabla^2w_2\nabla^2C_2\text{d}v$  \\
\hline
\multirow{4}{*}{$100\times 100$} &$\int_{\Omega}w_1\frac{\partial C_1}{\partial t}\text{d}v=\int_{\Omega}\nabla w_1\cdot\left(-15.31+41.25C_1+18.57C_2-41.25C_1^2-13.76C_2^2 \right)\nabla C_1\text{d}v$\\
& $+ \int_{\Omega}\nabla w_2\nabla\cdot\left(-9.27+18.55C_1+13.74C_2-27.48C_1C_2 \right)\nabla C_2\text{d}v+\int_{\Omega}-0.86\nabla^2w_1\nabla^2C_1\text{d}v $ \\
& $\int_{\Omega}w_2\frac{\partial C_2}{\partial t}\text{d}v=\int_{\Omega}\nabla w_2\cdot\left(-11.66+23.32C_1+17.27C_2-34.55C_1C_2 \right)\nabla C_1\text{d}v$   \\
& $+ \int_{\Omega}\nabla w_2\nabla\cdot\left(-13.18+17.24C_1+45.76C_2-17.24C_1^2-51.81C_2^2 \right)\nabla^2C_2\text{d}v+\int_{\Omega}-1.08\nabla^2w_2\nabla^2C_2\text{d}v$ \\
\hline
\multirow{4}{*}{$50\times 50$} &$\int_{\Omega}w_1\frac{\partial C_1}{\partial t}\text{d}v=\int_{\Omega}\nabla w_1\cdot\left(-15.06+40.57C_1+18.28C_2-40.57C_1^2-13.55C_2^2 \right)\nabla C_1\text{d}v$  \\
& $+ \int_{\Omega}\nabla w_2\nabla\cdot\left(-9.11+18.23C_1+13.5C_2-27C_1C_2 \right)\nabla C_2\text{d}v+\int_{\Omega}-0.846\nabla^2w_1\nabla^2C_1\text{d}v $ \\
& $\int_{\Omega}w_2\frac{\partial C_2}{\partial t}\text{d}v=\int_{\Omega}\nabla w_2\cdot\left(-13.71+27.42C_1+20.31C_2-40.62C_1C_2 \right)\nabla C_1\text{d}v$   \\
& $+ \int_{\Omega}\nabla w_2\nabla\cdot\left(-15.48+20.23C_1+53.81C_2-20.23C_1^2-60.94C_2^2 \right)\nabla^2C_2\text{d}v+\int_{\Omega}-1.27\nabla^2w_2\nabla^2C_2\text{d}v$  \\
\hline
\end{tabular}
\caption{Results using steady state and dynamic data generated from Model 2.}
\label{ta:tab:results_model2_steady_dynamic}
\end{table}

We remark that we have previously studied the effect of noise on VSI with full-field data \cite{WangCMAME2019}. Those results apply in entirety to VSI with steady state data as presented in this section. 

\begin{figure}
    \centering
    \subfigure[$C_1$ concentration: identified Model 2]{\includegraphics[scale=0.1]{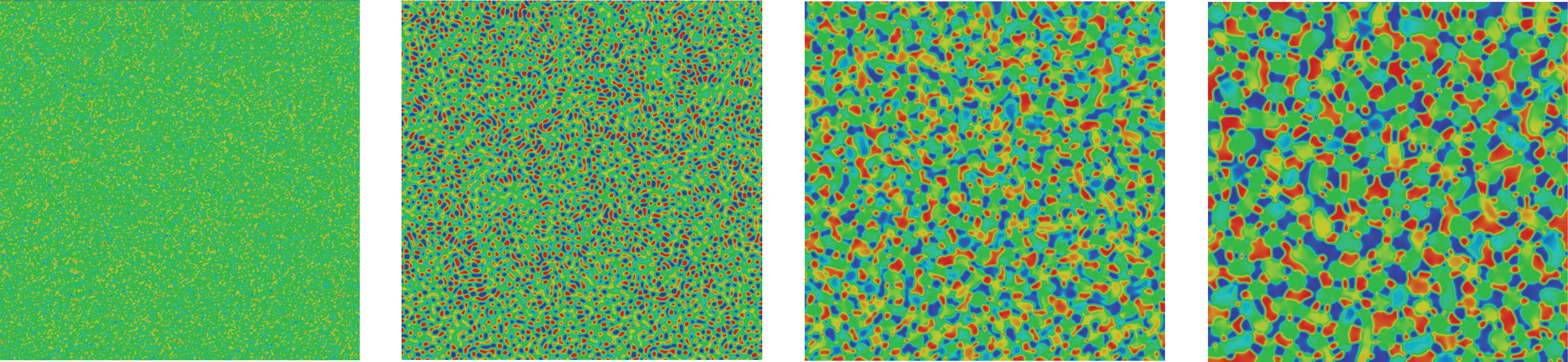} }
    \subfigure[$C_1$ concentration: true Model 2]{\includegraphics[scale=0.1]{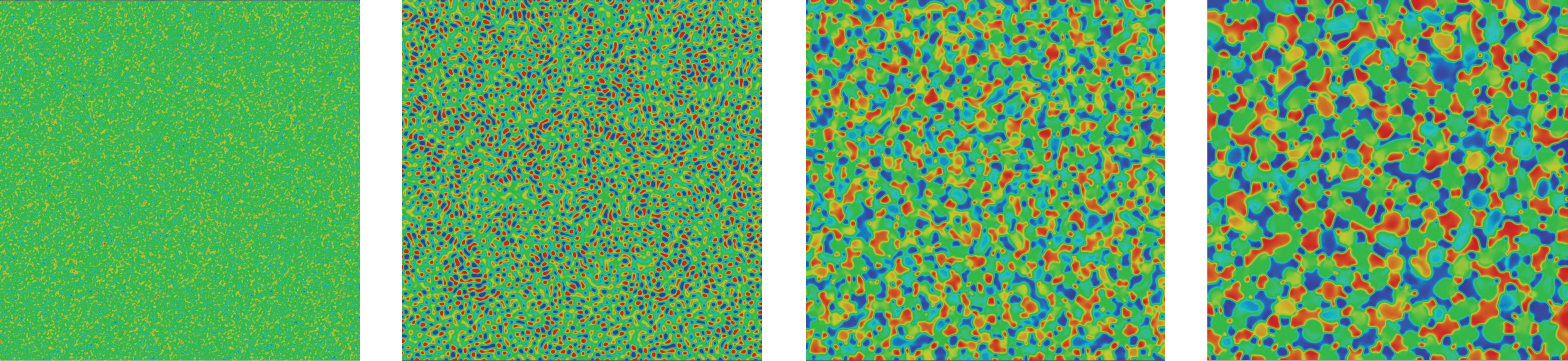} }
    \subfigure[First power ]{\includegraphics[scale=0.5]{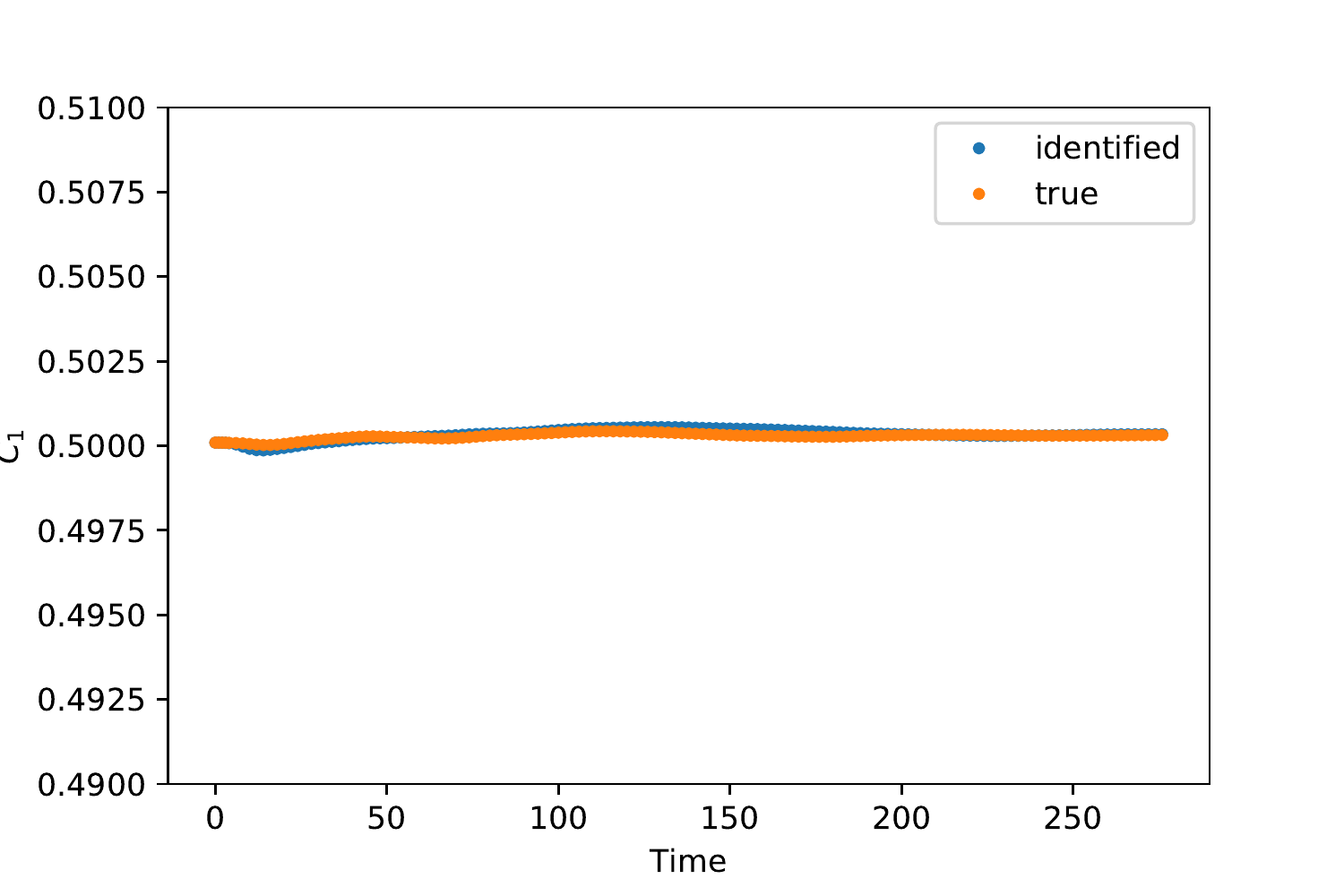} }
    \subfigure[Second power ]{\includegraphics[scale=0.5]{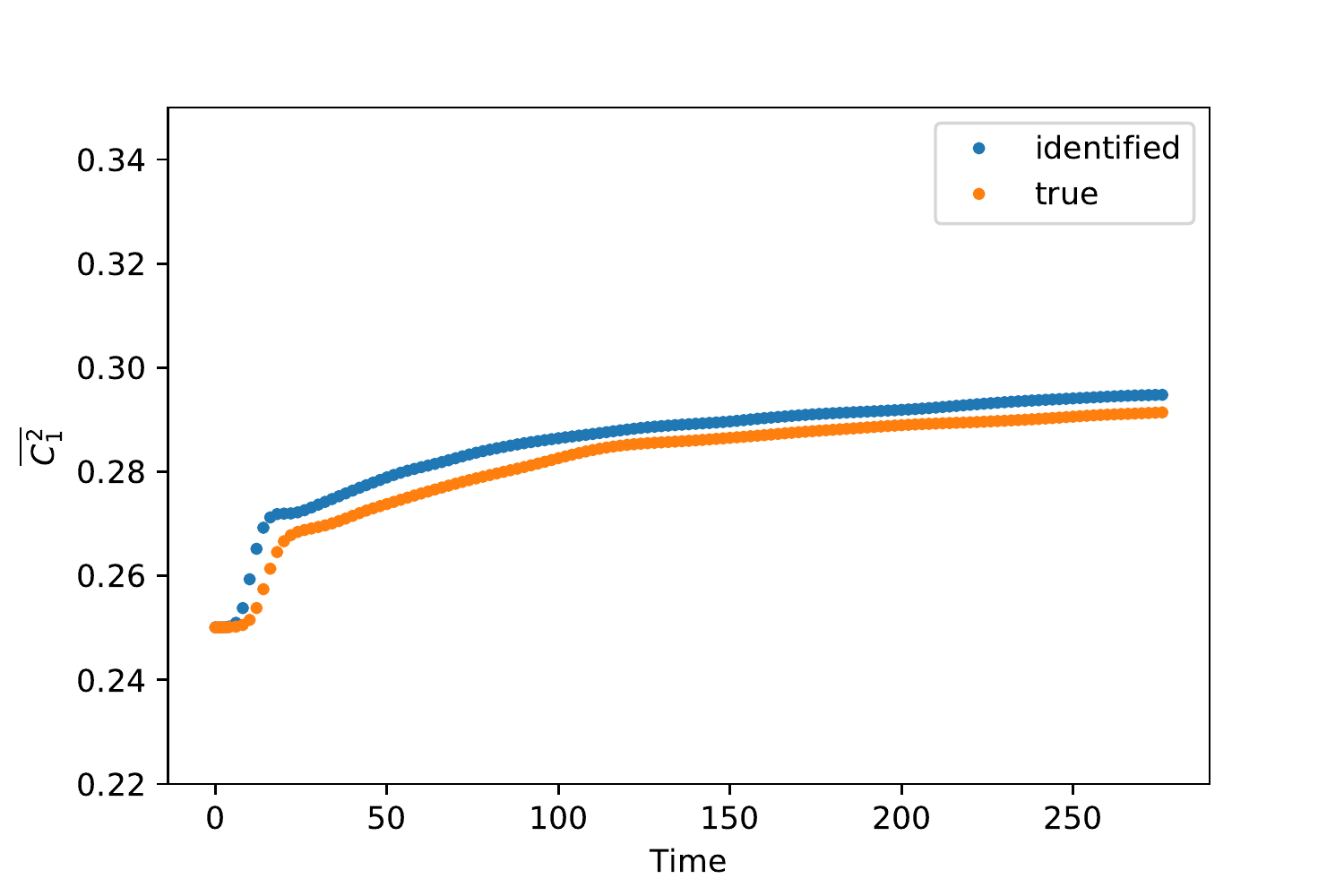} }
    \caption{Comparison of the $C_1$ concentration field simulated by the identified Model 2 (a), against the true Model 2 (b). The results are shown at $t=0, 10, 100, 276$. The results obtained from the identified model are similar to those obtained from the true model as shown by the good agreement between their first and second powers. } 
    \label{fig:comparison_cahn}
\end{figure}

The identified model is compared with the true model by solving with same initial conditions (see Figure (\ref{fig:comparison_cahn})). Although the results are different pointwise, being driven by the different kinetic coefficients that were inferred, the identified model produces similar patterns which are similar to the true model in terms of first and second powers of $C_1$.

\section{Discussion and conclusions}
\label{sec:concl}
The development of patterns in many physical phenomena is governed by a range of spatio-temporal PDEs. It is compelling to attempt to discover the analytic forms of these PDEs from data, because doing so immediately provides insight to the governing physics. System identification has been explored using the strong form \cite{SchmidtSCI2009, SchmidtPB2011,KutzPNAS2015, KutzIEEE2016, KutzSCIADV2017} and the weak form \cite{WangCMAME2019,Reinbold2020PRE,Bortz2020weakSINDY,Bortz2020weakSINDYPDE} of the PDEs as discussed in the Introduction. These techniques, however all need data that are spatially overlapping at distinct times in order to construct the time derivative operator. However, in materials physics in particular, the combination of processing and microscopy techniques results in datasets that are discordant with the PDE description of temporal evolution at fixed spatial locations. Microscopy data are only obtained over subsets of the entire domains, and the measured data at distinct times typically come from different experiments or specimens. This property of the datasets hinders the use of system identification methods including SINDy and VSI. While Bayesian methods, also discussed in the Introduction, are more flexible in working with such datasets, they require many forward solutions of the PDEs. Their computational expense makes it challenging to determine operators in PDEs from a comprehensive library of candidates.

For development of patterns in  materials physics, dynamic information can be inferred from data collected from different specimens and at different times, because of the similarity of the patterns across snapshots (see Figures \ref{fig:sample} and \ref{fig:diffusion_ini_pattern}). Many quantitative features, e.g. the powers of the concentration fields,  can be well approximated using such data. In the variational setting they can be recovered naturally by specific choices of weighting functions that allow a two-stage approach to separate the identification of algebraic and differential operators. VSI can thus be extended to incorporate spatially unrelated, non-overlapping and sparse information. Data available from larger snapshots; i.e., larger subsets of the entire domain yield better approximations of the global quantities as discussed in Section \ref{sec:statistical similarity}. The poor approximations using small snapshots pose challenges to system identification, resulting in inactive operators being incorrectly identified. This is particularly true for identification of operators that have much smaller contributions; e.g., operators with extremely small coefficients. In Appendix, we have shown how the VSI approach fares in this regime using dynamic data (see, Section \ref{sec:Appendix:small_parameter} and Tables \ref{ta:parametersDR}-\ref{ta:tab:results_model1_smallD_2}). In practice, however, we may have much higher resolution of data in experiments than the smallest snapshots of $50\times 50$ that were assumed here. This opens a door to improving the success of our methods.

We had already shown \cite{WangCMAME2019} that, given data from a few temporal snapshots but with high spatial resolution, VSI can pinpoint the complete governing equations of dynamical systems. Here, we have further leveraged steady state data that already provide information about most of the operators in the system, up to scaling factors corresponding to kinetic coefficients such as diffusivities and mobilities that determine the time scales. In Section \ref{sec:data at steady state without noise}, we demonstrated that, using data at steady state, we could identify all the operators in the PDEs up to a scaling factor. Without prior knowledge it is challenging to choose relevant operators to be the targets in our methods at steady state. However, by examining the consistency of the inferred operators using different candidates as the target, we were able to identify all the  governing equations at steady state. The associated Confirmation Test of Consistency with Operator Suppression has flavors of traditional approaches such as cross validation but is a mathematically stronger guarantee in that it confirms the identified result without prior knowledge or additional data. Identification of the single unknown in the original dynamic equations is then reduced to a straightforward exercise with two-stage VSI, which we have described for algebraic and differential operators, respectively. Combining the dynamic data with the high spatial resolution steady state data is also advantageous in identifying extreme cases such as very small parameters. We direct the reader to Section \ref{sec:Appendix:small_parameter} and Tables \ref{ta:parametersDR}-\ref{ta:tab:results_model1_smallD_2_steady} in the Appendix where the relevant results have been presented. Thus, in cases wherein steady state (or near-steady state) data are available, they serve as a ``zeroth'' step to be followed by only one of the stages that have been laid out for dynamic data in Section \ref{sec:incompdata}. 
If the dynamics remain far from steady state, the two-stage VSI is still applicable, given snapshots at sufficiently many time instants.

The challenge for system identification lies in selecting the correct operators from a large library of candidates. In our experience, the failure to identify the correct operators usually precedes the poor inference of coefficients, since the algorithms are then left with representing the data with the wrong model. In this communication we demonstrated our methods using synthetic data, and compared the identified models with the true models. In practice, quantifying the uncertainty becomes extremely important when using  real world data. In this work we also assumed that the physics is governed by first-order dynamics. The Confirmation Test may provide an avenue to system identification without this assumption, enabling the discernment of first-order in time, parabolic, PDEs from second-order, hyperbolic PDEs.

\res{We also note that system identification is significantly more challenging if greater numbers of candidate bases are retained. In this case, stepwise regression is roboust as it iteratively eliminates inactive operators instead of attempting to find a solution in ``one shot", thus mollifying the problem to a degree. Given the larger number of operator bases, the loss would be small and remain flat for several iterations before the model become underfit. However, VSI also suffers from missing key operators, similar to related methods such as SINDy \cite{KutzPNAS2015}. In this case, VSI would find the ``closest'' governing system where the missing basis are compensated by a larger number of other basis terms that are being considered. However difficulties may arise from non-unique combinations, and non-sparse solutions. We also anticipate that the loss will show many jumps indicating that operators that do contribute to the model have been eliminated in these iterations.}

Finally, we note that apart from assuming first-order dynamics, we have not imposed conditions to constrain the identified system of equations to be parabolic. This brings up an important question of stability. We have shown numerical comparisons in Figure \ref{fig:comparison_cahn} where the identified system predicts a solution field that is similar, but not identical to the original data
. Large errors in system identification can lead to instabilities in the solution. Such instabilities can be controlled by imposing constraints on combinations of coefficients in the library of operators. In other approaches (to appear soon) we have developed methods of staggering and ``round robin" operator suppression to eliminate system instabilities. A third approach worth exploration would be to incorporate forward trajectories computed from the inferred system in the loss function. 

\section*{Acknowledgements}
\label{sec:acknowledgements}
We acknowledge the support of Toyota Research Institute, Award \#849910, ``Computational framework for data-driven, predictive, multi-scale and multi-physics modeling of battery materials" (ZW and KG). Additional support: This material is based upon work supported by the Defense Advanced Research Projects Agency (DARPA) under Agreement No. HR0011199002, ``Artificial Intelligence guided multi-scale multi-physics framework for discovering complex emergent materials phenomena'' (ZW, XH and KG).

\bibliography{reference}

\begin{thebibliography}{45}
\providecommand{\natexlab}[1]{#1}
\providecommand{\url}[1]{\texttt{#1}}
\expandafter\ifx\csname urlstyle\endcsname\relax
  \providecommand{\doi}[1]{doi: #1}\else
  \providecommand{\doi}{doi: \begingroup \urlstyle{rm}\Url}\fi

\bibitem[Jiang et~al.(2016)Jiang, Rudraraju, Roy, der Ven, Garikipati, and
  Falk]{Jiangetal2016}
T.~Jiang, S.~Rudraraju, A.~Roy, A.~Van der Ven, K.~Garikipati, and M.~L. Falk.
\newblock Multi-physics simulations of lithiation-induced stress in litio
  electrode particles.
\newblock \emph{J. Phys. Chem. C}, 120, 2016.

\bibitem[Rudraraju et~al.(2016)Rudraraju, der Ven, and
  Garikipati]{Rudrarajuetal2016}
S.~Rudraraju, A.~Van der Ven, and K.~Garikipati.
\newblock Mechano-chemical spinodal decomposition: A phenomenological theory of
  phase transformations in multi-component crystalline solids.
\newblock \emph{Nature Computational Materials}, 2, 2016.

\bibitem[Teichert et~al.(2017)Teichert, Rudraraju, and
  Garikipati]{Teichertetal2017}
G.H. Teichert, S.~Rudraraju, and K.~Garikipati.
\newblock A variational treatment of material configurations with application
  to interface motion and microstructural evolution.
\newblock \emph{Journal of the Mechanics and Physics of Solids}, 99, 2017.

\bibitem[Cahn and Hilliard(1958)]{CahnHilliard1958}
J.~W. Cahn and J.~E. Hilliard.
\newblock Free energy of a nonuniform system. i interfacial energy.
\newblock \emph{J. Chem. Phys.}, 28, 1958.

\bibitem[Turing(1952)]{Turing1952}
A.~M. Turing.
\newblock The chemical basis of morphogenesis.
\newblock \emph{Phil. Trans. Roy. Soc. Lond. Ser. B.}, 237, 1952.

\bibitem[Gierer and Meinhardt(1972)]{Gierer1972}
A.~Gierer and H.~Meinhardt.
\newblock A theory of biological pattern formation.
\newblock \emph{Kybernetik}, 12, 1972.

\bibitem[Murray(1981)]{Murray1981}
J.~D. Murray.
\newblock On pattern formation mechanisms for lepidopteran wing patterns and
  mammalian coat markings.
\newblock \emph{Phil. Trans. Roy. Soc. Lond. Ser. B}, 295, 1981.

\bibitem[Dillon et~al.(1994)Dillon, Maini, and Othmer]{Dillon1994}
R.~Dillon, P.~K. Maini, and H.~G. Othmer.
\newblock Pattern formation in generalized turing systems i: Steady-state
  patterns in systems with mixed boundary conditions.
\newblock \emph{J. Math. Biol.}, 32, 1994.

\bibitem[Barrio et~al.(1999)Barrio, Varea, and Aragon]{Barrio1999}
R.~A. Barrio, C.~Varea, and J.~L. Aragon.
\newblock A two-dimensional numerical study of spatial pattern formation in
  interacting turing systems.
\newblock \emph{Bull. Math. Biol.}, 61, 1999.

\bibitem[Barrio et~al.(2009)Barrio, Baker, Vaughan, Tribuzy, de~Carvalho,
  Bassanezi, and Maini]{Barrio2009}
R.~A. Barrio, R.~E. Baker, B.~Vaughan, K.~Tribuzy, M.~R. de~Carvalho, Rodney
  Bassanezi, and P.~K. Maini.
\newblock Modeling the skin pattern of fishes.
\newblock \emph{Phys. Rev. E}, 79, 2009.

\bibitem[Maini et~al.(2012)Maini, Woolley, Baker, Gaffney, and
  Lee]{MainiByrne2012}
Philip~K. Maini, Thomas~E. Woolley, Ruth~E. Baker, Eamonn~A. Gaffney, and
  S.~Seirin Lee.
\newblock Turing's model for biological pattern formation and the robustness
  problem.
\newblock \emph{Interface Focus}, 2\penalty0 (4):\penalty0 487--496, 2012.
\newblock \doi{10.1098/rsfs.2011.0113}.

\bibitem[Spill et~al.(2015)Spill, Guerrero, Alarcon, Maini, and
  Byrne]{Spill2015}
F.~Spill, P.~Guerrero, T.~Alarcon, P.~K. Maini, and H.~Byrne.
\newblock Hybrid approaches for multiple-species stochastic reaction-diffusion
  models.
\newblock \emph{J. Comput. Phys.}, 299, 2015.

\bibitem[Korvasov\'{a} et~al.(2015)Korvasov\'{a}, Gaffney, Maini, Ferreira, and
  Klika]{Korvasova2015}
K.~Korvasov\'{a}, E.~A. Gaffney, P.~K. Maini, M.~A. Ferreira, and V.~Klika.
\newblock Investigating the turing conditions for diffusion-driven instability
  in the presence of a binding immobile substrate.
\newblock \emph{J. Theor. Biol.}, 367, 2015.

\bibitem[Garikipati(2017)]{GarikipatiJMPS2017}
K.~Garikipati.
\newblock Perspectives on the mathematics of biological patterning and
  morphogenesis.
\newblock \emph{J. Mech. Phys. Solids.}, 99, 2017.

\bibitem[Wise et~al.(2008)Wise, Lowengrub, Frieboes, and Cristini]{Wise2008}
S.~M. Wise, J.~S. Lowengrub, H.~B. Frieboes, and V.~Cristini.
\newblock Three-dimensional multispecies nonlinear tumor growth--model and
  numerical method.
\newblock \emph{J. Theor. Biol.}, 253, 2008.

\bibitem[Cristini et~al.(2009)Cristini, Li, Lowengrub, and Wise]{Cristini2009}
V.~Cristini, X.~Li, J.~S. Lowengrub, and S.~M. Wise.
\newblock Nonlinear simulations of solid tumor growth using a mixture model:
  invasion and branching.
\newblock \emph{J. Math. Biol.}, 58, 2009.

\bibitem[Lowengrub et~al.(2010)Lowengrub, Frieboes, Jin, Chuang, Li, Macklin,
  Wise, and Cristini]{Lowengrub2010}
J.~S. Lowengrub, H.~B. Frieboes, F~Jin, Y-L. Chuang, X.~Li, Macklin, S.~M.
  Wise, and V.~Cristini.
\newblock Nonlinear modelling of cancer: bridging the gap between cells and
  tumours.
\newblock \emph{Nonlinearity}, 23, 2010.

\bibitem[Lowengrub et~al.(2009)Lowengrub, R\"{a}tz, and Voigt]{Lowengrub2009}
J.~S. Lowengrub, A.~R\"{a}tz, and A.~Voigt.
\newblock Phase-field modeling of the dynamics of multicomponent vesicles:
  Spinodal decomposition, coarsening, budding, and fission.
\newblock \emph{Phys. Rev. E}, 79, 2009.

\bibitem[Vilanova et~al.(2013)Vilanova, Colominas, and Gomez]{Vilanova2013}
G.~Vilanova, I.~Colominas, and H.~Gomez.
\newblock Capillary networks in tumor angiogenesis: From discrete endothelial
  cells to phase-field averaged descriptions via isogeometric analysis.
\newblock \emph{Num. Meth. Biomed. Eng.}, 29, 2013.

\bibitem[Vilanova et~al.(2014)Vilanova, Colominas, and Gomez]{Vilanova2014}
G.~Vilanova, I.~Colominas, and H.~Gomez.
\newblock Coupling of discrete random walks and continuous modeling for
  three-dimensional tumor-induced angiogenesis.
\newblock \emph{Comput. Mech.}, 53, 2014.

\bibitem[Oden et~al.(2010)Oden, Hawkins, and Prudhomme]{Oden2010}
J.~T. Oden, A.~Hawkins, and S.~Prudhomme.
\newblock General diffuse-interface theories and an approach to predictive
  tumor growth modeling.
\newblock \emph{Math. Mod. Meth. App. Sci.}, 20, 2010.

\bibitem[Xu et~al.(2016)Xu, Vilanova, and Gomez]{Xu2016}
J.~Xu, G.~Vilanova, and H.~Gomez.
\newblock A mathematical model coupling tumor growth and angiogenesis.
\newblock \emph{PLoS ONE}, 11, 2016.

\bibitem[HilleRisLambers et~al.(2001)HilleRisLambers, Rietkerk, van~den Bosch,
  Prins, and de~Kroon]{HilleRisLambers2001}
R.~HilleRisLambers, M.~Rietkerk, F.~van~den Bosch, H.H.T. Prins, and
  H.~de~Kroon.
\newblock Vegetation pattern formation in semi-arid grazing systems.
\newblock \emph{Ecol.}, 82:\penalty0 50--61, 2001.

\bibitem[Rietkerk and van~de Koppel(2008)]{Rietkerk2008}
M.~Rietkerk and J.~van~de Koppel.
\newblock Regular pattern formation in real ecosystems.
\newblock \emph{Trends Ecol. Evol.}, 23:\penalty0 169--175, 2008.

\bibitem[Brooks et~al.(2011)Brooks, Gelman, Jones, and Meng]{Various2011}
Steve Brooks, Andrew Gelman, Galin Jones, and Xiao-Li Meng, editors.
\newblock \emph{{Handbook of Markov Chain Monte Carlo}}.
\newblock Chapman and Hall/CRC, 2011.
\newblock ISBN 978-1-4200-7941-8.
\newblock \doi{10.1201/b10905}.

\bibitem[Brunton et~al.(2016)Brunton, Proctor, and Kutz]{KutzPNAS2015}
S.~L. Brunton, J.~L. Proctor, and J.~N. Kutz.
\newblock Discovering governing equations from data by sparse identification of
  nonlinear dynamical systems.
\newblock \emph{Proc. Natl. Acad. Sci.}, 113, 2016.

\bibitem[Wang et~al.(2019)Wang, Huan, and Garikipati]{WangCMAME2019}
Z.~Wang, X.~Huan, and K.~Garikipati.
\newblock Variational system identification of the partial differential
  equations governing the physics of pattern-formation: Inference under varying
  fidelity and noise.
\newblock \emph{Computer Methods in Applied Mechanics and Engineering},
  356:\penalty0 44 -- 74, 2019.
\newblock ISSN 0045-7825.
\newblock \doi{https://doi.org/10.1016/j.cma.2019.07.007}.

\bibitem[Mangan et~al.(2016)Mangan, Brunton, Proctor, and Kutz]{KutzIEEE2016}
N.~M. Mangan, S.~L. Brunton, J.~L. Proctor, and J.~N. Kutz.
\newblock Inferring biological networks by sparse identification of nonlinear
  dynamics.
\newblock \emph{IEEE Trans. Mol. Biol. Multi-Scale Commun.}, 2, 2016.

\bibitem[Rudy et~al.(2017)Rudy, Brunton, Proctor, and Kutz]{KutzSCIADV2017}
S.~H. Rudy, S.~L. Brunton, J.~L. Proctor, and J.~N. Kutz.
\newblock Data-driven discovery of partial differential equations.
\newblock \emph{Sci. Adv.}, 3, 2017.

\bibitem[Raissi et~al.(2019)Raissi, Perdikaris, and Karniadakis]{Raissi2019}
M.~Raissi, P.~Perdikaris, and G.E. Karniadakis.
\newblock Physics-informed neural networks: A deep learning framework for
  solving forward and inverse problems involving nonlinear partial differential
  equations.
\newblock \emph{J. Comput. Phys.}, 378, 2019.

\bibitem[Wang et~al.(2020)Wang, Huan, and Garikipati]{TAML2020}
Z.~Wang, B.~Wu~X. Huan, and K.~Garikipati.
\newblock A perspective on regression and bayesian approaches for system
  identification of pattern formation dynamics.
\newblock \emph{Theoretical Appl. Mech. Lett.}, 2020.
\newblock To appear.

\bibitem[Atkinson et~al.(2019)Atkinson, Waad~Subber, Khan, Hawi, and
  Ghanem]{Ghanem2019}
Steven Atkinson, Liping~Wang Waad~Subber, Genghis Khan, Philippe Hawi, and
  Roger Ghanem.
\newblock Data-driven discovery of free-form governing differential equations.
\newblock \emph{Second Workshop on Machine Learning and the Physical Sciences},
  2019.

\bibitem[Yair et~al.(2017)Yair, Talmon, Coifman, and Kevrekidis]{YairE7865}
Or~Yair, Ronen Talmon, Ronald~R. Coifman, and Ioannis~G. Kevrekidis.
\newblock Reconstruction of normal forms by learning informed observation
  geometries from data.
\newblock \emph{Proceedings of the National Academy of Sciences}, 114\penalty0
  (38):\penalty0 E7865--E7874, 2017.
\newblock \doi{10.1073/pnas.1620045114}.

\bibitem[Reinbold et~al.(2020)Reinbold, Gurevich, and
  Grigoriev]{Reinbold2020PRE}
Patrick A.~K. Reinbold, Daniel~R. Gurevich, and Roman~O. Grigoriev.
\newblock Using noisy or incomplete data to discover models of spatiotemporal
  dynamics.
\newblock \emph{Phys. Rev. E}, 101:\penalty0 010203, Jan 2020.
\newblock \doi{10.1103/PhysRevE.101.010203}.

\bibitem[Messenger and Bortz(2020{\natexlab{a}})]{Bortz2020weakSINDY}
Daniel~A. Messenger and David~M. Bortz.
\newblock Weak sindy: Galerkin-based data-driven model selection.
\newblock \emph{arXiv:2005.04339}, 2020{\natexlab{a}}.

\bibitem[Messenger and Bortz(2020{\natexlab{b}})]{Bortz2020weakSINDYPDE}
Daniel~A. Messenger and David~M. Bortz.
\newblock Weak sindy for partial differential equations.
\newblock \emph{arXiv:2007.02848}, 2020{\natexlab{b}}.

\bibitem[Cottrell et~al.(2009)Cottrell, Hughes, and
  Bazilevs]{CottrellHughesBazilevs2009}
J.~Cottrell, T.~Hughes, and Y.~Bazilevs.
\newblock Isogeometric analysis: Toward integration of cad and fea.
\newblock \emph{Wiley, Chichester}, 2009.

\bibitem[Cand{\`{e}}s et~al.(2006)Cand{\`{e}}s, Romberg, and Tao]{Candes2006a}
Emmanuel~J. Cand{\`{e}}s, Justin Romberg, and Terence Tao.
\newblock {Robust Uncertainty Principles: Exact Signal Reconstruction From
  Highly Incomplete Frequency Information}.
\newblock \emph{IEEE Transactions on Information Theory}, 52\penalty0
  (2):\penalty0 489--509, 2006.
\newblock ISSN 0018-9448.
\newblock \doi{10.1109/TIT.2005.862083}.

\bibitem[Donoho(2006)]{Donoho2006a}
David~L. Donoho.
\newblock {Compressed sensing}.
\newblock \emph{IEEE Transactions on Information Theory}, 52\penalty0
  (4):\penalty0 1289--1306, 2006.
\newblock ISSN 00189448.
\newblock \doi{10.1109/Tit.2006.871582}.

\bibitem[James et~al.(2013)James, Witten, Hastie, and Tibshirani]{ISL}
G.~James, D.~Witten, T.~Hastie, and R.~Tibshirani.
\newblock An introduction to statistical learning.
\newblock \emph{Springer New York, Inc., New York, NY, USA.}, 2013.

\bibitem[Mangan et~al.(2017)Mangan, Kutz, Brunton, and Proctor]{KutzPRS2017}
N.~M. Mangan, J.~N. Kutz, S.~L. Brunton, and J.~L. Proctor.
\newblock Model selection for dynamical systems via sparse regression and
  information criteria.
\newblock \emph{Proceedings of the Royal Society A: Mathematical, Physical and
  Engineering Sciences}, 473\penalty0 (2204):\penalty0 20170009, 2017.
\newblock \doi{10.1098/rspa.2017.0009}.

\bibitem[Schnakenberg(1976)]{Schnakenberg1976}
J.~Schnakenberg.
\newblock Network theory of microscopic and macroscopic behavior of master
  equation systems.
\newblock \emph{Rev. Mod. Phys.}, 48, 1976.

\bibitem[Schmidt and Lipson(2009)]{SchmidtSCI2009}
M.~Schmidt and H.~Lipson.
\newblock Distilling free-form natural laws from experimental data.
\newblock \emph{Science}, 03, 2009.

\bibitem[Schmidt et~al.(2011)Schmidt, Vallabhajosyula, Jenkins, Hood, Soni,
  Wikswo, and Lipson]{SchmidtPB2011}
M.~D Schmidt, R.~R Vallabhajosyula, J.~W Jenkins, J.~E Hood, A.~S Soni, J~P
  Wikswo, and H.~Lipson.
\newblock Automated refinement and inference of analytical models for metabolic
  networks.
\newblock \emph{Phys. Biol.}, 8, 2011.

\bibitem[Allen and Cahn(1979)]{Allen1979}
Samuel~M. Allen and John~W. Cahn.
\newblock A microscopic theory for antiphase boundary motion and its
  application to antiphase domain coarsening.
\newblock \emph{Acta Metall.}, 27, 1979.

\end{thebibliography}
\bibliographystyle{unsrtnat}

\appendix

\section*{Appendix}

\section{Similarity of snapshots in Cahn-Hilliard and Allen-Cahn Equations}
\label{sec:Appendix:statistical_similarity}
In this section, we numerically validate our assumptions of similarity between snapshots  for the other two types of PDEs that commonly used to model pattern formation. For the ease of the reader, we restate the two assumptions we made:
\begin{equation}
    \left\vert\,\,\int\limits_{\overline{\Omega}^I}\boldsymbol{j}\cdot\boldsymbol{n} \text{d}S\,\right\vert \le \varepsilon_\text{b},\quad i = 1,2
    \label{eq:statsimboundaryeps_appendix}
\end{equation}
and
\begin{equation}
    \left\vert\frac{1}{\bar{V}^I}\int\limits_{\overline{\Omega}^I} C^k \text{d}V - \frac{1}{\bar{V}^J}\int\limits_{\overline{\Omega}^J} C^k \text{d}V\right\vert \le \varepsilon_\text{r}.
    \label{eq:statsimregioneps_appendix}
\end{equation}
Relation (\ref{eq:statsimboundaryeps_appendix}) indicates that the total flux vanishes and Relation (\ref{eq:statsimregioneps_appendix}) indicates that the powers of the concentration field within one snapshot can be approximated by the data over another spatially unrelated snapshot. 

We restate the Cahn-Hilliard equations (Model 2 in the main document):
\begin{align}
  \frac{\partial C_1}{\partial t}&=\nabla \cdot (M_1\nabla\mu_1)\label{eq:Cahn-C1_appendix}\\
    \frac{\partial C_2}{\partial t}&=\nabla \cdot (M_2\nabla\mu_2)\label{eq:Cahn-C2_appendix}\\
\mu_1&=\frac{\partial g}{\partial C_1}-k_1\nabla^2C_1 \label{eq:Cahn-mu_1_appendix}\\
\mu_2&=\frac{\partial g}{\partial C_2}-k_2\nabla^2C_2 \label{eq:Cahn-mu_2_appendix}\\
 \text{with}& \quad \nabla \mu_1\cdot\bn=0; \nabla C_1\cdot\bn = 0 \text{ on }\Gamma  \label{eq:CahnHillDirBC_appendix}\\
 & \quad \nabla \mu_2\cdot\bn=0; \nabla C_2\cdot\bn = 0  \text{ on }\Gamma
\end{align}
and $g$ is the non-convex, ``homogeneous'' free energy density function, whose form has been chosen as
\begin{align}
   g(C_1,C_2)&=\frac{3d}{2s^4}\left((2C_1-1)^2+(2C_2-1)^2\right)^2+\frac{d}{s^3}(2C_2-1)\left((2C_2-1)^2-3(2C_1-1)^2\right)\nonumber\\
      &-\frac{3d}{2s^2}\left((2C_1-1)^2+(2C_2-1)^2\right).
   \label{eq:freeEnergy_g_appendix}
 \end{align}
 
The Allen-Cahn equation \cite{Allen1979} is another type of parabolic PDE  that is commonly used to model pattern formation. This following is a coupled Cahn-Hilliard/Allen-Cahn equations system  where the field $C_2$ describes growth of alloy precipitates from nuclei created by a process of spinodal decomposition and Ostwald ripening that control the field $C_1$.
\begin{align}
  \frac{\partial C_1}{\partial t}&=\nabla \cdot (M_1\nabla\mu_1)\label{eq:Alen-C1_appendix}\\
    \frac{\partial C_2}{\partial t}&=-M_2\mu_2\label{eq:ALen-C2_appendix}\\
\mu_1&=\frac{\partial g}{\partial C_1}-\nabla\cdot k_1\nabla C_1 \label{eq:Alen-mu-eta_appendix}\\
\mu_2&=\frac{\partial g}{\partial C_2}-\nabla\cdot k_2\nabla C_2 \label{eq:Alen-J_appendix}\\
 \text{with}& \quad \nabla \mu_1\cdot\bn=0; \nabla C_1\cdot\bn = 0 \text{ on }\Gamma  \label{eq:AllenCahnDirBC_appendix}\\
 & \quad \nabla C_2\cdot\bn=0 \text{ on }\Gamma 
\end{align}
 The free energy density $g(C_1, C_2)$ couples these processes through $\mu_1$ and $\mu_2$. An important difference from Cahn-Hilliard equations is that $C_2$ is a non-conserved order parameter, which defines the identities of the precipitate, described by composition field $C_1$, and matrix phases. The parameters are smmarized in Table \ref{ta:parameters_appendix}.
\begin{table}[h]
\centering
 \begin{tabular}{|c|c|c|c|c|c|}
 \hline
$M_1$& $M_2$& $k_1$&$k_2$ &$d$ &$s$\\\hline
 0.1& 0.1& 10&10& 0.4& 0.7\\ \hline
\end{tabular}
\caption{parameters used in the simulations.}
\label{ta:parameters_appendix}
\end{table}

\begin{figure}[hbtp]
\centering
\subfigure[CH: $C_1$]{\includegraphics[scale=0.15]{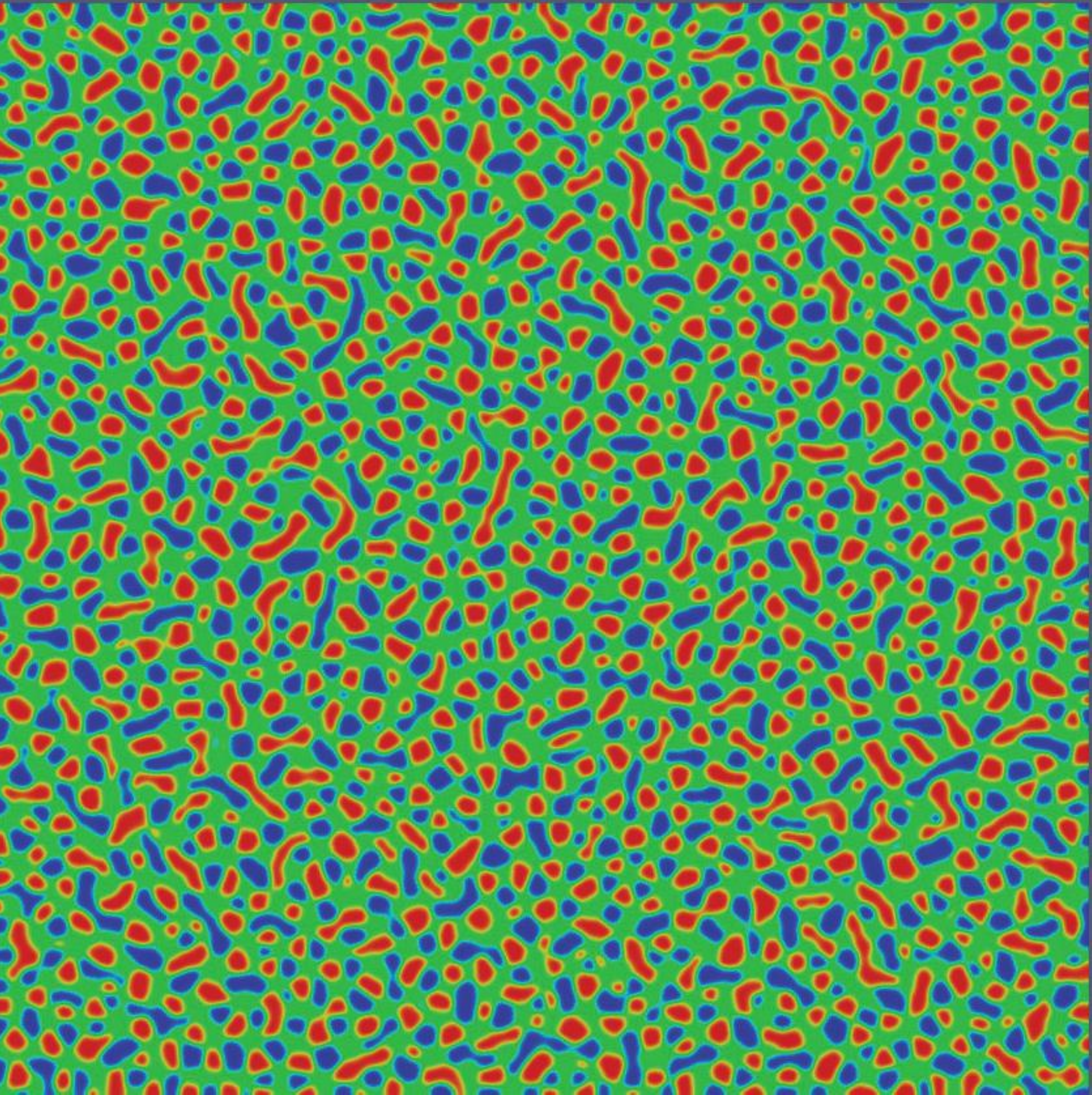}}
\subfigure[CH: $C_2$]{\includegraphics[scale=0.15]{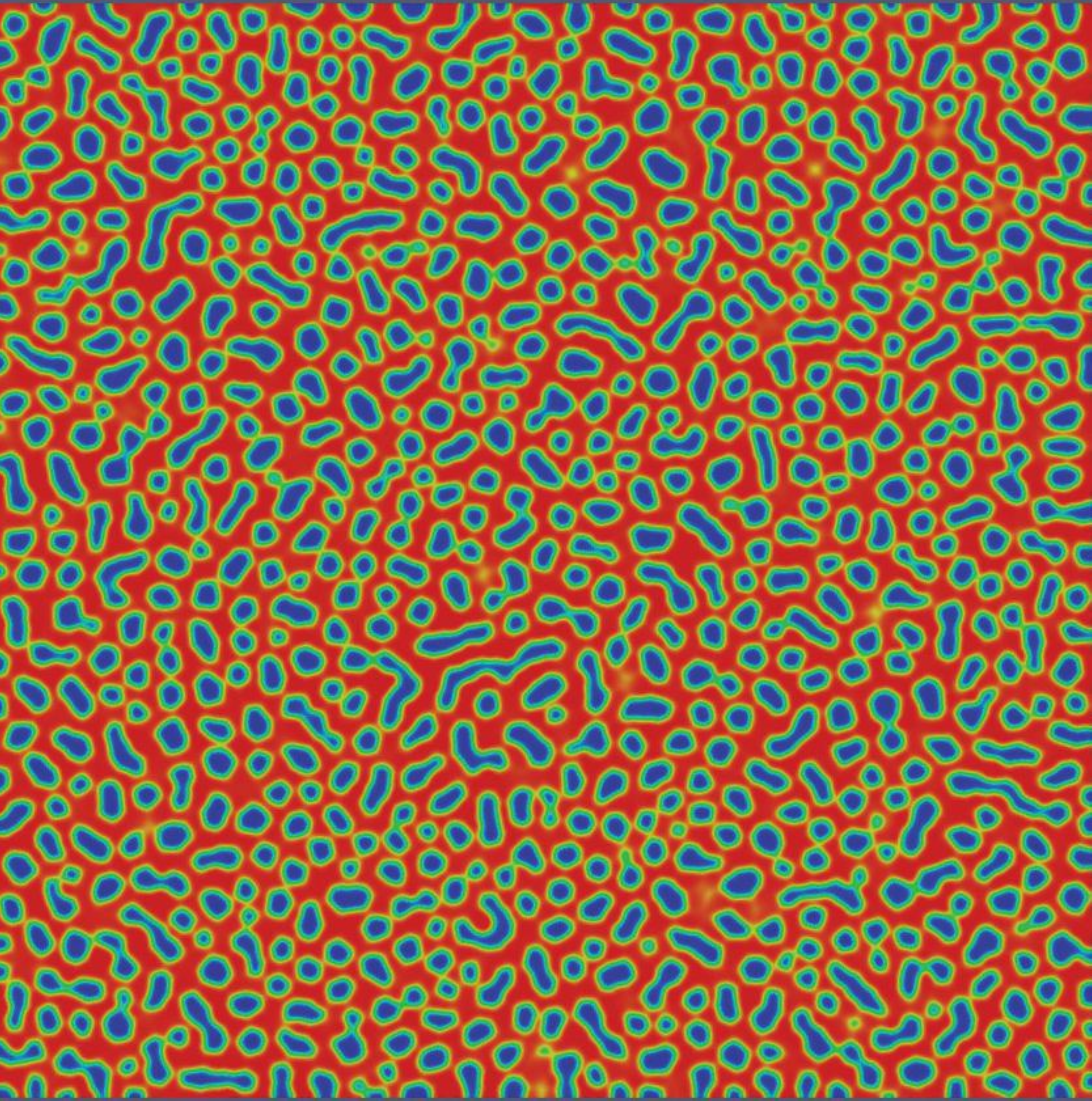}}\\
\subfigure[AC: $C_1$]{\includegraphics[scale=0.15]{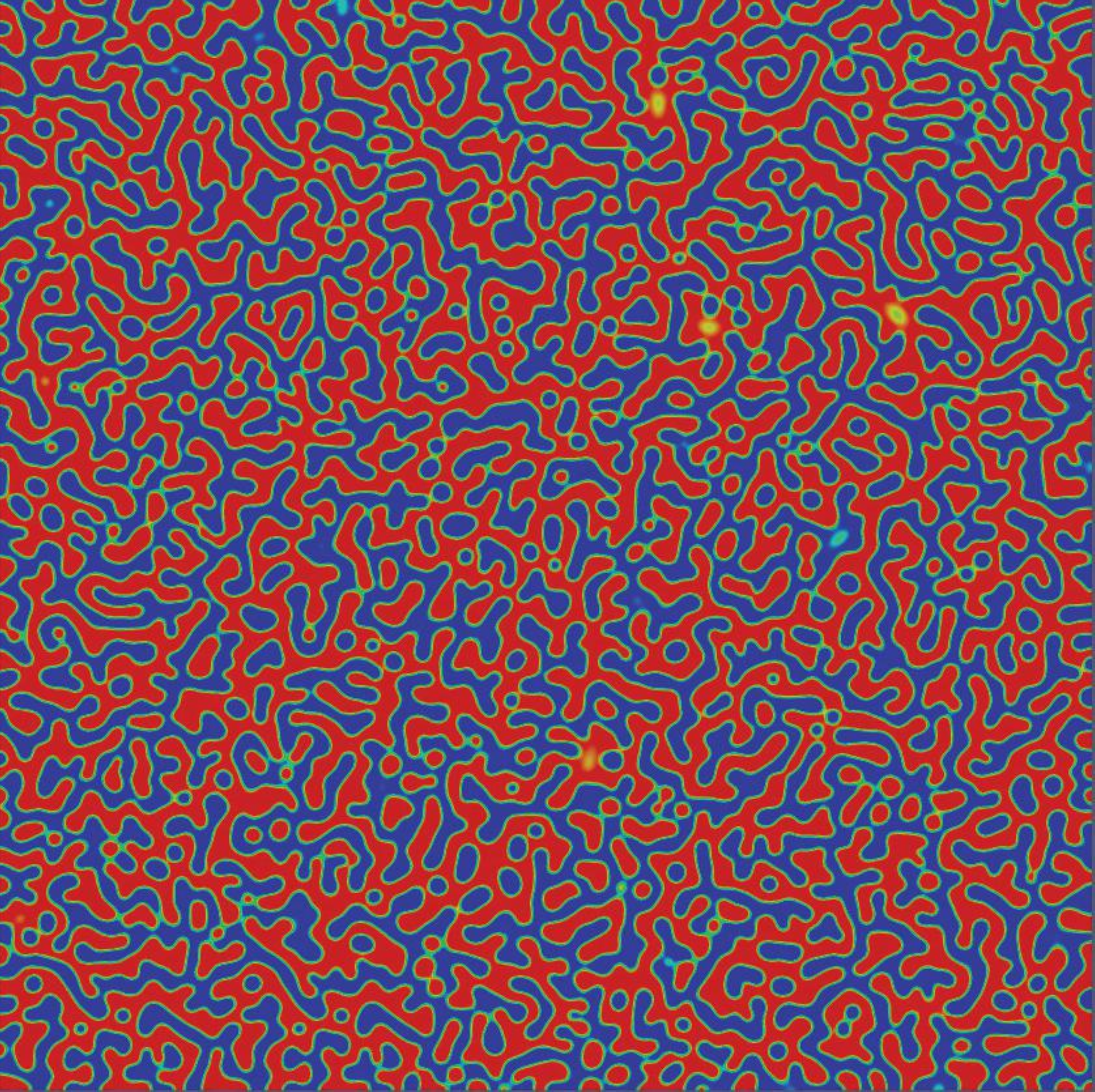}}
\subfigure[AC: $C_2$]{\includegraphics[scale=0.15]{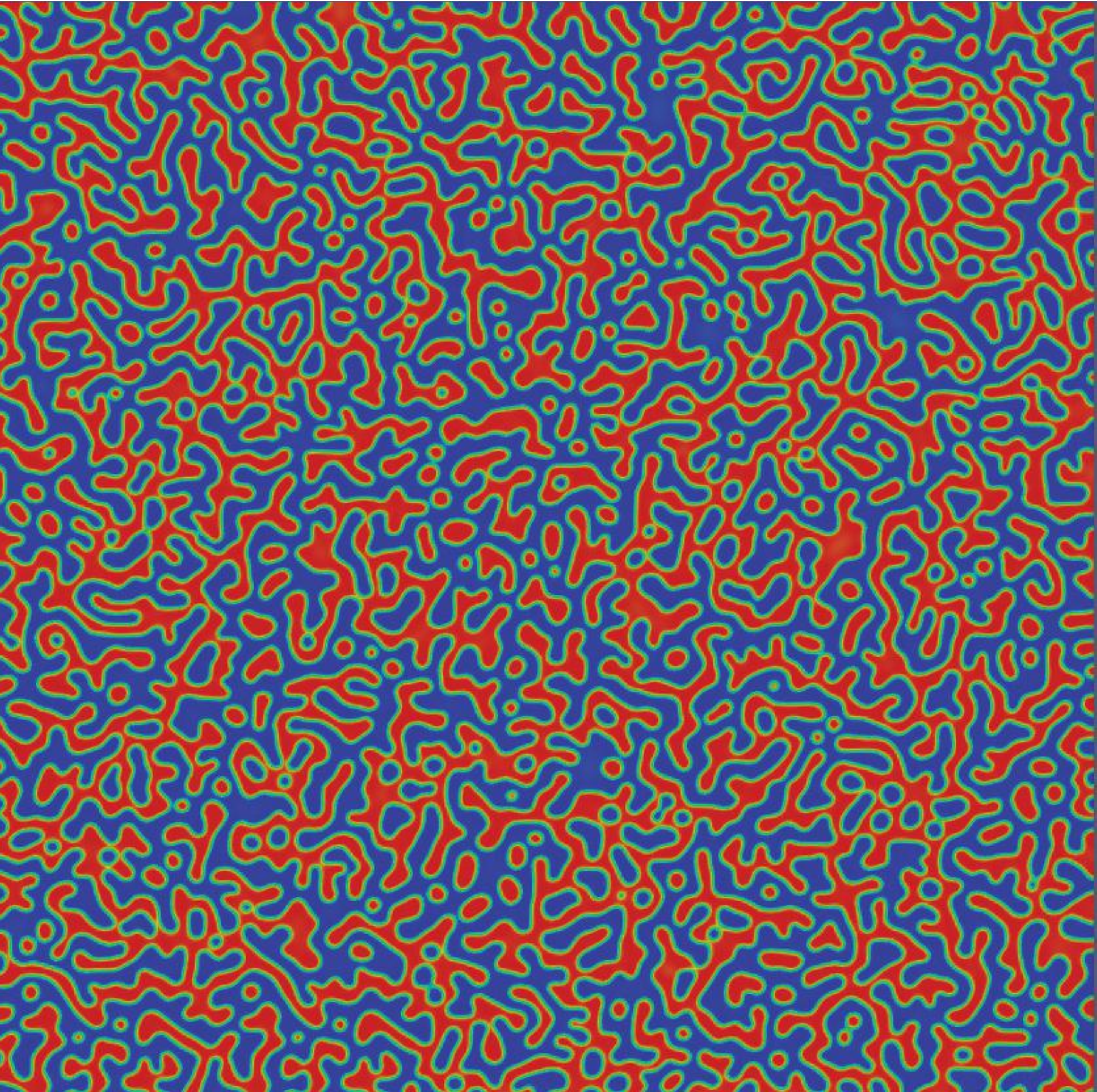}}
\caption{The upper plots show the patterning of ``particles" by Cahn-Hilliard equation, and the lower plots shows the patterning of short stripes by Allen-Cahn equations.}
\label{fig:patterns_CH_AC}
\end{figure}

\begin{figure}[hbtp]
\centering
\subfigure[CH: total flux for $C_1$ concentration]{\includegraphics[scale=0.45]{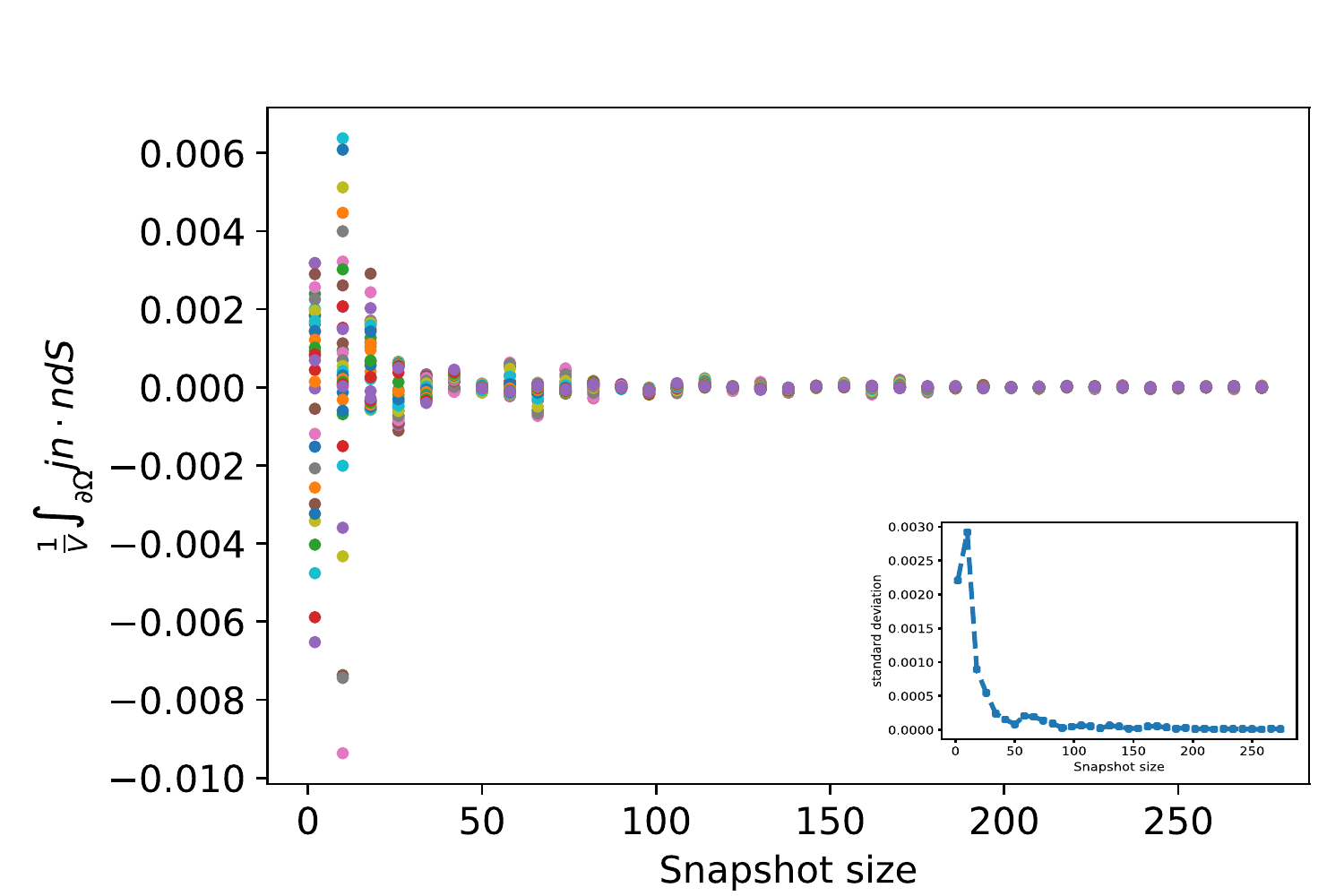} } 
\subfigure[CH: total flux for $C_2$ concentration]{\includegraphics[scale=0.45]{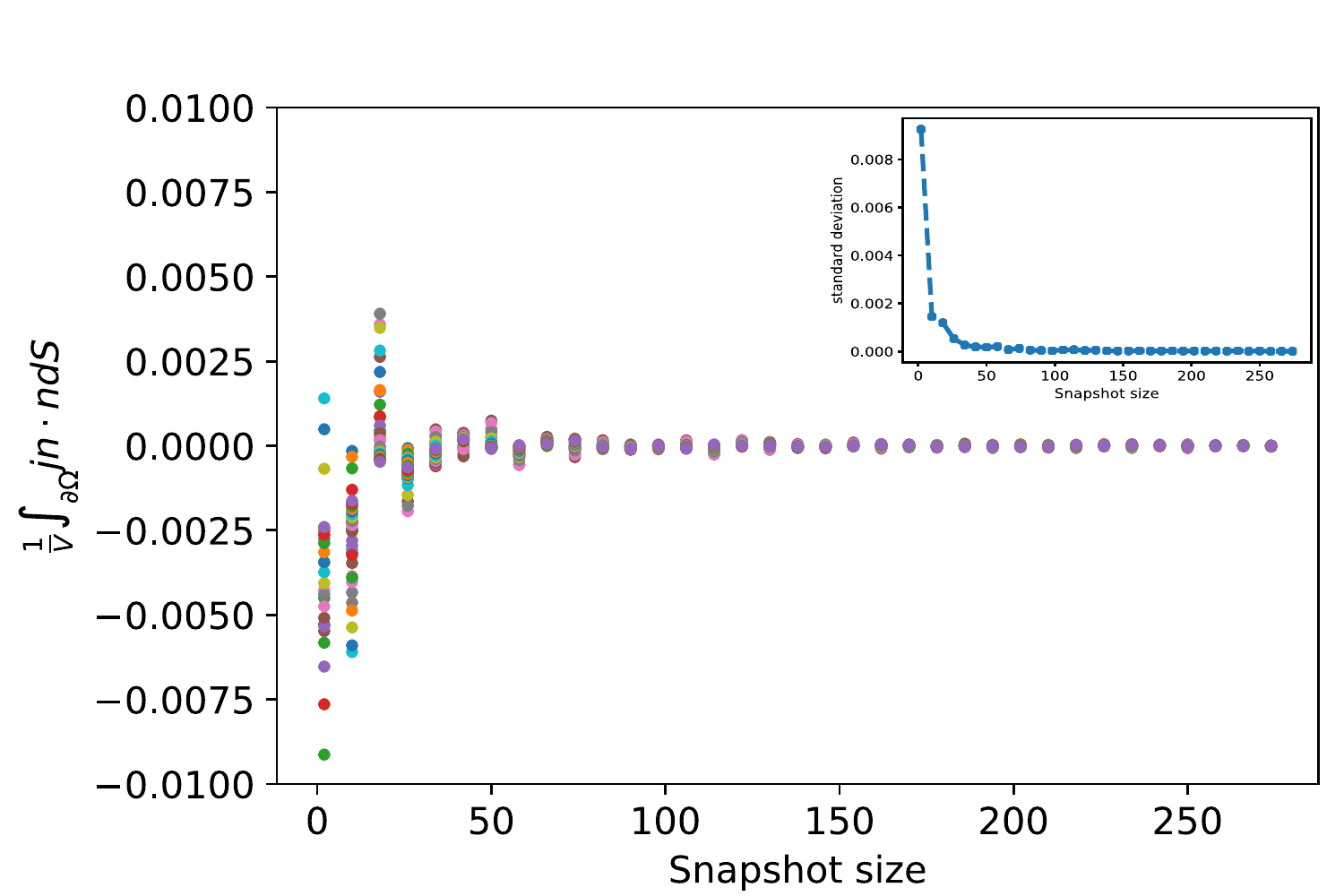} } 
\subfigure[AC: total flux for $C_1$ concentration]{\includegraphics[scale=0.45]{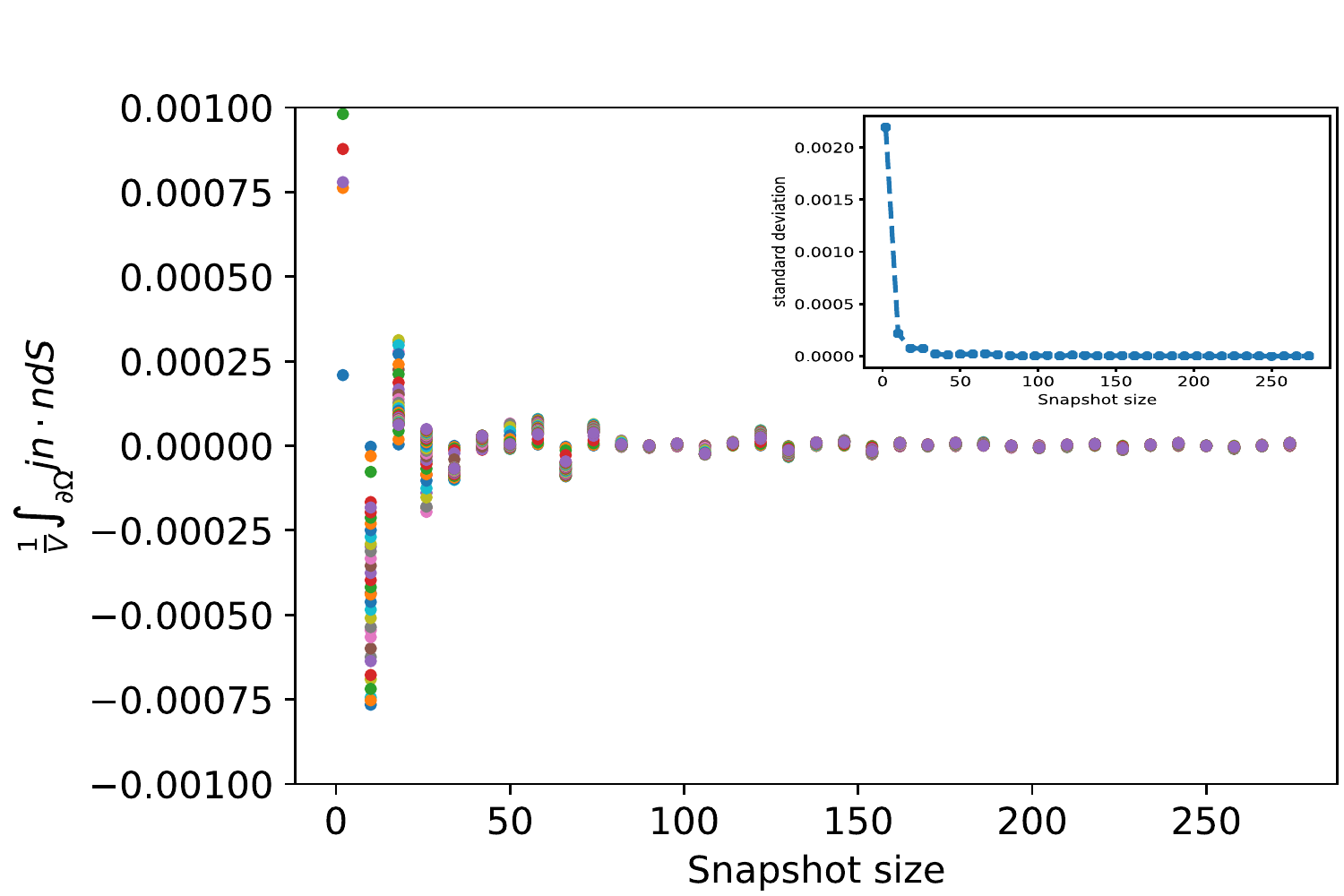} } 
\subfigure[AC: total flux for $C_2$ concentration]{\includegraphics[scale=0.45]{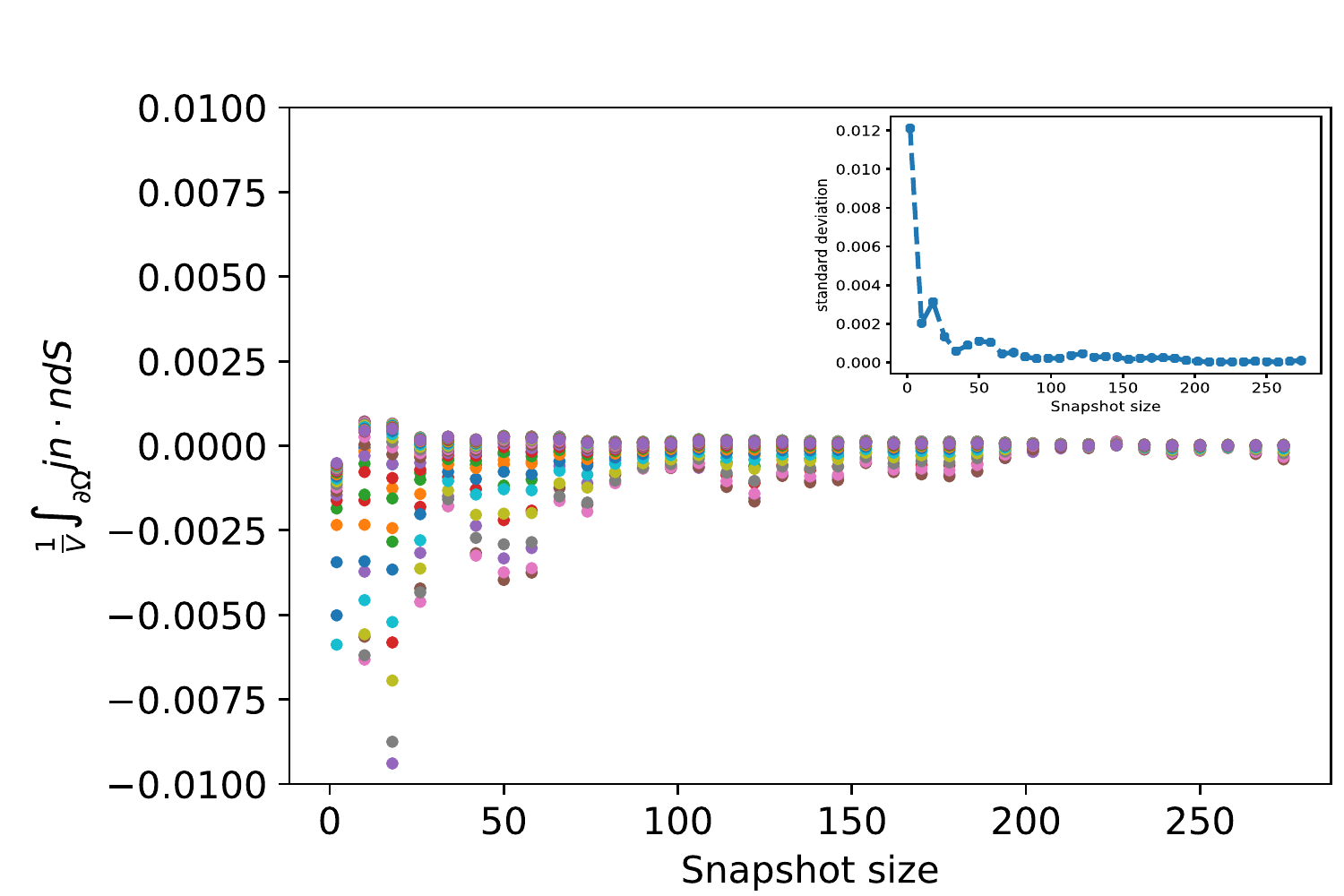} }
\caption{The total flux across the boundaries with different snapshot sizes with the Cahn-Hilliard (upper plots) and Allen-Cahn equations (lower plots) marked by different colors. The total flux converges to zero with increasing snapshot size. The embedded subplot shows the decreasing standard deviation. }
\label{fig:zero_flux_CH_AC}
\end{figure}

\begin{figure}[hbtp]
\centering
\subfigure[First power of $C_1$ ]{\includegraphics[scale=0.53]{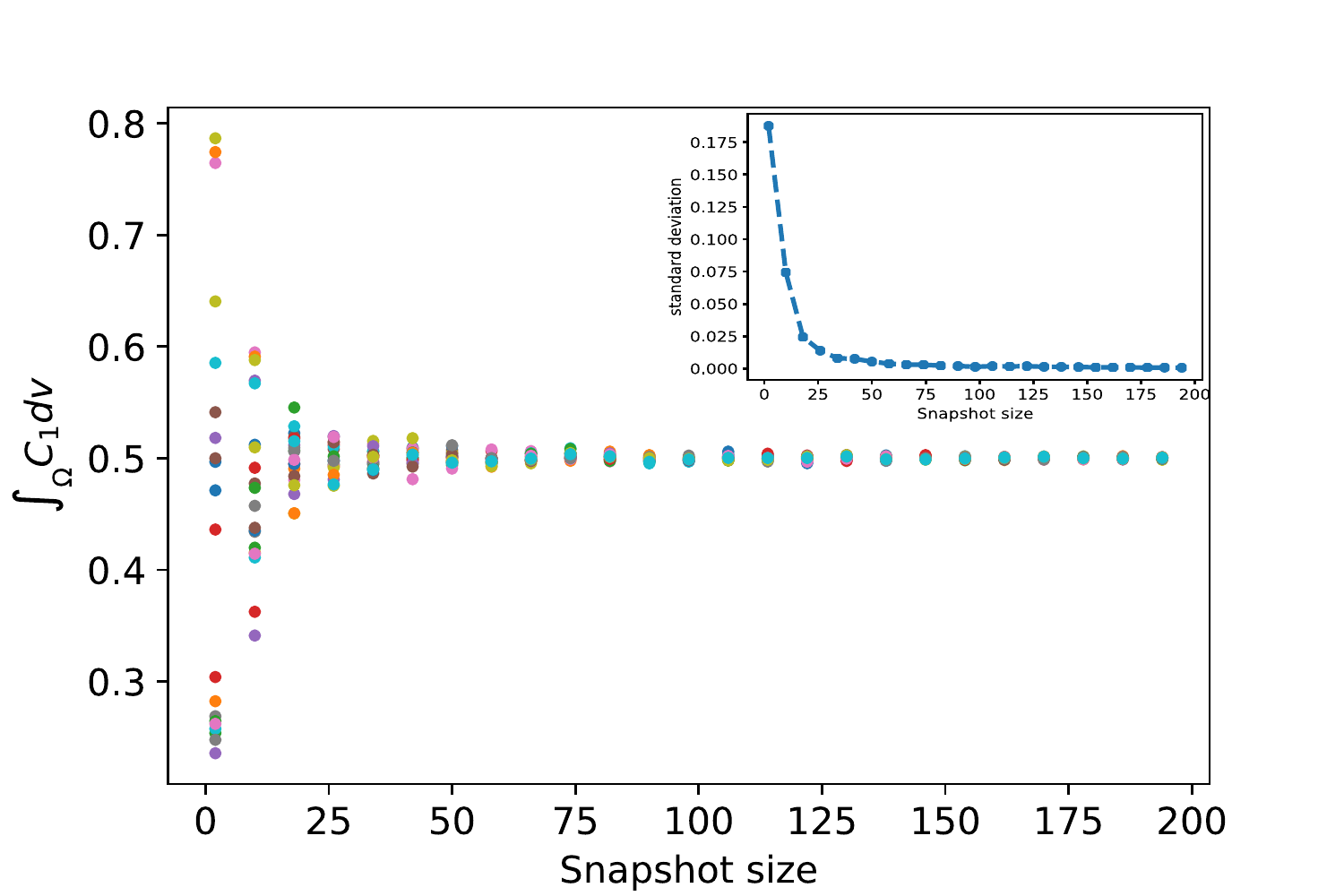}}
\subfigure[Second power  $C_1^2$ ]{\includegraphics[scale=0.53]{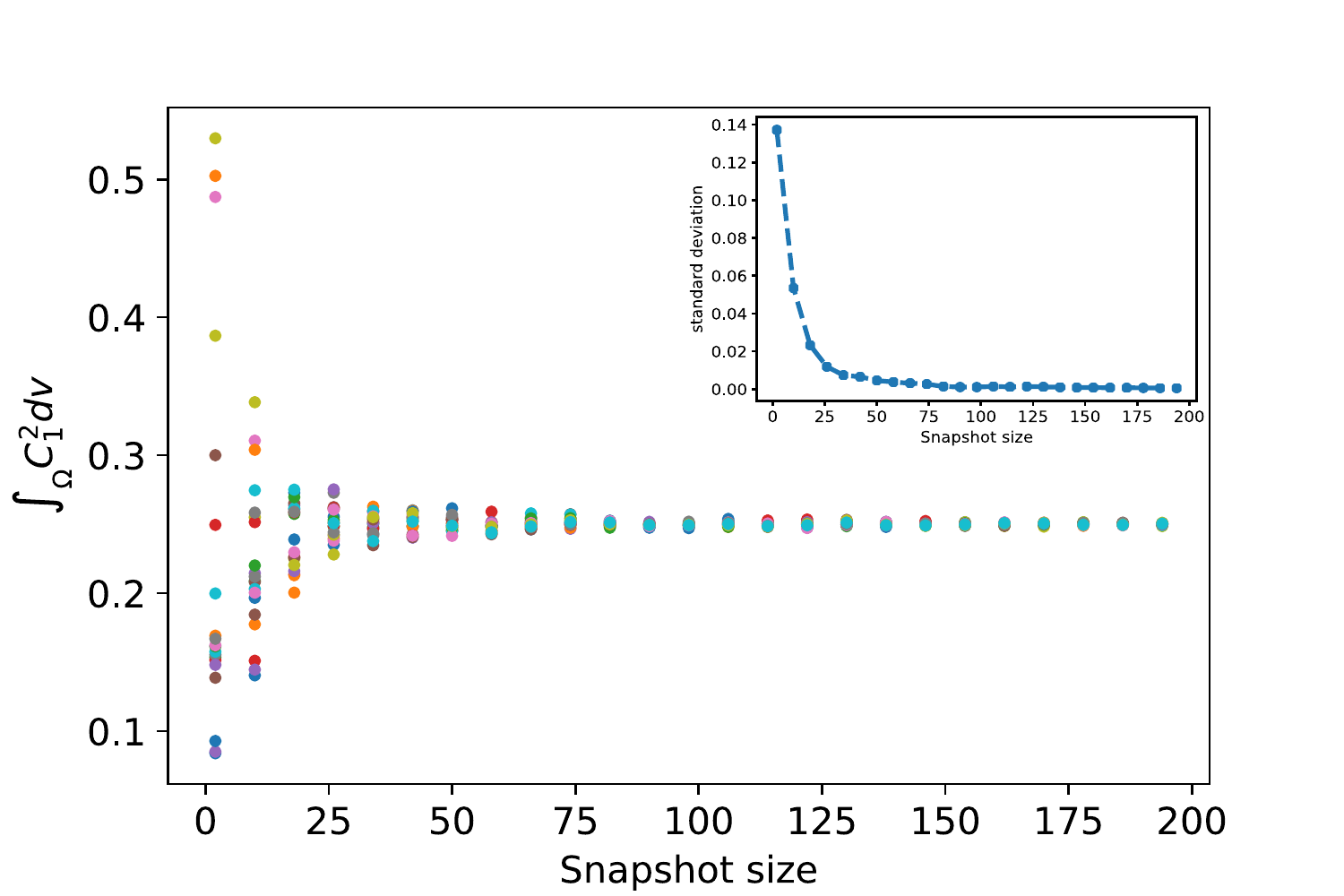}}
\subfigure[First power of $C_2$ ]{\includegraphics[scale=0.53]{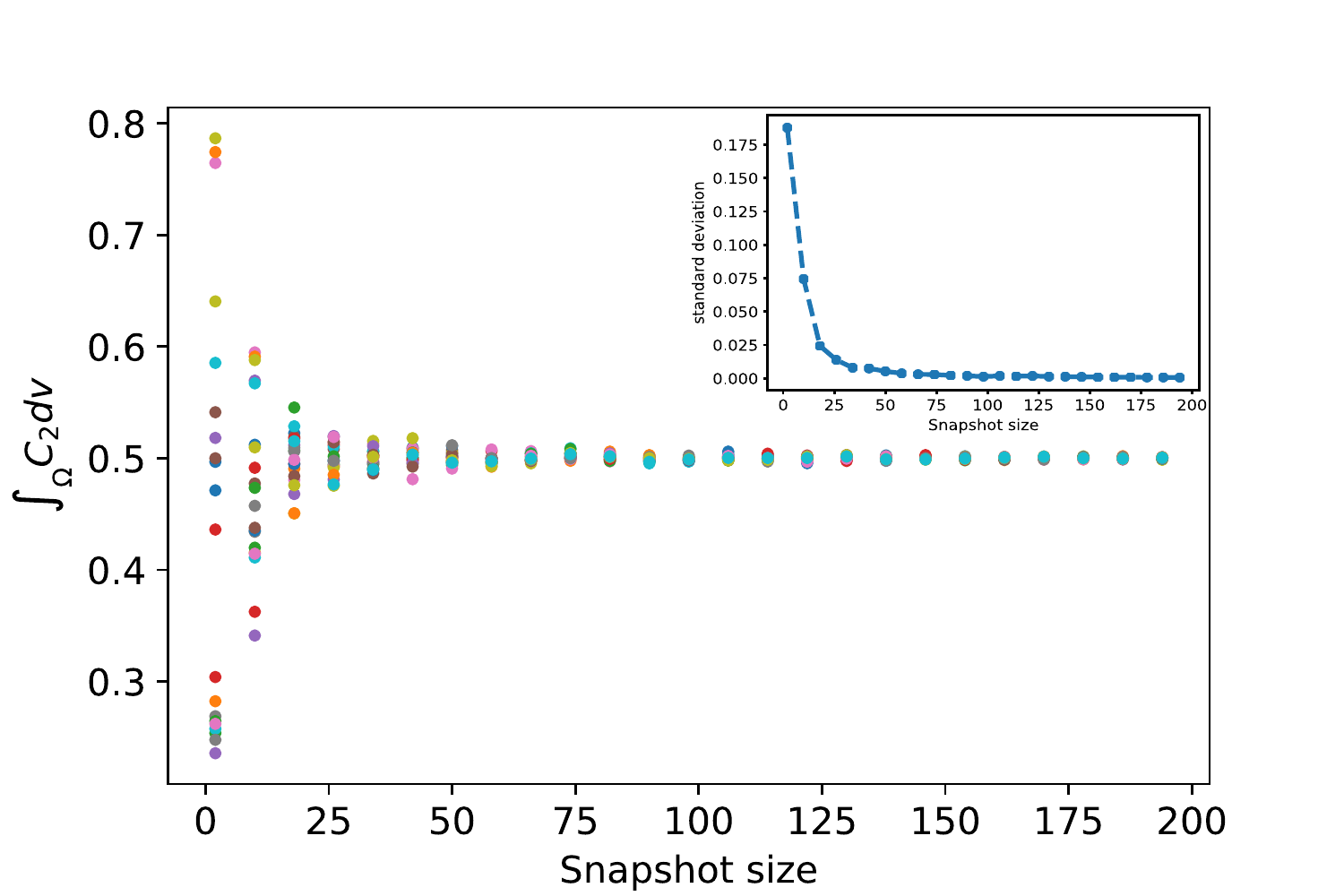}}
\subfigure[Second power  $C_2^2$]{\includegraphics[scale=0.53]{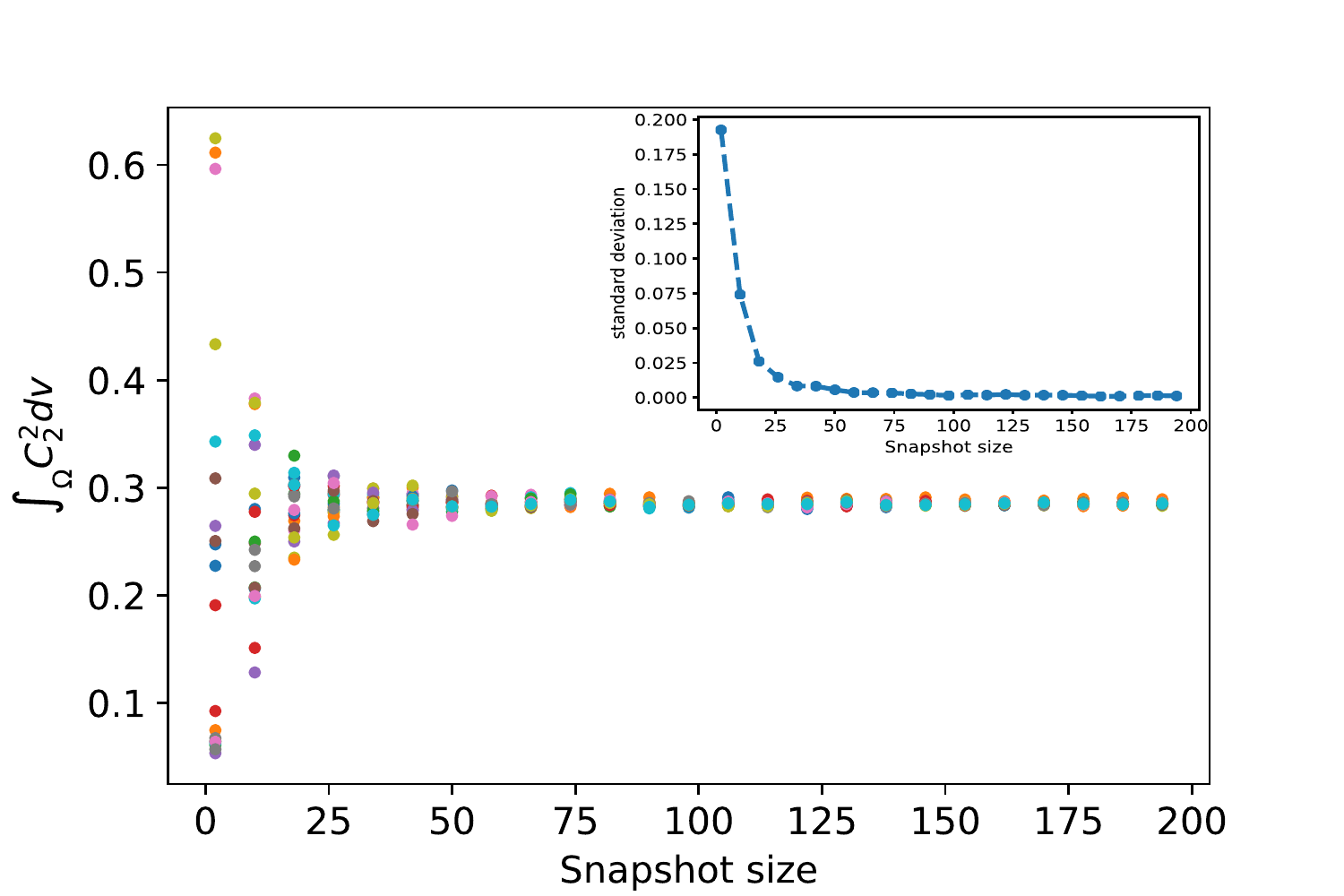}}
\caption{The powers evaluated using data generated by Cahn-Hilliard equations from 30 simulations, marked by different colors, with different randomized initial conditions modeling data from different specimens. The embedded subplot shows the decreasing standard deviation of moments. In the main and sub-plot, note the convergence with increasing snapshot size.  }
\label{fig:moments_multiple_sample_CH}
\end{figure}

\begin{figure}[hbtp]
\centering
\subfigure[First power of $C_1$ ]{\includegraphics[scale=0.53]{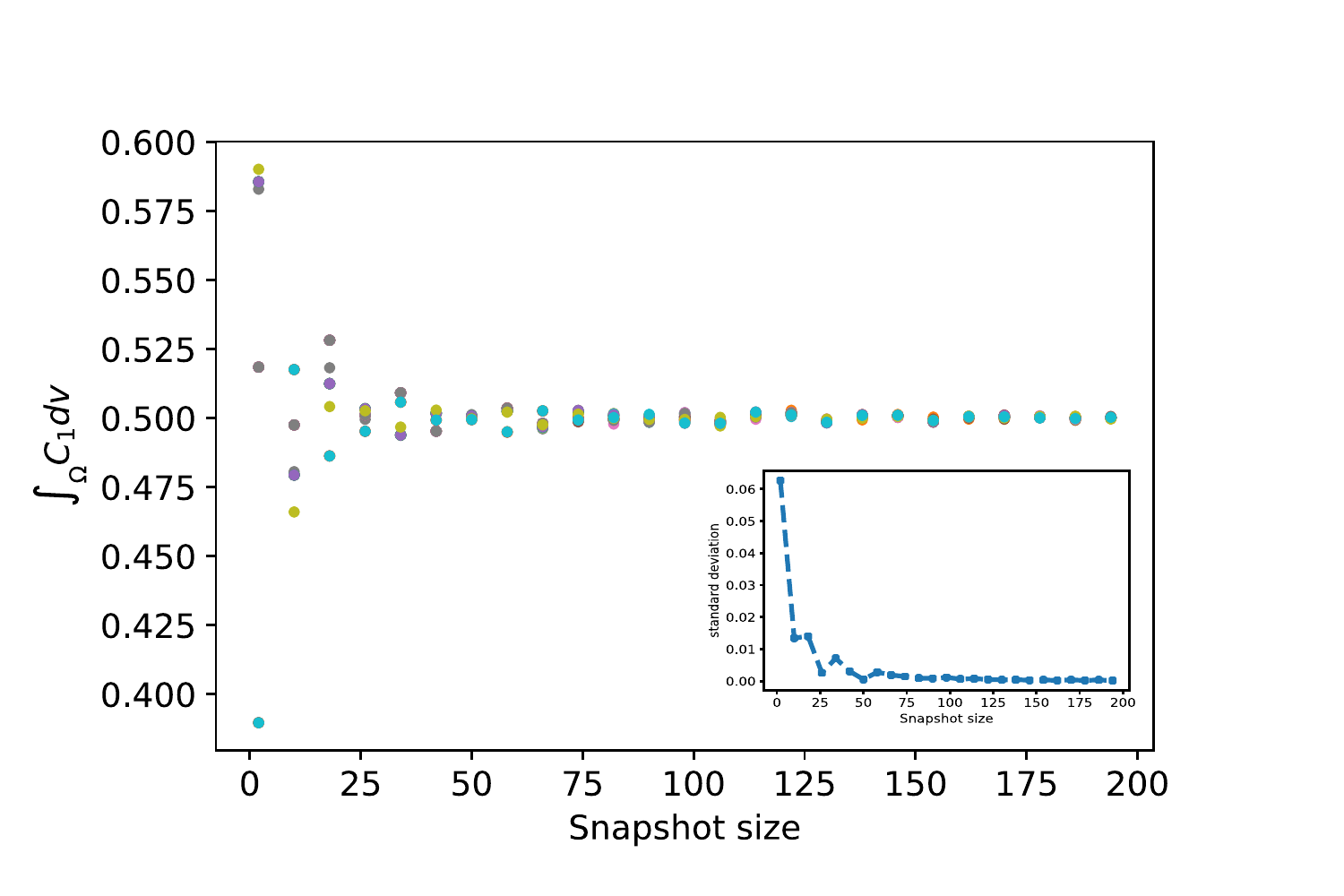}}
\subfigure[Second power  $C_1^2$ ]{\includegraphics[scale=0.53]{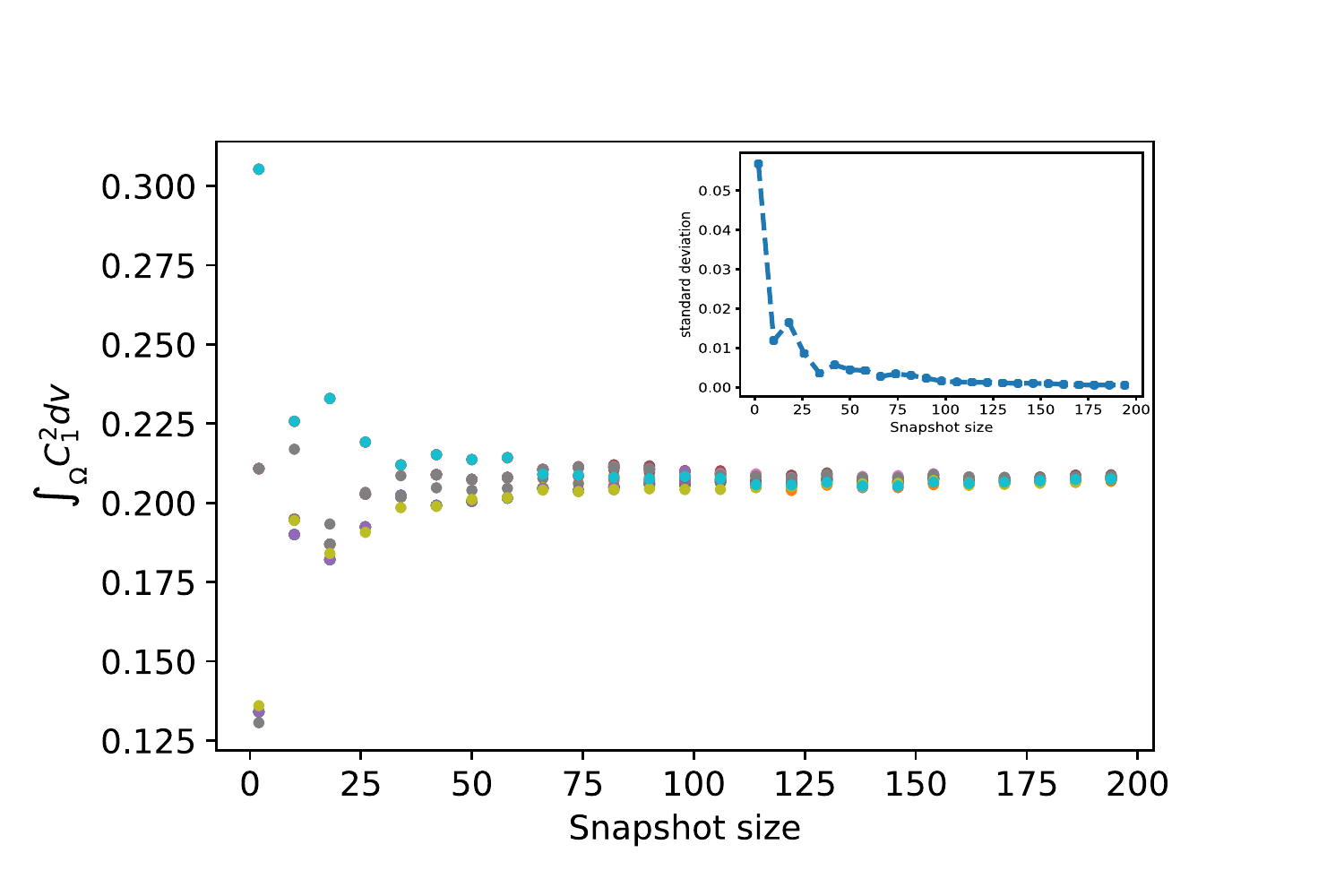}}
\subfigure[First power of $C_2$ ]{\includegraphics[scale=0.53]{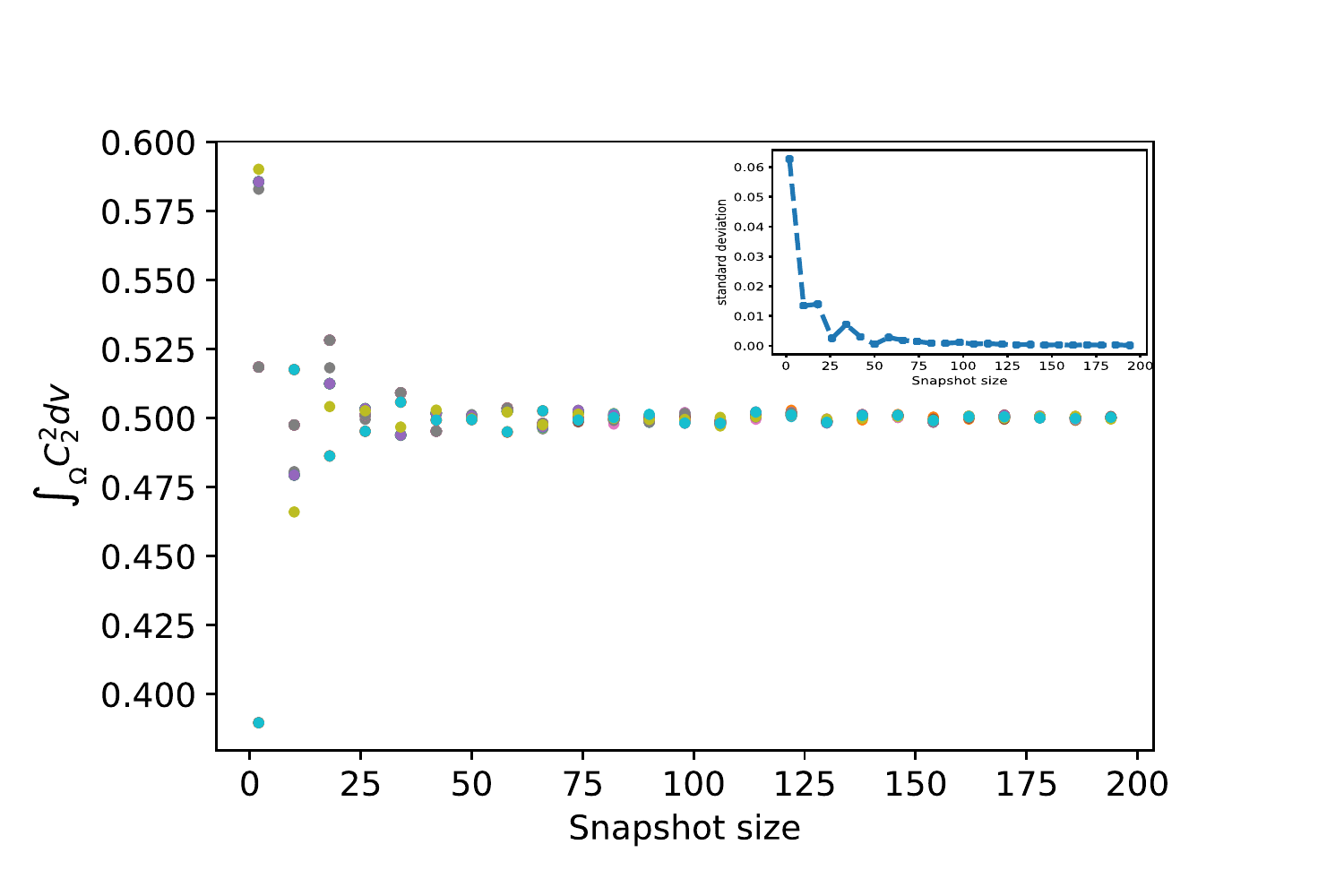}}
\subfigure[Second power  $C_2^2$]{\includegraphics[scale=0.53]{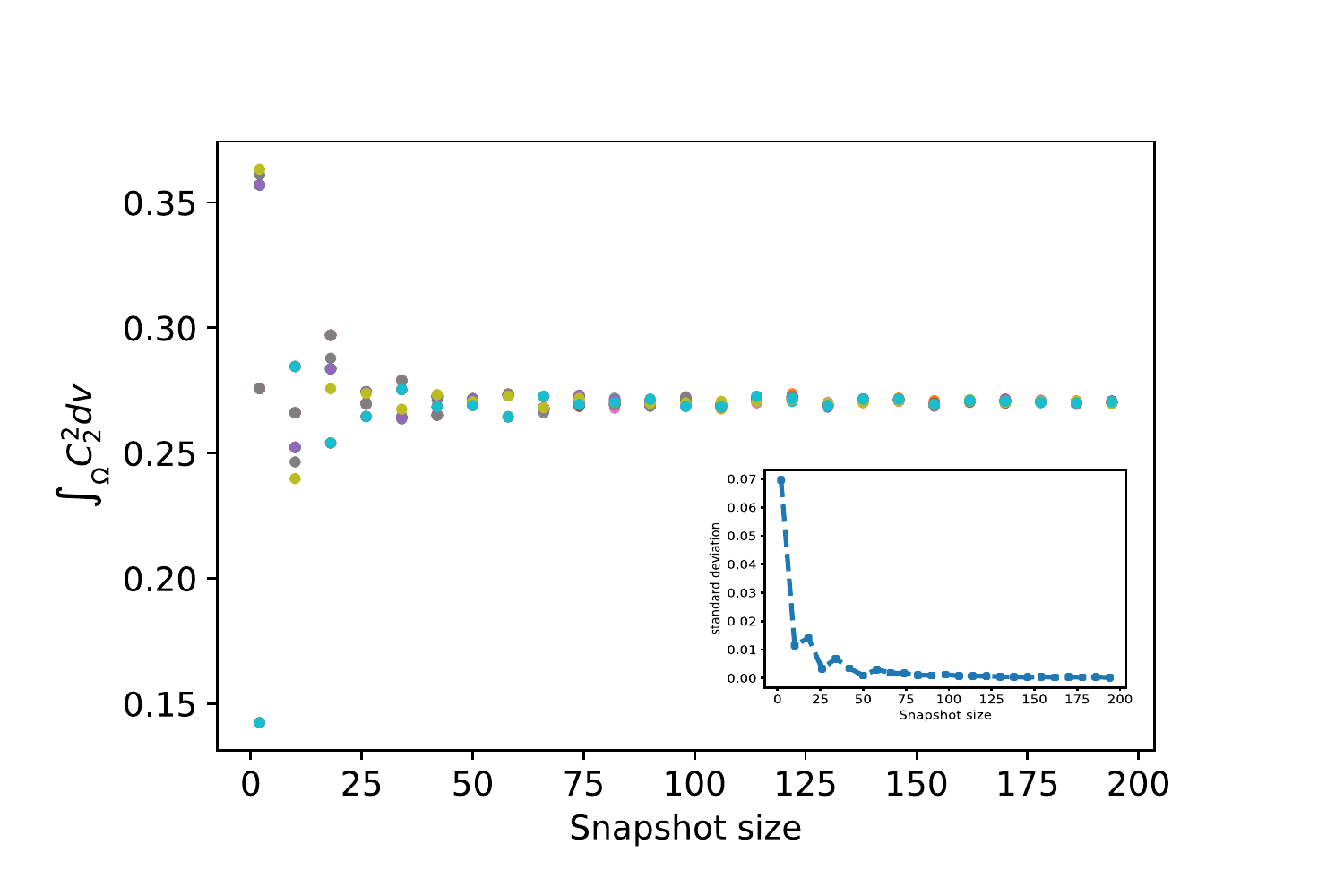}}
\caption{The powers evaluated using data generated by Allen-Cahn equations from 30 simulations, marked by different colors, with different randomized initial conditions modeling data from different specimens. The embedded subplot shows the decreasing standard deviation of moments. In the main and sub-plot, note the convergence with increasing snapshot size.  }
\label{fig:moments_multiple_sample_AC}
\end{figure}

Figure \ref{fig:patterns_CH_AC} shows the patterns formed by Cahn-Hilliard and Allen-Cahn equations, with ``particles" and short stripes randomly distributed over the domain.
We first evaluate the total flux of snapshots of different sizes shown in Figure \ref{fig:zero_flux_CH_AC}. As the snapshot size increases, the flux scaled by its volume vanishes for both Cahn-Hilliard and Allen-Cahn equations. This validates the relation (\ref{eq:statsimregioneps_appendix}).
Figures \ref{fig:moments_multiple_sample_CH} and \ref{fig:moments_multiple_sample_AC} show the first and second powers of $C_1$ and $C_2$, evaluated using data with different sizes of snapshots from separate simulations of Cahn-Hilliard and Allen-Cahn equations. Random perturbations in the initial conditions affect the local features of $C_1$ and $C_2$; however, with increasing size of snapshots their similarity becomes evident. Consequently powers evaluated using data from snapshots converge to the values obtained by using full field data on the entire domain. This validates Relation (\ref{eq:statsimregioneps_appendix}).

\section{System identification of models with small parameters}
\label{sec:Appendix:small_parameter}
In this section we studied the diffusion-reaction system (Model 1 in the main document) with the same reaction term, but smaller coefficients for diffusion terms. We restate the model here:
\begin{align}
\frac{\partial C_1}{\partial t}&=D_1\nabla^2C_1+R_{10}+R_{11}C_1+R_{13}C_1^2C_2\\
\frac{\partial C_2}{\partial t}&=D_2\nabla^2C_2+R_{20}+R_{21}C_1^2C_2\\
\text{with}& \quad \nabla C_1\cdot\bn=0; \quad \nabla C_2\cdot\bn=0 \text{ on }\Gamma 
\end{align}
where $\Gamma$ is the domain boundary. 

We generate  two data sets using models i) with $D_1=0.1$ and $D_2=4$ and ii) with $D_1=0.001$ and $D_2=0.04$. The coefficients used for reaction terms are summarized in Table \ref{ta:parametersDR}. The data are again collected at 30 time steps, each over a spatially unrelated snapshot from a different simulation. We also collect one steady state data set for the case with $D_1=0.001$ and $D_2=0.04$.
\begin{table}[h]
\centering
 \begin{tabular}{|c|c|c|c|c|}
 \hline
$R_{10}$ & $R_{11}$ & $R_{13}$ & $R_{20}$ &$R_{21}$\\\hline
0.1&-1&1&0.9&-1\\ \hline
\end{tabular}
\caption{True parameter values used in DNS for synthesizing test data.}
\label{ta:parametersDR}
\end{table}

Using the dynamic data, we successfully identify all the reaction terms for both cases in Stage 1 (See Tables \ref{ta:tab:results_model1_smallD_1} and \ref{ta:tab:results_model1_smallD_2}). This is because the two stage VSI eliminates the effects of gradient-dependent operators at the first stage. For the case with relatively higher diffusivities, we are able to identify the diffusion term using the $100\times 100$ and larger snapshots. However with the $50\times 50$ dataset, the Laplacian operator is incorrectly eliminated and another operator is selected eventually. For the case with $D_1=0.001, D_2=0.04$,  VSI failed to identify the Laplacian operator using all datasets. However, given access to steady state data, the Laplacian operator can be uncovered correctly using the Confirmation Test for the case with $D_1=0.001, D_2=0.04$ (See Table \ref{ta:tab:results_model1_smallD_2_steady}). This is because the high spatial resolution of the data at steady state can much more accurately resolve gradient operators.

It is challenging to identify operators that have much smaller contributions (e.g. the operators with extremely small coefficients). This is particularly true with noise that may wash out the true values of the operators constructed from the data. Splitting the process into two stages by the judicious choices of weighting functions can eliminate the effects of certain operators at different stages. Recently, it has been shown that variational system identification can successfully identify the model using data with large levels of noise (greater than 10$\%$) by exploiting the form of the weighting function \cite{Bortz2020weakSINDY,Bortz2020weakSINDYPDE}.  
\begin{table}[h]
\centering
\footnotesize
 \begin{tabular}{c|c}
 \hline
 mesh size& results\\
 \hline
\multirow{2}{*}{$400\times 400$} &$\int_{\Omega}w_1\frac{\partial C_1}{\partial t}\text{d}v=\int_{\Omega}-0.1004\nabla w_1 \cdot \nabla C_1\text{d}v
+\int_{\Omega}w_1 (0.10006-0.99995C_1+0.99994C_1^2C_2)\text{d}v$\\
 & $\int_{\Omega}w_2\frac{\partial C_2}{\partial t}\text{d}v=\int_{\Omega}-3.85\nabla w_2\cdot\nabla C_2\text{d}v+\int_{\Omega}w_2(0.9005-1.0004C_1^2C_2)\text{d}v$ \\
\hline
\multirow{2}{*}{$300\times 300$} &$\int_{\Omega}w_1\frac{\partial C_1}{\partial t}\text{d}v=\int_{\Omega}-0.1005\nabla w_1\cdot\nabla C_1\text{d}v
+\int_{\Omega}w_1 (0.09998-0.9999C_1+0.9999C_1^2C_2)\text{d}v$ \\ 
  & $\int_{\Omega}w_2\frac{\partial C_2}{\partial t}\text{d}v=\int_{\Omega}-3.85\nabla w_2\cdot \nabla C_2\text{d}v+\int_{\Omega}w_2(0.9007-0.9998C_1^2C_2)\text{d}v$  \\ \hline
\multirow{2}{*}{$200\times 200$} &$\int_{\Omega}w_1\frac{\partial C_1}{\partial t}\text{d}v=\int_{\Omega}-0.05\nabla \cdot w_1 C_1\nabla C_1\text{d}v
+\int_{\Omega}w_1 (0.10005-1.0003C_1+1.0001C_1^2C_2)\text{d}v$ \\ 
  & $\int_{\Omega}w_2\frac{\partial C_2}{\partial t}\text{d}v=\int_{\Omega}-3.86\nabla w_2\cdot \nabla C_2\text{d}v+\int_{\Omega}w_2(0.9005-0.9999C_1^2C_2)\text{d}v$\\ \hline
\multirow{2}{*}{$100\times 100$} & $\int_{\Omega}w_1\frac{\partial C_1}{\partial t}\text{d}v=\int_{\Omega}-0.1003\nabla w_1 \cdot \nabla C_1\text{d}v
+\int_{\Omega}w_1 (0.09975-0.9997C_1+0.9997C_1^2C_2)\text{d}v$  \\ 
&$\int_{\Omega}w_2\frac{\partial C_2}{\partial t}\text{d}v=\int_{\Omega}-1.34\nabla w_2\cdot C_1^2\nabla C_2\text{d}v+\int_{\Omega}w_2(0.9001-1.0001C_1^2C_2)\text{d}v$ \\
\hline
\multirow{2}{*}{$50\times 50$}  & $\int_{\Omega}w_1\frac{\partial C_1}{\partial t}\text{d}v=\int_{\Omega}0.0076\nabla w_1 \cdot C_1^2\nabla C_1\text{d}v
+\int_{\Omega}w_1 (0.0999-0.9991_1+0.9993C_1^2C_2)\text{d}v$  \\ 
&$\int_{\Omega}w_2\frac{\partial C_2}{\partial t}\text{d}v=\int_{\Omega}-1.48\nabla w_2\cdot C_1^2\nabla C_2\text{d}v+\int_{\Omega}w_2(0.8997-0.99994C_1^2C_2)\text{d}v$ \\ \hline
\end{tabular}
\caption{Results using noise-free dynamic data generated from Model 1 with $D_1=0.1, D_2=4$.}
\label{ta:tab:results_model1_smallD_1}
\end{table}


\begin{table}[h]
\centering
\footnotesize
 \begin{tabular}{c|c}
 \hline
 mesh size& results\\
 \hline
\multirow{2}{*}{$400\times 400$} &$\int_{\Omega}w_1\frac{\partial C_1}{\partial t}\text{d}v=\int_{\Omega}-0.00161\nabla w_1 \cdot C_1C_2^2\nabla C_1\text{d}v
+\int_{\Omega}w_1 (0.100038-1.00003C_1+1.000035C_1^2C_2)\text{d}v$\\
 & $\int_{\Omega}w_2\frac{\partial C_2}{\partial t}\text{d}v=\int_{\Omega}-0.043\nabla w_2\cdot C_1^2C_2\nabla C_2\text{d}v+\int_{\Omega}w_2(-0.8999-1.00005C_1^2C_2)\text{d}v$ \\
\hline
\multirow{2}{*}{$300\times 300$} &$\int_{\Omega}w_1\frac{\partial C_1}{\partial t}\text{d}v=\int_{\Omega}-0.00162\nabla w_1 \cdot C_1C_2^2\nabla C_1\text{d}v
+\int_{\Omega}w_1 (0.09995-0.99993C_1+1.00004C_1^2C_2)\text{d}v$ \\ 
  & $\int_{\Omega}w_2\frac{\partial C_2}{\partial t}\text{d}v=\int_{\Omega}-0.044\nabla w_2\cdot C_1^2C_2\nabla C_2\text{d}v+\int_{\Omega}w_2(-0.9001-1.0002C_1^2C_2)\text{d}v$  \\ \hline
\multirow{2}{*}{$200\times 200$} &$\int_{\Omega}w_1\frac{\partial C_1}{\partial t}\text{d}v=\int_{\Omega}-0.00164\nabla w_1 \cdot C_1C_2^2\nabla C_1\text{d}v
+\int_{\Omega}w_1 (0.0997-0.9995C_1+0.9996C_1^2C_2)\text{d}v$ \\ 
  & $\int_{\Omega}w_2\frac{\partial C_2}{\partial t}\text{d}v=\int_{\Omega}-0.044\nabla w_2\cdot C_1^2C_2\nabla C_2+\int_{\Omega}w_2(-0.9001-1.0002C_1^2C_2)\text{d}v$\\ \hline
\multirow{2}{*}{$100\times 100$} & $\int_{\Omega}w_1\frac{\partial C_1}{\partial t}\text{d}v=\int_{\Omega}-0.00064\nabla w_1 \cdot C_1^2C_2\nabla C_1\text{d}v
+\int_{\Omega}w_1 (0.09975-1.0012C_1+1.0014C_1^2C_2)\text{d}v$  \\ 
&$\int_{\Omega}w_2\frac{\partial C_2}{\partial t}\text{d}v=\int_{\Omega}-0.044\nabla w_2\cdot C_1\nabla C_2\text{d}v+\int_{\Omega}w_2(-0.901-1.0003C_1^2C_2)\text{d}v$ \\
\hline
\multirow{2}{*}{$50\times 50$}  & $\int_{\Omega}w_1\frac{\partial C_1}{\partial t}\text{d}v=\int_{\Omega}-0.0017\nabla w_1 \cdot C_1^2C_2\nabla C_1\text{d}v
+\int_{\Omega}w_1 (-0.0999-0.9991C_1+0.9993C_1^2C_2)\text{d}v$  \\ 
&$\int_{\Omega}w_2\frac{\partial C_2}{\partial t}\text{d}v=\int_{\Omega}-0.024\nabla w_2\cdot C_1^2\nabla C_1\text{d}v+\int_{\Omega}w_2(-0.898-0.9991C_1^2C_2)\text{d}v$ \\ \hline
\end{tabular}
\caption{Results using noise-free dynamic data generated from Model 1 with $D_1=0.001, D_2=0.04$.}
\label{ta:tab:results_model1_smallD_2}
\end{table}


\begin{table}[h]
\centering
\footnotesize
 \begin{tabular}{c|c}
 \hline
 mesh size& results\\
 \hline
\multirow{2}{*}{$400\times 400$} &$\int_{\Omega}w_1\frac{\partial C_1}{\partial t}\text{d}v=\int_{\Omega}-0.00100005\nabla w_1 \cdot \nabla C_1\text{d}v
+\int_{\Omega}w_1 (0.100038-1.00003C_1+1.000035C_1^2C_2)\text{d}v$\\
 & $\int_{\Omega}w_2\frac{\partial C_2}{\partial t}\text{d}v=\int_{\Omega}-0.040002\nabla w_2\cdot\nabla C_2\text{d}v+\int_{\Omega}w_2(-0.8999-1.00005C_1^2C_2)\text{d}v$ \\
\hline
\multirow{2}{*}{$300\times 300$} &$\int_{\Omega}w_1\frac{\partial C_1}{\partial t}\text{d}v=\int_{\Omega}-0.0010003\nabla w_1 \cdot \nabla C_1\text{d}v
+\int_{\Omega}w_1 (0.09995-0.99993C_1+1.00004C_1^2C_2)\text{d}v$ \\ 
  & $\int_{\Omega}w_2\frac{\partial C_2}{\partial t}\text{d}v=\int_{\Omega}-0.04000867\nabla w_2\cdot \nabla C_2\text{d}v+\int_{\Omega}w_2(-0.9001-1.0002C_1^2C_2)\text{d}v$  \\ \hline
\multirow{2}{*}{$200\times 200$} &$\int_{\Omega}w_1\frac{\partial C_1}{\partial t}\text{d}v=\int_{\Omega}-0.00099984\nabla w_1 \cdot \nabla C_1\text{d}v
+\int_{\Omega}w_1 (0.0997-0.9995C_1+0.9996C_1^2C_2)\text{d}v$ \\ 
  & $\int_{\Omega}w_2\frac{\partial C_2}{\partial t}\text{d}v=\int_{\Omega}-0.04000868\nabla w_2\cdot \nabla C_2+\int_{\Omega}w_2(-0.9001-1.0002C_1^2C_2)\text{d}v$\\ \hline
\multirow{2}{*}{$100\times 100$} & $\int_{\Omega}w_1\frac{\partial C_1}{\partial t}\text{d}v=\int_{\Omega}-0.00100208\nabla w_1 \cdot \nabla C_1\text{d}v
+\int_{\Omega}w_1 (0.09975-1.0012C_1+1.0014C_1^2C_2)\text{d}v$  \\ 
&$\int_{\Omega}w_2\frac{\partial C_2}{\partial t}\text{d}v=\int_{\Omega}-0.04001184\nabla w_2\cdot \nabla C_2\text{d}v+\int_{\Omega}w_2(-0.901-1.0003C_1^2C_2)\text{d}v$ \\
\hline
\multirow{2}{*}{$50\times 50$}  & $\int_{\Omega}w_1\frac{\partial C_1}{\partial t}\text{d}v=\int_{\Omega}-0.00099981\nabla w_1 \cdot\nabla C_1\text{d}v
+\int_{\Omega}w_1 (-0.0999-0.9991C_1+0.9993C_1^2C_2)\text{d}v$  \\ 
&$\int_{\Omega}w_2\frac{\partial C_2}{\partial t}\text{d}v=\int_{\Omega}-0.0399638\nabla w_2\cdot \nabla C_1\text{d}v+\int_{\Omega}w_2(-0.898-0.9991C_1^2C_2)\text{d}v$ \\ \hline
\end{tabular}
\caption{Results using noise-free steady state and dynamic data generated from Model 1 with $D_1=0.001, D_2=0.04$.}
\label{ta:tab:results_model1_smallD_2_steady}
\end{table}

\end{document}